# A System of Interaction and Structure

Alessio Guglielmi

University of Bath
Bath BA2 7AY
United Kingdom
http://alessio.guglielmi.name

**Abstract** *This paper introduces a logical system, called* BV*, which extends multiplicative linear logic by a non-commutative self-dual logical operator. This extension is particularly challenging for the sequent calculus, and so far it is not achieved therein. It becomes very natural in a new formalism, called the* calculus of structures*, which is the main contribution of this work. Structures are formulae subject to certain equational laws typical of sequents. The calculus of structures is obtained by generalising the sequent calculus in such a way that a new top-down symmetry of derivations is observed, and it employs inference rules that rewrite inside structures at any depth. These properties, in addition to allowing the design of* BV*, yield a modular proof of cut elimination.*

**Table of Contents**

## 1 Introduction

In the beginning of this enterprise I only wanted to define a deductive system mixing commutative and non-commutative logical operators. The specific variant I was looking for is a conservative extension of multiplicative linear logic plus mix, obtained by adding a self-dual non-commutative connective. This problem has been studied mainly by Retoré [26, 27]: he has proof nets for his logic, which is called *pomset logic*, and cut elimination therein, but, despite many efforts, nobody has been able so far to provide a sequent system for that logic. My challenge was to bring pomset logic into the realm of the sequent calculus, for two reasons: 1) It is interesting to see *why* such a simple logic requires so much effort to be expressed in the most universal proof-theoretical formalism. 2) If one wants to extend pomset logic to more expressive logics, then the sequent calculus usually is a better formalism than proof nets, because it is more versatile, for example with exponentials and additives.

What my colleagues and I found in this research has been surprising: there are deep reasons for this kind of logic not to be expressible in the sequent calculus, and there is a simple formalism, which we call the *calculus of structures*, that is instead able to express self-dual non-commutativity with great ease. In fact, self-dual non-commutative operators naturally generate a class of provable formulae whose proofs depend on accessing subformulae at arbitrary depths, what makes for an impossible challenge to the sequent calculus. The new formalism is more general than the sequent calculus for logics with involutive negation, like classical and linear logic, and allows a much more refined proof theory than possible in the sequent calculus, without sacrificing simplicity. In particular, rules in the calculus of structures can access arbitrarily nested subformulae.

This paper represents the beginning of our research program in the calculus of structures, and focuses on designing a deductive system with the characteristics mentioned above for pomset logic. It is still open whether the logic in this paper, called BV, is the same as pomset logic. We conjecture that it is actually the same logic, but one crucial step is still missing, at the time of this writing, in the equivalence proof. This paper is the first in a planned series of three papers dedicated to BV. In the present, first part, I obtain BV from a certain semantic idea, and I show cut elimination for it. In the second part, Alwen Tiu will show why BV cannot be defined in any sequent system [35, 36]. In the third part, some of my colleagues will hopefully show the equivalence of BV and pomset logic, this way explaining why it was impossible to find a sequent system for pomset logic. Some of the results in this paper have been already shown by Straßburger and myself in [13, 14] and used by Bruscoli in [10].

Pomset logic naturally derives from the study of coherence spaces for multiplicative linear logic (see Girard's [12]), and its self-dual operator has a close correspondence to sequential operators as defined, for example, in process algebras. The cut elimination procedure in the proof nets of pomset logic gets sequentialised by the non-commutative links; this naturally induces a computational model where sequentiality plays a role as important as parallelism, which is interesting in the light of the Curry-Howard correspondence. Non-commutative logics are also important in linguistics, their use dating back to the Lambek calculus [19]. As well as from its semantic origins, BV can be motivated by the same reasons as for pomset logic: essentially, non-commutativity brings sequentiality, and sequentiality is an archetypal algebraic composition both for human and computer languages. Self-dual non-commutativity captures very precisely the sequentiality notion of many process algebras, CCS [21] in the first place, as Bruscoli shows in [10]. So, at a very concrete level, this paper is about a simple logic that can help understand several linguistic phenomena. For it, I provide a simple deductive system and prove cut elimination.

On the other hand, at a different level, this paper only uses BV as an experiment for tuning up a workbench where other ambitious proof-theoretical problems could be studied. I present here a new formalism, the calculus of structures, and some of the proof-theoretical methods associated to it. The calculus of structures is a generalisation of the one-sided sequent calculus, which is also called the *Gentzen-Schütte* calculus, (see, for example, [37]). Many logics with involutive negation and De Morgan laws can be defined in the one-sided sequent calculus: in that case, translating them into the calculus of structures is a trivial (and also uninteresting) operation. What makes the calculus of structures interesting is the possibility of defining logics by employing concepts fundamentally different from those in the sequent calculus. We can isolate two ideas:

1 *Deep inference*: inference rules in the calculus of structures operate anywhere inside expressions, they are not confined to the outermost subformulae around the roots of formula trees.

2 *Top-down symmetry*: contrary to the sequent calculus, natural deduction and other formalisms where derivations are essentially based on trees, in our new formalism derivations can be flipped upside-down and negated, and they remain valid (this corresponds to $A \rightarrow B$ being equivalent to $\bar{B} \rightarrow \bar{A}$).

These ideas find a simple technical realisation by employing the unifying notion of 'structure'.

I borrowed the terminology from Belnap's [23], and in general from the tradition in philosophical and substructural logics [25]: a *structure* is an expression intermediate between a (one-sided) sequent and a formula. More precisely, it is an ordinary logical formula modulo an equational theory of the kind typically imposed on sequents. From a practical viewpoint, logical connectives disappear (in particular connectives at the root of formula trees), and all logical rules become *structural* (as opposed to *logical*), in the sense that they deal with the relative position of substructures in a structure. Structures are the only kind of expression allowed, and inference rules are simply rewriting rules on structures, whence the name 'calculus of structures'. Of course, the freedom allowed by this formalism is dangerous. One could use it in a wild way, and lose every proof-theoretical property. I will be very cautious in defining rules, and actually the main part of our research is understanding and defining the methodologies necessary for making an educated and profitable use of the new freedom.

The main point to note is that in the calculus of structures it is possible to do proof theory: we can define a cut rule, and cut elimination makes sense, just as in the sequent calculus. Moreover, there is an analogue to the subformula property, which guarantees that any given rule (except for the cut and perhaps other special rules) has finite applicability. These features make the calculus of structures closer to the sequent calculus than to any other formalism. On the other hand, cut elimination in the calculus of structures is completely different from that in the sequent calculus, at least when systems are designed by using the deep inference feature. In fact, in the sequent calculus the cut elimination procedure depends crucially on the existence of a root connective, and this is not the case in the calculus of structures. The complication comes essentially from deep inference; luckily, the top-down symmetry contributes to simplifying the situation considerably. As a consequence of the symmetry, in fact, we have that the cut rule can always be reduced trivially to its atomic version, i.e., the principal formulae are atoms. This is the perfect dual of the typical sequent calculus fact that generic identity rules can be replaced by their atomic counterparts.

In this paper I introduce one of the two existing techniques for proving cut elimination in the calculus of structures, and I call it *splitting* (the other one is called *decomposition*, see [13, 14], and it is best used in conjunction with splitting). An important difference of the calculus of structures with respect to the sequent calculus is that the cut rule can be equivalently divided into several rules. Often, only one of these rules is infinitary, the atomic cut, but all the rules in the decomposition of cut can be shown admissible. An advantage is that one can have a wide range of equivalent systems (one for each combination of admissible rules), with varying degrees of expressiveness in proof construction. All the admissible rules are in fact independent, and their admissibility can be shown independently by way of splitting. For big logical systems, like linear logic, one can easily get tens of thousands of equivalent systems without much effort [31]. This *modularity* is ultimately made possible by the top-down symmetry, and modularity is of course important both conceptually and for the typical computer science applications.

The ideas of structure, deep inference and top-down symmetry, all come from a certain combinatorial, graph-like representation of logical formulae that I call *relation webs*. Relation webs are used in this paper as both a semantics and a very abstract, distributed computational model for $\mathsf{BV}$. They make for a compact representation of formulae that: 1) independently justifies the equational theory imposed on structures; 2) induces an extremely natural notion of rewriting, which actually generates $\mathsf{BV}$ independently of coherence spaces or any other formalism. Relation webs play a crucial role in Tiu's proof of inadequacy of the sequent calculus for system $\mathsf{BV}$ [36], where deep inference is shown to be an essential feature for representing $\mathsf{BV}$—a feature that the sequent calculus does not possess. I will argue about relation webs having a broad applicability. In fact, a certain characterisation theorem for relation webs, which is crucial in our treatment, scales up to the generic case of a logic made by any number of different multiplicative logical relations.

Relation webs justify the *structure fragment* of $\mathsf{BV}$, i.e., the rules responsible for the combinatorial behaviour of the logic, independently of negation. The *interaction fragment* is made by the rules corresponding to identity and cut in the sequent calculus, i.e., the rules which depend on negation. The latter fragment also enjoys a nice interpretation in relation webs, but, at least at present, the integration of interaction and structure fragments is best studied in the calculus of structures, owing to the relatively poor development of relation webs so far. I decided to underscore the importance of interaction and structure in the paper's title because I have reasons to believe that they are actually a key to understand in a new way many proof-theoretical phenomena. I will not give any hard evidence of this belief in this paper, because we are still in a phase in which we have to collect evidence from the study of several deductive systems before attempting a more philosophical explanation of our observations. On the other hand, we studied already a number of systems and they all fit in the interaction/structure pattern outlined here; I invite the reader to form a personal opinion on this matter by reading the papers on classical [9] and linear logic [33, 31], as well as the extensions of system BV [14].

In summary, these are the contributions of this paper:

1 A new proof-theoretical formalism, the calculus of structures, which allows to express new logical systems and has proof-theoretical properties of special interest for computer science; the formalism comes with new, non-trivial techniques for proving proof-theoretical results.

2 A deductive system, called BV, for a logic which extends multiplicative linear logic with a logical operator whose interpretation is a broad notion of sequentiality.

3 A combinatorial representation of formulae, called relation web, which allows an independent, semantic justification for BV.

Section 2 introduces structures together with relation webs and an informal representation for structures that should help understanding them intuitively. Section 3 introduces the calculus of structures and derives system BV together with its cut rule starting from relation webs. System BV and cut elimination are discussed in Section 4. Finally, in Section 5, I will show how BV is a conservative extension of linear logic plus mix.

## 2 Structures and Relation Webs

The notion of structure is not complicated, but it is useful to have an intuitive idea of how it relates to traditional proof-theoretical notions. Very roughly, a structure is both a formula and a sequent, but, as we will see soon, it also captures aspects of proof nets. A good way to understand structures, at least for what matters to this paper, is through their graph representation. The first subsection is dedicated to structures and their representation, the second one to relation webs.

### 2.1 Structures

Since structures also capture some aspects of proofs, it should be appropriate to start introducing them by an example from proof construction in multiplicative linear logic, whose management by structures is also of independent interest. In a one-sided sequent presentation, the multiplicative conjunction $\otimes$ (*times*) of linear logic is defined by:

$$\otimes\,\frac{\vdash A, \Phi \quad \vdash B, \Psi}{\vdash A \otimes B, \Phi, \Psi} \quad .$$

Instances of $\otimes$ can be seen as elementary steps in a computation that corresponds to the bottom-up building of a proof. This perspective is called the *proof search as computation*, or *proof construction*, paradigm. From this point of view the rule $\otimes$ above has a serious flaw: when the rule is applied we have to decide how to split the context of the main formula, and if $n$ formulae are in the multiset $\Phi, \Psi$, there are $2^n$ ways to assign them to $\Phi$ and $\Psi$. This is an exponential source of probably unwanted non-determinism.

Of course, we could adopt an approach external to the sequent calculus, an *implementation* that controls the use of the rule *a posteriori*: no commitment is made at the time the rule is applied, and resources in the context are used based on necessity along the search for the proof, ideally updating the actual instance of the rule $\otimes$ until a valid proof is formed. This 'lazy' approach is similar to the use of most general unifiers when having to decide how to instantiate quantified variables. It is of course a very reasonable solution, already adopted by Hodas and Miller in the case of intuitionistic linear logic [15].

There still remains the curiosity of whether this problem could be solved *inside* a deductive system. We should create clusters of formulae, corresponding to contexts to be split, whence each formula could be picked up on demand. In the traditional sequent calculus this solution is not achievable by simple means. What really counts, while building a proof in the bottom-up way in the sequent calculus, is the frontier of hypotheses, whose natural structure is simply a *multiset*. Instead, we need to keep a *tree*, whose structure takes care of the nesting of clusters. Consider for example the following proof in multiplicative linear logic, where $\bar{a}$ is the negation of the atom $a$ and the connective $\parr$ is called *par*:

$$
\parr\,\dfrac{\parr\,\dfrac{\otimes\,\dfrac{\mathsf{id}\,\dfrac{}{\vdash b,\bar b}\qquad \parr\,\dfrac{\otimes\,\dfrac{\mathsf{id}\,\dfrac{}{\vdash a,\bar a}\quad \mathsf{id}\,\dfrac{}{\vdash c,\bar c}}{\vdash a,\bar a\otimes c,\bar c}}{\vdash a,(\bar a\otimes c)\parr\bar c}}{\vdash a,b,\bar b\otimes((\bar a\otimes c)\parr\bar c)}}{\vdash a,b\parr(\bar b\otimes((\bar a\otimes c)\parr\bar c))}}{\vdash a\parr(b\parr(\bar b\otimes((\bar a\otimes c)\parr\bar c)))}\quad .
$$

In this case the appropriate partitions of contexts have been found, and we got a proof.

Consider now the following two graphs:

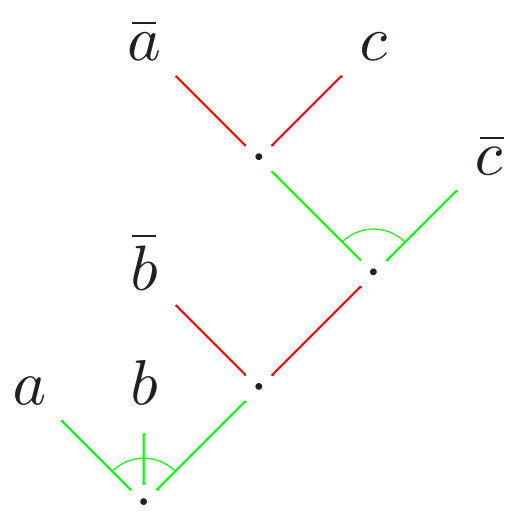

and

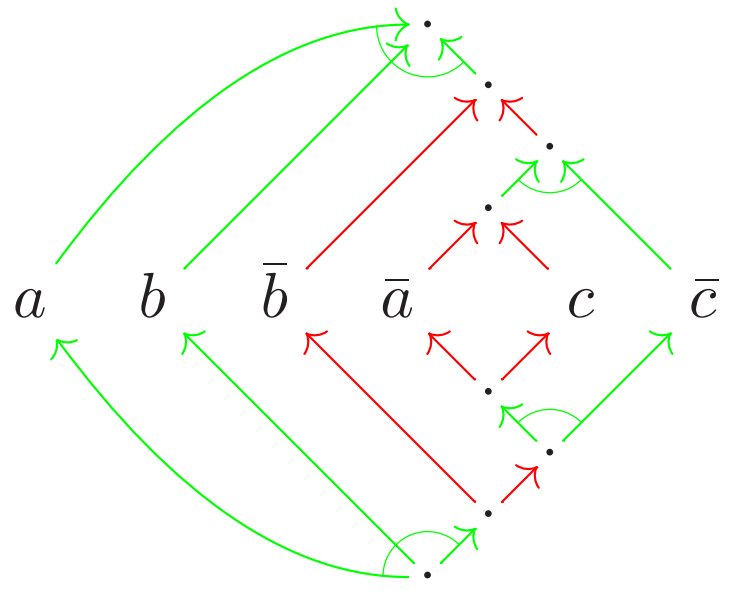

.

The tree-like structure at the left is a representation of the conclusion of the proof above. Times-connected formulae are linked to the roots by simple arcs; these formulae are in a relation that I call *copar* (instead of times). Par-connected formulae are connected to the roots by arcs that are in a special relation, represented by the bow. The only purpose of the bows is to distinguish pars from copars; I could have equivalently put bows on copar arcs. The graph at the left is more or less a proof net without the axiom links, but we can also consider it as a partial representation of the proof above, which only deals with the ⅋ rule instances. This example is important and I invite the reader to get back to it while reading this paper. The deductive mechanism we are going to see will take care of the ⊗ rule instances which are present in the multiplicative linear logic proof and absent from the graphs above.

The graph at the right is obtained from the left one by taking the horizontal mirror image of the lower part above the row of atoms and by directing arcs. Nodes that are not atoms have the only meaning of keeping things together. Being redundant, this kind of structure is probably silly to use with linear logic; and anyway, my main interest here is not in the partitioning of the times context. I will slightly generalise these structures in order to include a new, non-commutative logical relation called *seq*, for which the top-down symmetry will not be trivial any more.

Formally to deal with these graphs, I will now introduce a bit of syntax, alternative to the one of linear logic. In this language the formula to be proved above can be written $[a, b, (\bar{b}, [(\bar{a}, c), \bar{c}])]$ and expressions like this are called *structures*. Instead of using binary connectives for defining logical relations, I consider relations induced by a context, meaning that, for example, $[R_1, \dots, R_h]$ is a par structure, where the structures inside it are to be considered connected by pars. Since par is commutative and associative (and the commutativity and associativity equivalences are decidable), I will not distinguish structures based on ordering or on the nesting of a par structure into another par. The same is true for copar: a copar structure $(R_1, \dots, R_h)$ has the same properties as a par one.

**2.1.1 Definition** There are infinitely many *positive atoms* and infinitely many *negative atoms*. Atoms, no matter whether positive or negative, are denoted by $a$, $b$, $c$, .... *Structures*, denoted by $S$, $P$, $Q$, $R$, $T$, $U$, $V$ and $X$, are generated by

$$S ::= \circ \mid a \mid \langle \underbrace{S; \dots; S}_{>0} \rangle \mid [\underbrace{S, \dots, S}_{>0}] \mid (\underbrace{S, \dots, S}_{>0}) \mid \bar{S} \quad ,$$

where $\circ$, the *unit*, is not an atom. $\langle S_1; \dots; S_h \rangle$ is called a *seq structure*, $[S_1, \dots, S_h]$ is called a *par structure* and $(S_1, \dots, S_h)$ is called a *copar structure*; $\bar{S}$ is the *negation* of the structure $S$; a negated atom $\bar{a}$ is a negative atom if $a$ is positive and a positive one if $a$ is negative. We can designate a special atom as a *hole*, denoted by $\{\ \}$, whose purpose is to indicate a specific place in a given structure, where sometimes other structures are plugged in; structures with a hole that does not appear in the scope of a negation are denoted as in $S\{\ \}$, and are called *structure contexts*. The structure $R$ is a *substructure* of $S\{R\}$, and $S\{\ \}$ is the *context* of $R$ therein.

**2.1.2 Notation** When structural brackets of any kind surround the content of a hole, hole braces will be omitted. For example, $S[a, b]$ stands for $S\{[a, b]\}$.

**2.1.3 Notation** A letter implicitly establishes the class to which an object belongs: for example, when we write $S$, we denote a structure, without necessarily saying explicitly that $S$ is a structure.

The structures $S\langle R_1; \dots; R_h \rangle$, $S[R_1, \dots, R_h]$ and $S(R_1, \dots, R_h)$ are respectively rep-

resented as

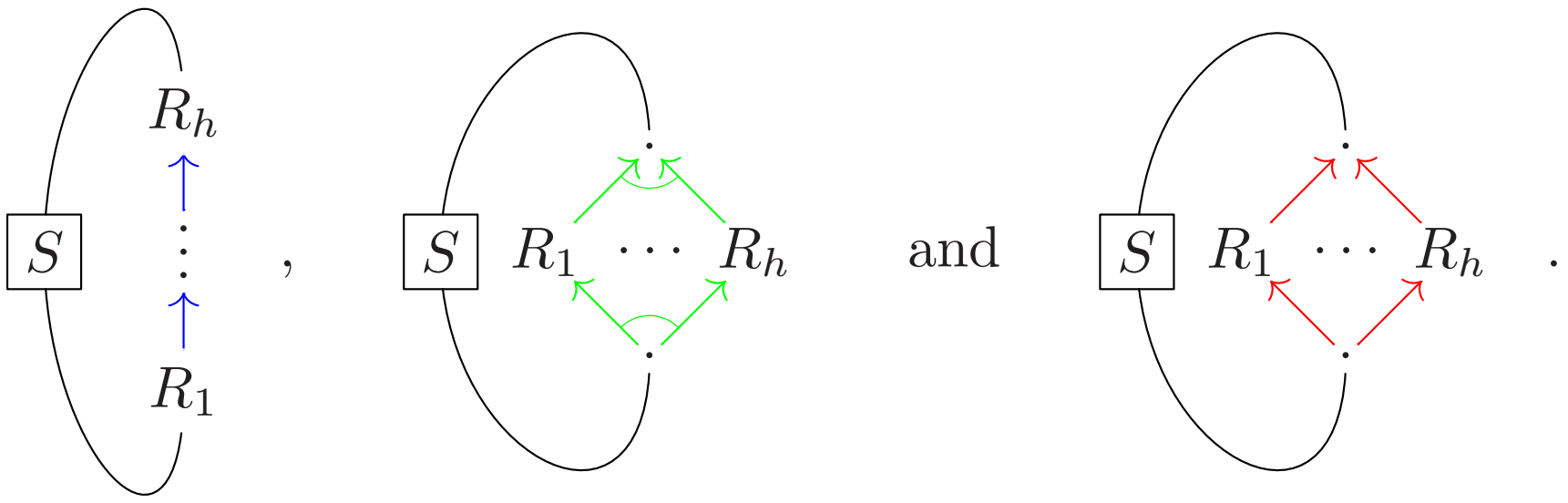

Arrows between non-seq structures in a seq structure context will be omitted.

Associativity holds for seq structures, but commutativity does not. Contrarily to their commutative counterparts, which live in what we can consider the horizontal line of space, the seq structures can make use of time. For example, consider the three structures in Figure 8; they correspond, from left to right, to the expressions

$$[\langle [P,U];(Q,R)\rangle ,V] \quad , \qquad [\langle P;(Q,R)\rangle ,U,V] \qquad \text{and} \qquad [\langle P;([Q,U],R)\rangle ,V] \quad .$$

An intuitive understanding is the following: Nodes represent points in space-time, and the order induced by arcs can be mapped consistently into a total order, which we can consider time. The structures in a par can interact at all times through the span of the par arc that connects them to the rest of the structure. The structures in a copar, instead, cannot interact. I could have also considered the 'cointerpretation' in which the structures in a par cannot interact and the structures in a copar can. No matter the interpretation, the structures in a seq should not be allowed to interact, because they stay in different time spans. Figure 8, in Subsection 3.2, shows two possible rewritings of the central structure, which respect our space-temporal intuition. I will get back to that figure later on.

Let us now turn our attention to negation. Negation is involutive and obeys the usual De Morgan laws for par and copar: it switches them. It is natural to state that $\overline{\langle S_1;\dots;S_h\rangle} = \langle \bar{S}_1;\dots;\bar{S}_h\rangle$: time is not reversed through negation. The behaviour of negation with respect to seq is in accord with the similar case in Retoré's pomset logic [27], where a commutative/non-commutative linear logic is derived from coherence semantics. By applying these laws, negation can always be pushed inward to atoms.

I will equip structures with a decidable equivalence relation that takes care of associativity, commutativity and negation. This places the burden of realising these properties on our pattern-matching abilities instead of on structural rules of the system. The same happens in sequent systems where the exchange rule becomes commutativity in sequents, or the contraction rule becomes idempotency, etc. All the three kinds of structures enjoy associativity, but only par and copar are commutative.

**2.1.4 Definition** Structures are considered equivalent modulo the relation $=$, which is the minimal equivalence relation defined by the axioms in Figure 1. There, $\vec{R}$, $\vec{T}$ and $\vec{U}$ stand for finite, non-empty sequences of structures. A structure, or a structure context, is said to be *in normal form* when the only negated structures appearing in it are atoms, no unit $\circ$ appears in it if atoms appear and no brackets can be eliminated while maintaining equivalence. If structures $R$ and $T$ are such that $R \neq \circ \neq T$, then the structure $\langle R;T\rangle$ is a *proper seq*, the structure $[R,T]$ is a *proper par* and the structure $(R,T)$ is a *proper*

**Associativity**

$$\langle \vec{R}; \langle \vec{T} \rangle; \vec{U} \rangle = \langle \vec{R}; \vec{T}; \vec{U} \rangle$$
$$[\vec{R}, [\vec{T}]] = [\vec{R}, \vec{T}]$$
$$(\vec{R}, (\vec{T})) = (\vec{R}, \vec{T})$$

**Singleton**

$$\langle R \rangle = [R] = (R) = R$$

**Unit**

$$\langle \circ; \vec{R} \rangle = \langle \vec{R}; \circ \rangle = \langle \vec{R} \rangle$$
$$[\circ, \vec{R}] = [\vec{R}]$$
$$(\circ, \vec{R}) = (\vec{R})$$

**Commutativity**

$$[\vec{R}, \vec{T}] = [\vec{T}, \vec{R}]$$
$$(\vec{R}, \vec{T}) = (\vec{T}, \vec{R})$$

**Negation**

$$\bar{\circ} = \circ$$
$$\overline{\langle R; T \rangle} = \langle \bar{R}; \bar{T} \rangle$$
$$\overline{[R, T]} = (\bar{R}, \bar{T})$$
$$\overline{(R, T)} = [\bar{R}, \bar{T}]$$
$$\bar{\bar{R}} = R$$

**Contextual Closure**

$$\text{if } R = T \text{ then } S\{R\} = S\{T\} \text{ and } \bar{R} = \bar{T}$$

**Fig. 1** *Syntactic equivalence* $=$

*copar*. A structure context $S\{\ \}$ is a *proper seq context* (a *proper par context*, a *proper copar context*) if, for all $X \neq \circ$, the structure $S\{X\}$ is a proper seq (a proper par, a proper copar). The structures whose normal forms do not contain seq structures are called *flat* (if one normal form does not contain seq, none does).

The set of equations in Figure 1 is not minimal, of course, but there is no real reason to minimise it. The structures $\overline{[a, \circ, b]}$, $((\overline{[\circ, b]}), \langle \bar{a} \rangle)$ and $(\bar{a}, \bar{\circ}, \bar{b})$ are all equivalent by $=$, but they are not in normal form; $(\bar{a}, \bar{b})$ is equivalent to them and normal, as well as $(\bar{b}, \bar{a})$; so, all the previous structures are flat. Since structures are considered equivalent under $=$, the structure $[\circ, \langle a; b \rangle]$ is a proper seq, but not a proper par or a proper copar; $\langle a; [\circ, b] \rangle$ is a proper seq, and $\langle a; [\{\ \}, b] \rangle$ is a proper seq context, while $[\{\ \}, b]$ is a proper par context.

**2.1.5 Remark** All structures can equivalently be considered in normal form, since negations can always be pushed inward to atoms by using the negation axioms, and units can be removed, as well as extra brackets (by associativity and singleton laws). Every structure can only be equivalent either to the unit, or to an atom, or, mutually exclusively, to a proper seq, or a proper par, or a proper copar.

Please note that negation obeys De Morgan-like equivalences for par and copar; seq is self-dual and ordering is maintained through negation. The reader should not be alarmed by the equation $\bar{\circ} = \circ$. The use of the unit $\circ$ is slightly different from the traditional use of logical units, like $\mathsf{t}$ and $\mathsf{f}$. In our case $\circ$ is just a convenient, syntactic mark that allows us to get a compact form for inference rules; I do not plan to interpret it semantically. There are anyway consistent formal systems where the units collapse into a unique one for all logical operators: for example, multiplicative linear logic with mix and nullary mix, a logic of which $\mathsf{BV}$ is a conservative extension (See Remark 5.9).

**2.1.6 Definition** Given a structure $S$, we talk about *atom occurrences* when considering all the atoms appearing in $S$ as distinct (for example, by indexing them so that two atoms which are equal get different indices); therefore, in $\langle a; a\rangle$ there are two atom occurrences, those of the atom $a$; in the structure $\bar{a}$, there is only one atom occurrence, the negative atom $\bar{a}$. The notation $\mathsf{occ}\, S$ indicates the set of all the atom occurrences appearing in $S$. The *size* of $S$ is the cardinality of the set $\mathsf{occ}\, S$. We will use the expression 'atom occurrences' also outside of structures, whenever we consider distinct possibly multiple appearances of the same atom.

The set $\mathsf{occ}\, S$ could be defined as the multiset of the atoms of $S$, or of $S$ in normal form. Note that $\mathsf{occ}\, \circ = \varnothing$; also note that $\mathsf{occ}\, \langle S; S'\rangle = \mathsf{occ}\, [S, S'] = \mathsf{occ}\, (S, S') = \mathsf{occ}\, S \cup \mathsf{occ}\, S'$ is true only if $\mathsf{occ}\, S$ and $\mathsf{occ}\, S'$ are disjoint, and we can always assume this without loss of generality.

## 2.2 Relation Webs

The informal graphic representation of structures seen above partly justifies the choice of equations we imposed on structures. It is a weak justification, which only relies on the representation being intuitively appealing. In this subsection I will offer a strong justification. We will see another representation of structures, this time formal: *relation webs*. Contrary to the 'space-time' representation, relation webs do not have an immediate correspondence to the inference rules we are going to see later, but they have other important features:

1 There is a unique relation web for every equivalence class of structures under $=$.

2 The formal system $\mathsf{BV}$ will be derived from relation webs by asking for a certain conservation property to hold while we manipulate relation webs.

Relation webs are used as a sort of semantics, which gives independent and faithful account of inference rules. I will develop them only in relation to the structure fragment of $\mathsf{BV}$. This is enough to establish Tiu's results on the inadequacy of the sequent calculus for $\mathsf{BV}$ in [36], where relation webs are used to show that certain structures are not provable.

The readers who wish only to understand the behaviour of system $\mathsf{BV}$, possibly also in relation to the sequent calculus of linear logic, need not read this subsection. It is safe to ignore everything about relation webs and still read the rest of the paper, with the exception of Subsection 3.1.

Consider $[R_1, \dots, R_h]$: let us establish that for distinct $i$ and $j$ and for all the atoms $a$ appearing in $R_i$ and $b$ appearing in $R_j$ the relation $a \downarrow b$ holds (so $\downarrow$ is symmetric). Analogously, two distinct structures $R$ and $T$ in a copar induce on their atoms a relation $a \uparrow b$, where $a$ is in $R$ and $b$ is in $T$. For example, in $[a, b, (\bar{b}, [(\bar{a}, c), \bar{c}])]$ these relations are determined: $a \downarrow b$, $a \downarrow \bar{b}$, $a \downarrow \bar{a}$, $a \downarrow c$, $a \downarrow \bar{c}$, $b \downarrow \bar{b}$, $b \downarrow \bar{a}$, $b \downarrow c$, $b \downarrow \bar{c}$, $\bar{b} \uparrow \bar{a}$, $\bar{b} \uparrow c$, $\bar{b} \uparrow \bar{c}$, $\bar{a} \uparrow c$, $\bar{a} \downarrow \bar{c}$, $c \downarrow \bar{c}$ (the symmetric relations have been omitted, for example $\bar{a} \uparrow \bar{b}$). Let us add that $\langle S_1; \dots; S_h\rangle$ induces the relation $a \lhd b$ for all the atoms $a$ in $S_i$ and $b$ in $S_j$ such that $1 \leqslant i < j \leqslant h$ (so $\lhd$ is not symmetric).

**2.2.1 Definition** Given a structure $S$ in normal form, the four *structural relations* $\lhd_S$ (*seq*), $\rhd_S$ (*coseq*), $\downarrow_S$ (*par*) and $\uparrow_S$ (*copar*) are the minimal sets such that $\lhd_S, \rhd_S, \downarrow_S, \uparrow_S \subset (\mathsf{occ}\, S)^2$ and, for every $S'\{\ \}$, $U$ and $V$ and for every $a$ in $U$ and $b$ in $V$, the following hold:

**1** if $S = S'\langle U; V\rangle$ then $a \lhd_S b$ and $b \rhd_S a$;

**2** if $S = S'[U, V]$ then $a \downarrow_S b$;

**3** if $S = S'(U, V)$ then $a \uparrow_S b$.

To a structure that is not in normal form we associate the structural relations obtained from any of its normal forms, since they yield the same. The quadruple $(\mathsf{occ}\, S, \triangleleft_S, \downarrow_S, \uparrow_S)$ is called the *relation web* (or simply *web*) of $S$, written $\mathsf{w}\, S$. We can abolish the subscripts in $\triangleleft_S$, $\triangleright_S$, $\downarrow_S$ and $\uparrow_S$ when they are not necessary. Given two sets of atom occurrences $\mu$ and $\nu$, we write $\mu \triangleleft \nu$, $\mu \triangleright \nu$, $\mu \downarrow \nu$ and $\mu \uparrow \nu$ to indicate situations where, for every $a$ in $\mu$ and for every $b$ in $\nu$, the following hold, respectively: $a \triangleleft b$, $a \triangleright b$, $a \downarrow b$ and $a \uparrow b$. We represent structural relations between occurrences of atoms by drawing $a \leadsto b$ when $a \triangleleft b$ (and $b \triangleright a$), $a \leftrightsquigarrow b$ when $a \triangleleft b$ or $a \triangleright b$, $a$ —— $b$ when $a \downarrow b$ and $a \sim b$ when $a \uparrow b$. Dashed arrows represent negations of structural relations.

For example, in $(\overline{\langle \bar{a}; b\rangle}, \overline{(c, \bar{d})}) = (\langle a; \bar{b}\rangle, [\bar{c}, d])$ these relations are determined: $a \triangleleft \bar{b}$, $a \uparrow \bar{c}$, $a \uparrow d$, $\bar{b} \triangleright a$, $\bar{b} \uparrow \bar{c}$, $\bar{b} \uparrow d$, $\bar{c} \uparrow a$, $\bar{c} \uparrow \bar{b}$, $\bar{c} \downarrow d$, $d \uparrow a$, $d \uparrow \bar{b}$, $d \downarrow \bar{c}$. The relation web for $\circ$ is $(\varnothing, \varnothing, \varnothing, \varnothing)$.

The graphical notation for structural relations will be extensively used in the following: please note that the graphs obtained from structural relations are not the same as the 'space-time' ones representing structures!

**2.2.2 Remark** A structure $S$ such that $\mathsf{w}\, S = (\mathsf{occ}\, S, \triangleleft, \downarrow, \uparrow)$ is flat iff $\triangleleft = \varnothing$.

It is easy to see from the definitions that all the atoms in a substructure are in the same structural relation with respect to each of the atoms in the context surrounding the substructure:

**2.2.3 Proposition** *Given a structure $S\{R\}$ and two atom occurrences $a$ in $S\{\ \}$ and $b$ in $R$, if $a \triangleleft b$ (respectively, $a \triangleright b$, $a \downarrow b$, $a \uparrow b$) then $a \triangleleft c$ (respectively, $a \triangleright c$, $a \downarrow c$, $a \uparrow c$) for all the atom occurrences $c$ in $R$.*

The given syntax of structures, and the use of the equivalence $=$, help to focus the system more on meaning than on representation. A structure should picture, with the minimum amount of ink and ambiguity, a certain situation, where what really matters are atoms and their mutual structural relations: the web of a structure displays this information. However, taking a bunch of atoms and assigning structural relations to couples of them does not guarantee to produce a valid structure. Two questions are then in order: 1) when does an assignment of relations to atoms actually give a structure, and 2) whether the structures having a given web are equivalent by $=$, or not.

This subsection answers both questions above. At this point the reader in a hurry knows enough to jump immediately to the next section, after reading the statement of Theorem 2.2.9. What follows is a characterisation of structures in terms of seven properties of structural relations. The most intriguing (and less expected) of them is what I call the 'square property'. It has been found with much help from a paper by Möhring, [22], where a simpler case about series-parallel orders is studied. The next theorem shows that certain properties are necessary for structures.

**2.2.4 Theorem** *Given $S$ and its associated structural relations $\triangleleft$, $\triangleright$, $\downarrow$ and $\uparrow$, the following properties hold, where $a$, $b$, $c$ and $d$ are distinct atom occurrences in $S$:*

$\mathsf{s}_1$ *None of $\triangleleft$, $\triangleright$, $\downarrow$ and $\uparrow$ is reflexive: $\neg(a \triangleleft a)$, $\neg(a \triangleright a)$, $\neg(a \downarrow a)$, $\neg(a \uparrow a)$.*

$\mathsf{s}_2$ *One and only one among* $a \triangleleft b$, $a \triangleright b$, $a \downarrow b$ *and* $a \uparrow b$ *holds.*

$\mathsf{s}_3$ *The relations* $\triangleleft$ *and* $\triangleright$ *are mutually inverse*: $a \triangleleft b \Leftrightarrow b \triangleright a$.

$\mathsf{s}_4$ *The relations* $\triangleleft$ *and* $\triangleright$ *are transitive*: $(a \triangleleft b) \wedge (b \triangleleft c) \Rightarrow a \triangleleft c$ *and* $(a \triangleright b) \wedge (b \triangleright c) \Rightarrow a \triangleright c$.

$\mathsf{s}_5$ *The relations* $\downarrow$ *and* $\uparrow$ *are symmetric*: $a \downarrow b \Leftrightarrow b \downarrow a$ *and* $a \uparrow b \Leftrightarrow b \uparrow a$.

$\mathsf{s}_6$ Triangular property: *for* $\sigma_1, \sigma_2, \sigma_3 \in \{\triangleleft \cup \triangleright, \downarrow, \uparrow\}$, *it holds*

$$(a\ \sigma_1\ b) \wedge (b\ \sigma_2\ c) \wedge (c\ \sigma_3\ a) \Rightarrow (\sigma_1 = \sigma_2) \vee (\sigma_2 = \sigma_3) \vee (\sigma_3 = \sigma_1) \quad .$$

$\mathsf{s}_7$ Square property:

$\mathsf{s}_7^{\triangleleft}$
$$(a \triangleleft b) \wedge (a \triangleleft d) \wedge (c \triangleleft d) \Rightarrow (a \triangleleft c) \vee (b \triangleleft c) \vee (b \triangleleft d) \vee (c \triangleleft a) \vee (c \triangleleft b) \vee (d \triangleleft b) \quad ,$$

$\mathsf{s}_7^{\downarrow}$
$$(a \downarrow b) \wedge (a \downarrow d) \wedge (c \downarrow d) \Rightarrow (a \downarrow c) \vee (b \downarrow c) \vee (b \downarrow d) \quad ,$$

$\mathsf{s}_7^{\uparrow}$
$$(a \uparrow b) \wedge (a \uparrow d) \wedge (c \uparrow d) \Rightarrow (a \uparrow c) \vee (b \uparrow c) \vee (b \uparrow d) \quad .$$

**Proof** The properties $\mathsf{s}_1$, $\mathsf{s}_2$, $\mathsf{s}_3$, $\mathsf{s}_4$ and $\mathsf{s}_5$ are readily proved using the relevant definitions. Let us consider the more challenging cases of $\mathsf{s}_6$ and $\mathsf{s}_7$.

$\mathsf{s}_6$ Suppose that $a \triangleleft b$ and $b \downarrow c$: the only possible cases are $S\langle P\{a\}; T[Q\{b\}, R\{c\}]\rangle$ (so $a \triangleleft c$) and $S[T\langle P\{a\}; Q\{b\}\rangle, R\{c\}]$ (so $a \downarrow c$), for some structure contexts $P\{\ \}$, $Q\{\ \}$, $R\{\ \}$, $S\{\ \}$ and $T\{\ \}$. Other combinations of $\sigma_1$ and $\sigma_2$ generate similar possibilities.

$\mathsf{s}_7$ Let us proceed by structural induction on $S$. Every structure with less than four atom occurrences trivially satisfies $\mathsf{s}_7$, since four distinct occurrences are requested, therefore let us consider situations where at least four atom occurrences are present. Let $U$ and $V$ be any two structures such that $U \neq \circ \neq V$ and either $S = \langle U; V\rangle$ or $S = [U, V]$ or $S = (U, V)$; let us choose $a$, $b$, $c$ and $d$ in $S$. If $a$, $b$, $c$ and $d$ are either all in $U$ or all in $V$ then we can apply the induction hypothesis; let us then consider the cases when they are not all in $U$ or all in $V$. Consider $\mathsf{s}_7^{\triangleleft}$ (Figure 2 should help). Since $S = [U, V]$ and $S = (U, V)$ falsify the hypothesis of $\mathsf{s}_7^{\triangleleft}$, the only situation we have to consider is $S = \langle U; V\rangle$. Suppose that the conclusion of $\mathsf{s}_7^{\triangleleft}$ is false and suppose that $a$ is in $U$; then $c$ must be in $U$ (otherwise $a \triangleleft c$ would be true), and then $b$ and then $d$ must be in $U$, but this contradicts our assumption. Analogously, if $a$ is in $V$ then $c$ must be in $V$ and then $b$ and then $d$ must be in $V$. In the end, if the hypothesis of $\mathsf{s}_7^{\triangleleft}$ is true when $a$, $b$, $c$ and $d$ are scattered between $U$ and $V$, then its conclusion is true. The same argument, simplified by the holding of symmetry, applies to $\mathsf{s}_7^{\downarrow}$ and $\mathsf{s}_7^{\uparrow}$. □

The triangular property says that there are no structures such that the following configuration may be found in them:

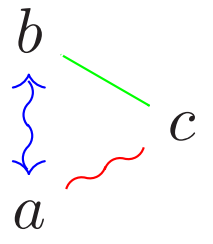

;

in every triangle at least two sides must represent the same structural relation.

The square property for $\triangleleft$ may be represented like in Figure 2, where transitivity has been taken into account and an example structure is shown under each diagram. Enjoying commutativity, the cases of par and copar are simpler. See, in Figure 3, what happens with par. We can say, informally, that no square has exactly three sides or diagonals of the same nature and forming a simple path (disregarding orientation).

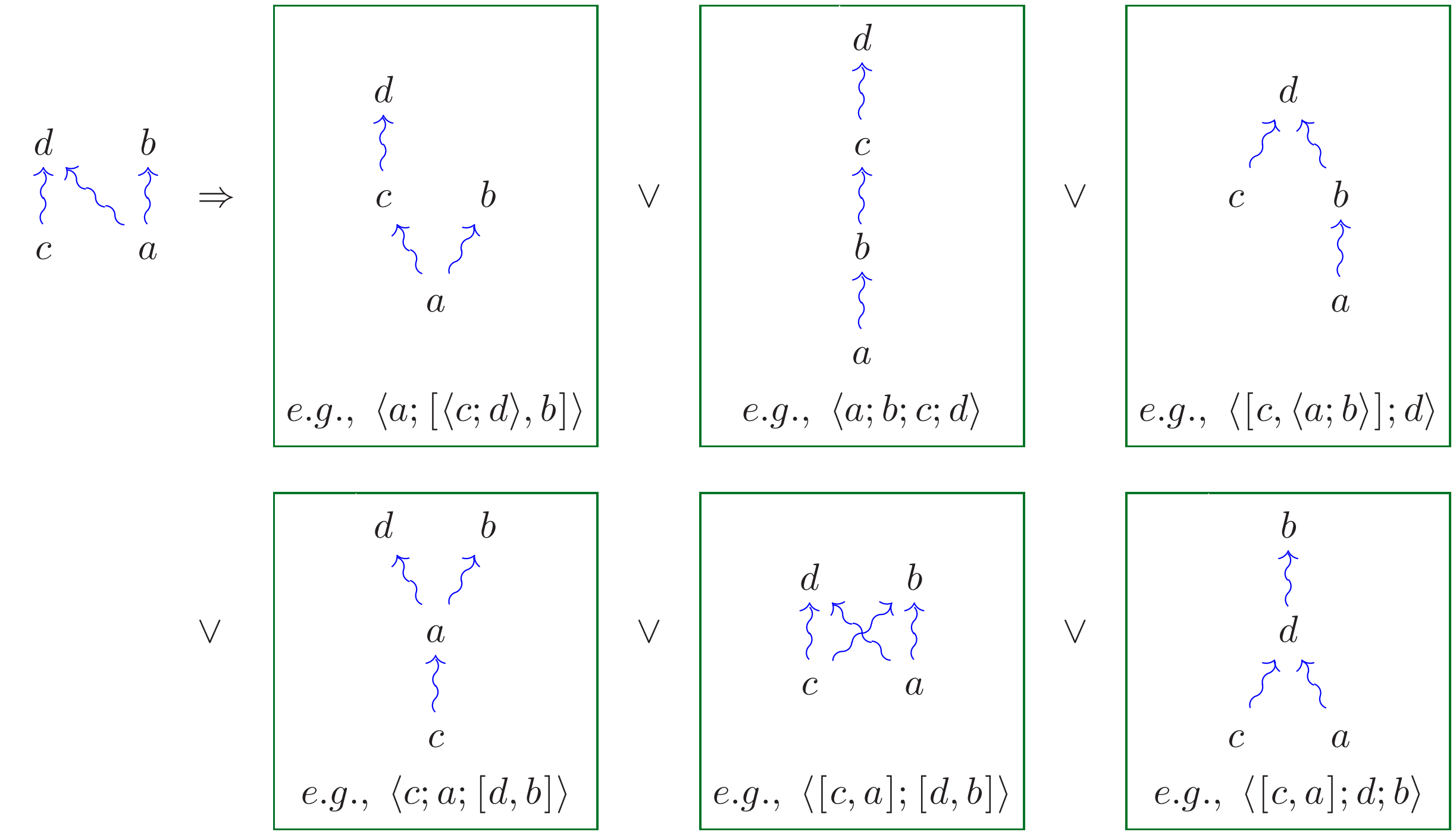

**Fig. 2** *Square property for* $\triangleleft$

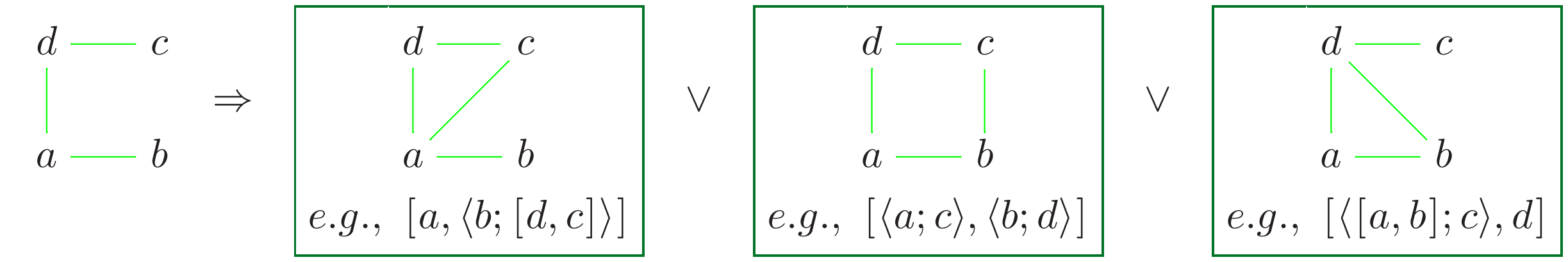

**Fig. 3** *Square property for* $\downarrow$

**2.2.5 Remark** The relations $\downarrow$ and $\uparrow$ are not transitive: consider $[\langle a; c\rangle, b]$ and $(\langle a; c\rangle, b)$.

We saw that the conditions $\mathsf{s}_1$–$\mathsf{s}_7$ are necessary for a structure. I am going to show that they are sufficient. The proof consists in recursively finding suitable partitions of a given set of atom occurrences $\xi$, where $\xi$ is equipped with structural relations. If we are able to partition $\xi$ into two parts $\mu$ and $\nu$ such that one and only one among $\mu \triangleleft \nu$, $\mu \downarrow \nu$ and $\mu \uparrow \nu$ holds, and we can go ahead like this recursively, then we can recover a structure. I need to establish some straightforward preliminary notions.

**2.2.6 Definition** A *web candidate* is a quadruple $\zeta = (\xi, \triangleleft, \downarrow, \uparrow)$, where $\xi$ is a set of atom occurrences and $\triangleleft, \downarrow, \uparrow \subseteq \xi^2$. Given the web candidates $\zeta_\mu = (\mu, \triangleleft_\mu, \downarrow_\mu, \uparrow_\mu)$ and $\zeta_\nu = (\nu, \triangleleft_\nu, \downarrow_\nu, \uparrow_\nu)$, such that $\mu \neq \varnothing \neq \nu$, $\mu \cup \nu = \xi$ and $\mu \cap \nu = \triangleleft_\mu \cap \triangleleft_\nu = \downarrow_\mu \cap \downarrow_\nu = \uparrow_\mu \cap \uparrow_\nu = \varnothing$, the couple $(\zeta_\mu, \zeta_\nu)$ can be:

**1** a $\triangleleft$-*partition* of $\zeta$ iff $\downarrow = \downarrow_\mu \cup \downarrow_\nu$, $\uparrow = \uparrow_\mu \cup \uparrow_\nu$ and

$$\triangleleft = \triangleleft_\mu \cup \triangleleft_\nu \cup \{\, (a, b) \mid a \in \mu,\ b \in \nu \,\} \quad ;$$

**2** a *$\downarrow$-partition* of $\zeta$ iff $\triangleleft = \triangleleft_\mu \cup \triangleleft_\nu$, $\uparrow = \uparrow_\mu \cup \uparrow_\nu$ and

$$\downarrow = \downarrow_\mu \cup \downarrow_\nu \cup \{\, (a,b) \mid (a \in \mu \wedge b \in \nu) \vee (a \in \nu \wedge b \in \mu) \,\} \quad ;$$

**3** an *$\uparrow$-partition* of $\zeta$ iff $\triangleleft = \triangleleft_\mu \cup \triangleleft_\nu$, $\downarrow = \downarrow_\mu \cup \downarrow_\nu$ and

$$\uparrow = \uparrow_\mu \cup \uparrow_\nu \cup \{\, (a,b) \mid (a \in \mu \wedge b \in \nu) \vee (a \in \nu \wedge b \in \mu) \,\} \quad .$$

For every web candidate, the relation $\triangleright = \{\, (a,b) \mid b \triangleleft a \,\}$ may be defined, and we will do so implicitly.

Of course, webs are (successful!) web candidates.

I am ready to show that $\mathsf{s}_1$–$\mathsf{s}_7$ constitute an adequate characterisation of structures. It is important to note that these conditions can be checked over at most four atoms at a time, and this means that an algorithm that checks the conditions might be implemented locally in a network of processors representing a structure. François Lamarche proposes a different, more compact axiomatisation than $\mathsf{s}_1$–$\mathsf{s}_7$ in [18].

I encourage the reader to check the following proof by trying to think at it as the task of a distributed algorithm.

**2.2.7 Theorem** *If the conditions $\mathsf{s}_1$–$\mathsf{s}_7$ hold for a web candidate $\zeta$ then there is a structure whose web is $\zeta$.*

**Proof** Let $\zeta = (\xi, \triangleleft, \downarrow, \uparrow)$: We will proceed by induction on the cardinality $|\xi|$ of $\xi$ to build a structure $S$ such that $\mathsf{w}\, S = \zeta$. If $\xi = \varnothing$ then $S = \circ$. If $\xi = \{a\}$ then $\triangleleft = \downarrow = \uparrow = \varnothing$ (by $\mathsf{s}_1$) and $S = a$. Let us consider then the case where at least two atom occurrences are in $\xi$. We will see that the conditions $\mathsf{s}_1$–$\mathsf{s}_7$ enforce the existence of a $\triangleleft$-, $\downarrow$- or $\uparrow$-partition of $\zeta$. Suppose that a $\triangleleft$-partition of $\zeta$ exists, consisting of $\zeta_\mu$ and $\zeta_\nu$. The conditions $\mathsf{s}_1$–$\mathsf{s}_7$ hold for $\zeta_\mu$ and $\zeta_\nu$, therefore, by induction hypothesis, two structures $U$ and $V$ exist such that $\mathsf{w}\, U = \zeta_\mu$ and $\mathsf{w}\, V = \zeta_\nu$. But then we can take $S = \langle U; V\rangle$, and, by definition and by $\mathsf{s}_2$ and $\mathsf{s}_3$, we have $\mathsf{w}\, S = \zeta$. We can proceed analogously when $\zeta_\mu$ and $\zeta_\nu$ form a $\downarrow$-partition (take $S = [U, V]$) or an $\uparrow$-partition (take $S = (U, V)$); $\mathsf{s}_5$ has a role here in ensuring the correct formation of a partition.

We have to show that a $\triangleleft$-, $\downarrow$- or $\uparrow$-partition of $\zeta$ exists, in the given hypotheses, consisting of $\zeta_\mu = (\mu, \triangleleft_\mu, \downarrow_\mu, \uparrow_\mu)$ and $\zeta_\nu = (\nu, \triangleleft_\nu, \downarrow_\nu, \uparrow_\nu)$. We will construct the $\mu$ and $\nu$ sets of atom occurrences incrementally, starting from $\mu_2 = \{a\}$ and $\nu_2 = \{b\}$, for some $a$ and $b$ in $\xi$, and building a family of couples $\{(\mu_i, \nu_i)\}_{2 \leqslant i \leqslant |\xi|}$ such that at each step one element of $\xi$ is added to the union of $\mu_i$ and $\nu_i$ that was not added before; at each step $\mu_i \neq \varnothing \neq \nu_i$ and either $\mu_i \triangleleft \nu_i$ or $\mu_i \downarrow \nu_i$ or $\mu_i \uparrow \nu_i$. The final step gives the partition, i.e., $\mu = \mu_{|\xi|}$ and $\nu = \nu_{|\xi|}$. Here is a non-deterministic algorithm.

**First Step**

Take $\mu_2 = \{a\}$ and $\nu_2 = \{b\}$, where $a$ and $b$ are distinct atom occurrences that are randomly chosen in $\xi$ and such that one of $\mu_2 \triangleleft \nu_2$, $\mu_2 \downarrow \nu_2$ or $\mu_2 \uparrow \nu_2$ holds (the conditions $\mathsf{s}_2$ and $\mathsf{s}_3$ apply).

**Iterative Step**

We have two disjoint and non-empty sets of occurrences $\mu_i$ and $\nu_i$ such that all the atom occurrences in $\mu_i$ are in the same structural relation $\sigma \in \{\triangleleft, \downarrow, \uparrow\}$ with the atom occurrences in $\nu_i$, i.e., either $\mu_i \triangleleft \nu_i$ or $\mu_i \downarrow \nu_i$ or $\mu_i \uparrow \nu_i$. Pick any $c$ in $\xi \setminus (\mu_i \cup \nu_i)$. If $d \;\sigma\; c$ for every $d$ in $\mu_i$ then let $\mu_{i+1} = \mu_i$ and $\nu_{i+1} = \nu_i \cup \{c\}$; if $c \; \sigma \; e$ for every $e$ in $\nu_i$ then let $\mu_{i+1} = \mu_i \cup \{c\}$ and $\nu_{i+1} = \nu_i$; in both cases $\mu_{i+1} \; \sigma \; \nu_{i+1}$. Otherwise we have to rearrange $\mu_i$ and $\nu_i$ in order to meet our requirements. Let us proceed by cases.

**1** $\mu_i \triangleleft \nu_i$ and there are $a$ in $\mu_i$ and $b$ in $\nu_i$ such that $\neg(a \triangleleft c)$ and $\neg(c \triangleleft b)$: this situation is represented at the left (where $a \rightsquigarrow c$ stands for $\neg(a \triangleleft c)$):

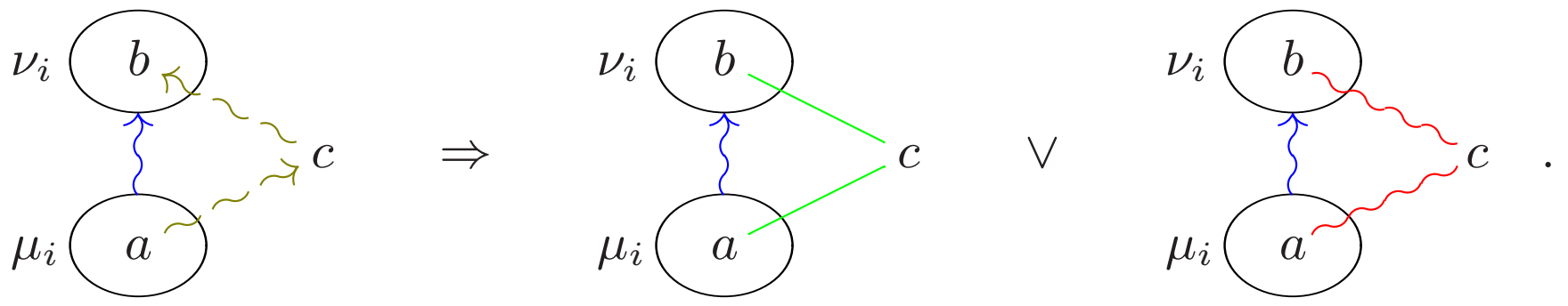

Since $a \triangleleft b$, by transitivity of $\triangleleft$ ($\mathsf{s}_4$), by symmetry of $\downarrow$ and $\uparrow$ ($\mathsf{s}_5$) and by the triangular property ($\mathsf{s}_6$), only two cases are possible: either $a \downarrow c$ and $c \downarrow b$, or $a \uparrow c$ and $c \uparrow b$ (the former case is represented in the central diagram, the latter at the right). Let us only consider the first case, the other one being similar. Again by $\mathsf{s}_4$, $\mathsf{s}_5$ and $\mathsf{s}_6$, either $d \downarrow c$ or $d \triangleleft c$, for each element $d$ in $\mu_i$, and either $c \downarrow e$ or $c \triangleleft e$, for each element $e$ in $\nu_i$. We can then partition $\mu_i$ into the two disjoint sets $\mu_i^{\downarrow}$ and $\mu_i^{\triangleleft}$ and $\nu_i$ into the two disjoint sets $\nu_i^{\downarrow}$ and $\nu_i^{\triangleright}$ in such a way that $\mu_i^{\downarrow} \downarrow \{c\}$, $\mu_i^{\triangleleft} \triangleleft \{c\}$ and $\{c\} \downarrow \nu_i^{\downarrow}$, $\{c\} \triangleleft \nu_i^{\triangleright}$; of course, $a \in \mu_i^{\downarrow}$ and $b \in \nu_i^{\downarrow}$. This situation is represented at the left:

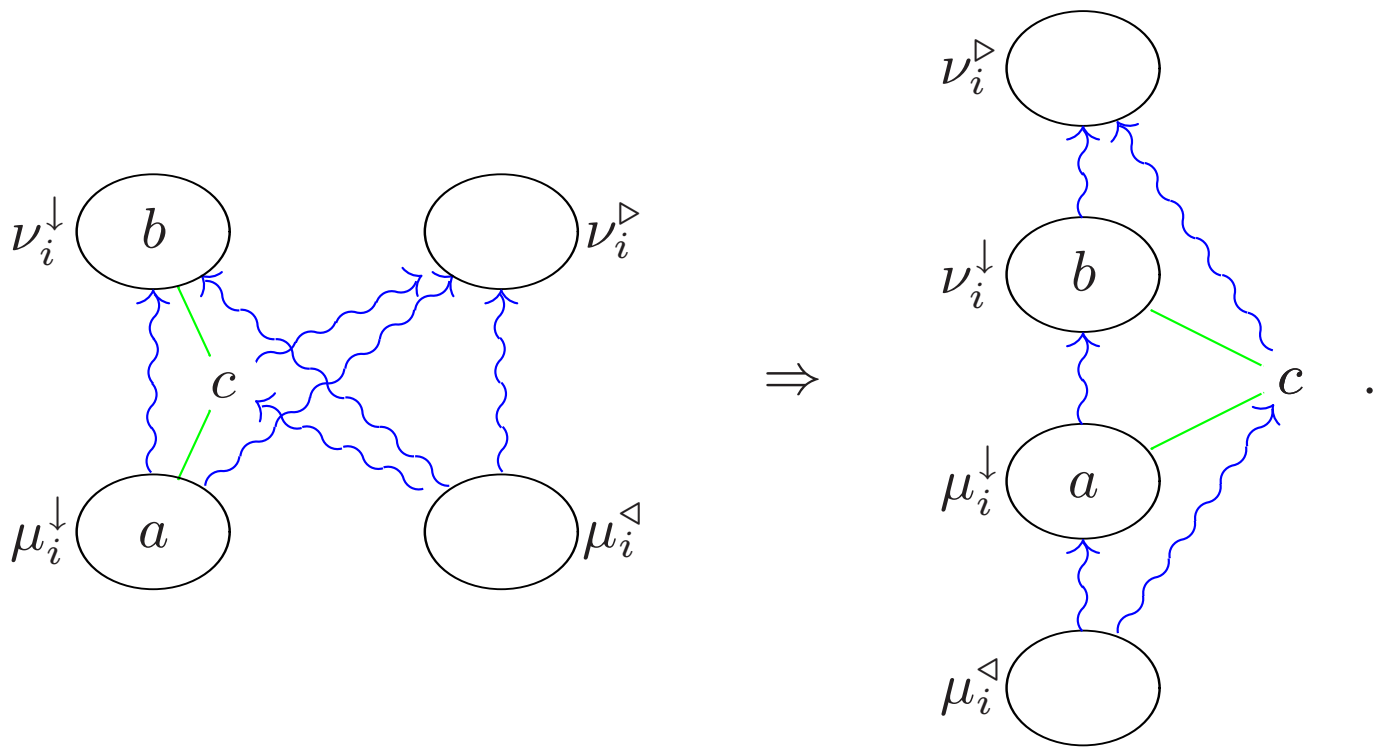

It must be the case that $\mu_i^{\triangleleft} \triangleleft \mu_i^{\downarrow}$, because of $\mathsf{s}_4$, $\mathsf{s}_5$, $\mathsf{s}_6$ and the square property for $\downarrow$ ($\mathsf{s}_7^{\downarrow}$) between $\mu_i^{\downarrow}$, $\mu_i^{\triangleleft}$, $\nu_i^{\downarrow}$ and $c$. Analogously, it must be $\nu_i^{\downarrow} \triangleleft \nu_i^{\triangleright}$. The resulting situation, simplified by transitivity, is shown at the right. If $\mu_i^{\triangleleft} \neq \varnothing$ then take $\mu_{i+1} = \mu_i^{\triangleleft}$ and $\nu_{i+1} = \mu_i^{\downarrow} \cup \nu_i^{\downarrow} \cup \nu_i^{\triangleright} \cup \{c\}$: in this case $\mu_{i+1} \triangleleft \nu_{i+1}$. Otherwise, if $\nu_i^{\triangleright} \neq \varnothing$ then take $\mu_{i+1} = \mu_i^{\downarrow} \cup \nu_i^{\downarrow} \cup \{c\}$ and $\nu_{i+1} = \nu_i^{\triangleright}$: again $\mu_{i+1} \triangleleft \nu_{i+1}$. If both $\mu_i^{\triangleleft}$ and $\nu_i^{\triangleright}$ are empty, take $\mu_{i+1} = \mu_i^{\downarrow} \cup \nu_i^{\downarrow}$ and $\nu_{i+1} = \{c\}$: in this case $\mu_{i+1} \downarrow \nu_{i+1}$.

**2** $\mu_i \downarrow \nu_i$ and there are $a$ in $\mu_i$ and $b$ in $\nu_i$ such that $\neg(a \downarrow c)$ and $\neg(b \downarrow c)$: By an analogous argument to that in Case 1, we have that this situation, represented at the left in the following diagram (where $a$ — — $c$ stands for $\neg(a \downarrow c)$), entails the possibilities at the right, and those only:

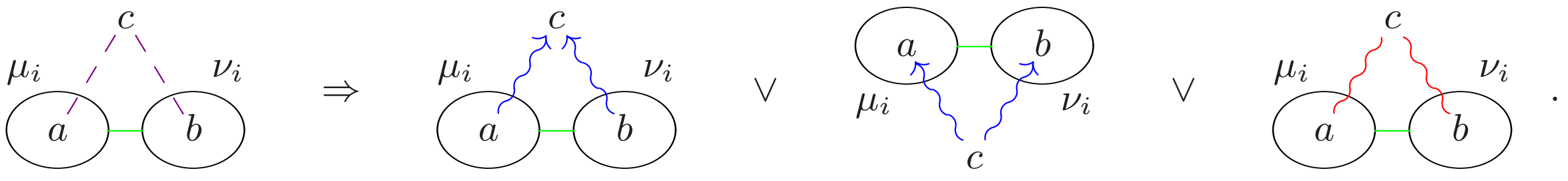

Let us consider the first case, where $a \triangleleft c$ and $b \triangleleft c$. By use of the triangular property, we can partition $\mu_i$ into $\mu_i^{\downarrow}$ and $\mu_i^{\triangleleft}$ and $\nu_i$ into $\nu_i^{\downarrow}$ and $\nu_i^{\triangleleft}$ in such a way that we have the situation represented at the left:

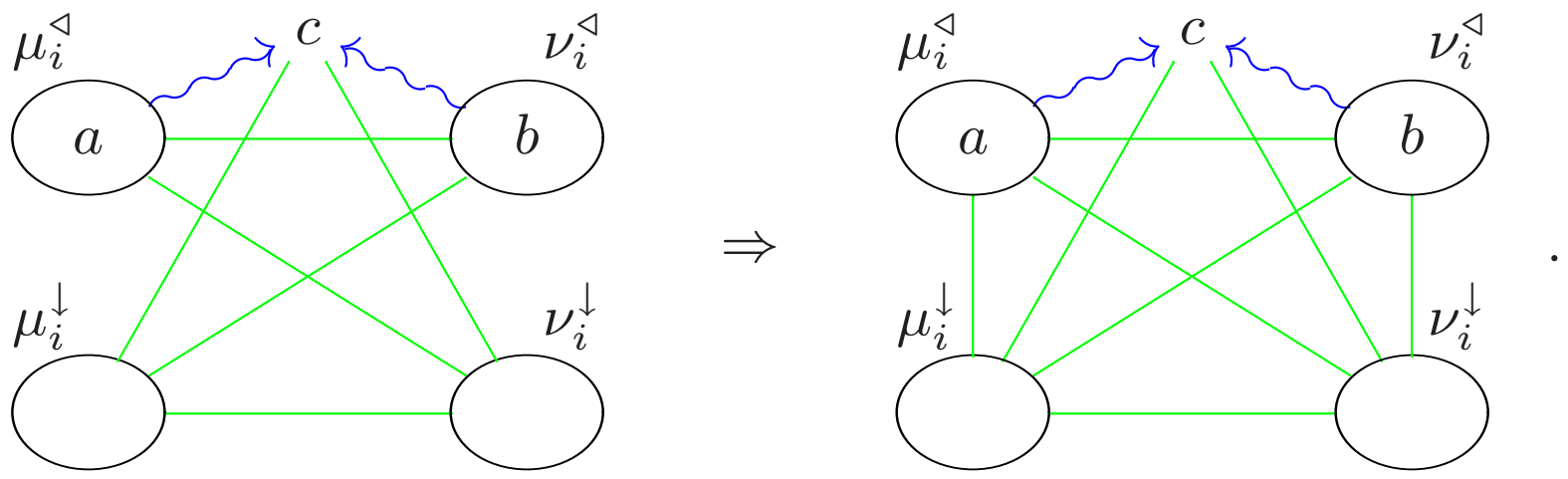

The square property enforces the situation at the right, where we can now define an appropriate partition. If $\mu_i^{\downarrow} \neq \varnothing$ then take $\mu_{i+1} = \mu_i^{\downarrow}$ and $\nu_{i+1} = \mu_i^{\lhd} \cup \{c\} \cup \nu_i^{\lhd} \cup \nu_i^{\downarrow}$: in this case $\mu_{i+1} \downarrow \nu_{i+1}$. Otherwise, if $\nu_i^{\downarrow} \neq \varnothing$ then take $\mu_{i+1} = \mu_i^{\lhd} \cup \{c\} \cup \nu_i^{\lhd}$ and $\nu_{i+1} = \nu_i^{\downarrow}$: again $\mu_{i+1} \downarrow \nu_{i+1}$. If both $\mu_i^{\downarrow}$ and $\nu_i^{\downarrow}$ are empty, take $\mu_{i+1} = \mu_i^{\lhd} \cup \nu_i^{\lhd}$ and $\nu_{i+1} = \{c\}$: in this case $\mu_{i+1} \lhd \nu_{i+1}$. The other cases above, where $a \rhd c$ and $b \rhd c$ and where $a \uparrow c$ and $b \uparrow c$, are managed in an analogous way.

**3** $\mu_i \uparrow \nu_i$ and there are $a$ in $\mu_i$ and $b$ in $\nu_i$ such that $\neg(a \uparrow c)$ and $\neg(b \uparrow c)$: this case is similar to Case 2.

The final step of the algorithm occurs when no atom occurrences are left in $\xi$ that are not already in the partition. □

In the previous proof, the reader might be puzzled by not seeing a ‘bottom-up’ construction where atoms are gathered together into bigger and bigger structures, starting from those that correspond to the smallest substructures of the structure to be built. The problem with that approach is that it is not symmetric: contrary to the approach above, only a few atoms can participate at any given time to the building of the structure. I feel that the symmetry in our construction might be important in expected future developments, aside from being a mathematically respectable property.

The core of the proof above relies on the combined action of the triangular and square properties. The property $\mathsf{s}_6$ reduces the problem to a case in which only two structural relations are involved, then $\mathsf{s}_7$ is used to decide the remaining undecided side of a square. This procedure can then be generalised to the case in which more than three different kinds of context are allowed in structures. In fact, going back to the proof of Theorem 2.2.4, it is easy to see that $\mathsf{s}_6$ and $\mathsf{s}_7$ do not depend on the number and quality of different structural relations; rather they make use of the uniqueness property $\mathsf{s}_2$, which is in turn guaranteed by the inherently unambiguous constitution of structures.

The algorithm given in the proof above is non-deterministic, therefore it cannot reasonably be used to answer the following, inevitable question: is a structure corresponding to a web unique? It turns out that it is, modulo equivalence of course, as the following theorem shows.

**2.2.8 Lemma** *Given a structure $T$, if $(\zeta_\mu, \zeta_\nu)$ is a $\lhd$-partition of $\mathsf{w}\,T$ (respectively, a $\downarrow$-partition, an $\uparrow$-partition) then there are two structures $U$ and $V$ such that $\mathsf{w}\,U = \zeta_\mu$, $\mathsf{w}\,V = \zeta_\nu$ and $T = \langle U; V\rangle$ (respectively, $T = [U, V]$, $T = (U, V)$).*

**Proof** Let $\mathsf{w}\,T = (\mathsf{occ}\,T, \lhd, \downarrow, \uparrow)$; the web candidates $\zeta_\mu = (\mu, \lhd_\mu, \downarrow_\mu, \uparrow_\mu)$ and $\zeta_\nu = (\nu, \lhd_\nu, \downarrow_\nu, \uparrow_\nu)$ form a $\lhd$-partition of $\mathsf{w}\,T$. Since $\mu \neq \varnothing \neq \nu$, the structure $T$ falls in one of the three cases:

**1** $T = \langle T_1; \ldots; T_h\rangle$, where $h > 1$ and, for $1 \leqslant i \leqslant h$, it holds $T_i \neq \circ$ and $T_i$ is not a proper seq: It must be the case that $\mathsf{occ}\,T_i \subseteq \mu$ or $\mathsf{occ}\,T_i \subseteq \nu$, for every $i$. In fact, suppose the contrary, and suppose that $T_i = [T'_i, T''_i]$ for some $T'_i$ and $T''_i$ such that $T'_i \neq \circ \neq T''_i$ (the same argument holds when $T_i = (T'_i, T''_i)$, in the same conditions). It is then possible to find $a$ in $T'_i$ and $b$ in $T''_i$, or $a$ in $T''_i$ and $b$ in $T'_i$, such that $a$ is in $\mu$ and $b$ is in $\nu$. But then $a \downarrow b$, and this violates the hypothesis. Then, for every $i$, the atom occurrences in $T_i$ come either from $\mu$ or from $\nu$, but not from both. It must also be the case that there are $k$ and $k+1$ in $1, \dots, h$ such that $T_1, \dots, T_k$ have all their atom occurrences in $\mu$ and $T_{k+1}, \dots, T_h$ have all their atom occurrences in $\nu$, otherwise there would be cases of $b \lhd a$ for some $a$ in $\mu$ and $b$ in $\nu$. Then take $U = \langle T_1; \dots; T_k \rangle$ and $V = \langle T_{k+1}; \dots; T_h \rangle$.

**2** $T = [T', T'']$, where $T' \neq \circ \neq T''$: There must be $a$ in $T'$ and $b$ in $T''$, or $a$ in $T''$ and $b$ in $T'$, such that $a$ is in $\mu$ and $b$ is in $\nu$. But then $a \downarrow b$, and this violates the hypothesis. Therefore, this case is actually impossible.

**3** $T = (T', T'')$, where $T' \neq \circ \neq T''$: The argument is the same as for Case 2.

Therefore, the lemma is proved for any $\lhd$-partition of $\mathsf{w}\,T$. A similar argument holds for $\downarrow$- and $\uparrow$-partitions, made simpler by the fact that we should not worry about seq orders, as we had to do in Case 1 above. □

**2.2.9 Theorem** *Two structures are equivalent if and only if they have the same web.*

**Proof** The 'only if' part is trivial, so let us concentrate on the 'if' one. Let $S$ and $T$ be two structures in normal form; we have to prove that if $\mathsf{w}\,S = \mathsf{w}\,T$ then $S = T$. Let us proceed by structural induction on $S$. In the base cases when $S = \circ$ or $S = a$ we trivially get $S = T$. Suppose now that there are $P$ and $Q$ such that $S = \langle P; Q \rangle$ and $P \neq \circ \neq Q$. The couple $(\mathsf{w}\,P, \mathsf{w}\,Q)$ is then a $\lhd$-partition of $\mathsf{w}\,S$, and therefore of $\mathsf{w}\,T$. By Lemma 2.2.8 there are $U$ and $V$ such that $T = \langle U; V \rangle$ and $\mathsf{w}\,U = \mathsf{w}\,P$ and $\mathsf{w}\,V = \mathsf{w}\,Q$, and then the induction hypothesis applies. Similar arguments hold when $S = [P, Q]$ and $S = (P, Q)$, where $P \neq \circ \neq Q$. □

## 3 Synthesis of a Formal System

In this section we will synthesise the formal system SBV, which is, as we will see, approximately a symmetric closure of system BV. Before plunging into the technicalities, and at the cost of some repetition, I believe it is useful to provide an informal account of what we will see later. Many of the intuitions valid in general for the calculus of structures can be seen here at work in the special case of the definition of system SBV.

We have by now a convenient syntax and an intuitive, albeit vague, 'space-time' interpretation. Getting back to our previous example, let us consider the problem of designing a system, equivalent to, or conservatively extending, multiplicative linear logic. We would also like this system to deal with structures without having the problem of the non-deterministic partitioning of the context seen above for $\otimes$. In the partitioning into two branches, double-premiss inference rules are doomed to lose some of the possibilities left open in the conclusion, which we want to retain. Let us see then if we can do this job with single-premiss rules. Dealing with copar under this constraint requires a big departure from Gentzen's sequents style: In the sequent calculus a formula is decomposed in its main connective, and its context is split. In the calculus of structures a formula is moved into another formula in a way determined by their local structure, and their context stays fixed. The underlying claim, for which this paper provides some evidence, is that Gentzen's sequent systems, and the tree structure of derivations in them, are perfectly fit for traditional logics, like LK for classical logic, but not necessarily for new, exotic ones, like the multiplicative core of linear logic.

I will better address this issue in the concluding remarks. For now let us just notice that the calculus of structures is rather radical regarding logical rules, i.e., rules applying to connectives: they completely disappear in favour of structural rules. Structures are expressive enough to internalise the tree organisation of a sequent derivation. Inference rules become more capable of controlling what happens in premisses, with respect to what rules can do in the sequent calculus, and having more control in inference rules yields a more efficient management of resources.

**3.1 Definition** An *inference rule* is a scheme of the kind

$$\rho\,\frac{T}{R}\quad ,$$

where $\rho$ is the *name* of the rule, $T$ is its *premiss* and $R$ is its *conclusion*; rule names are denoted by $\rho$ and $\pi$. In an inference rule, either the premiss or the conclusion can be missing, but not both. When premiss and conclusion in an instance of an inference rule are equivalent, that instance is said *trivial*, otherwise it is said *non-trivial*. A *(formal) system* is a set of inference rules; formal systems are denoted by $\mathscr{S}$. A *derivation* in a certain formal system is a finite sequence of instances of inference rules in the system, and it can consist of just one structure; derivations are denoted by $\Delta$. The premiss of the topmost inference rule instance in a derivation, if present, is called the *premiss* of the derivation; if present, the conclusion of the bottommost inference rule instance is called the *conclusion* of the derivation; the premiss and conclusion of derivations consisting of one structure is that structure. A derivation $\Delta$ whose premiss is $T$, conclusion is $R$, and whose inference rules are in $\mathscr{S}$ is indicated by $\begin{matrix} T \\ \Delta\Vert\mathscr{S} \\ R \end{matrix}$ (the name $\Delta$ can be omitted). The *length* of a derivation is the number of instances of inference rules appearing in it. Two systems $\mathscr{S}$ and $\mathscr{S}'$ are *strongly equivalent* if for every derivation $\begin{matrix} T \\ \Vert\mathscr{S} \\ R \end{matrix}$ there exists a derivation $\begin{matrix} T \\ \Vert\mathscr{S}' \\ R \end{matrix}$, and vice versa.

There are, in the sequent calculus, two complementary, dynamic views of derivations, which we can adapt to our case:

1 The *top-down* view: premisses join (in trees) to form new conclusions, and the derivation grows toward its conclusion; this can be called the *deductive* viewpoint.

2 The *bottom-up* view: the conclusion is the starting point and inference rules are used to reach the desired premisses; this can be called the *proof-construction* viewpoint.

For the moment, we should try not to assign any traditional, special meaning to the action of making a derivation grow upward or downward. Symmetry will be broken at last by the introduction of a logical axiom and, with it, of the concept of *proof*, which in this paper stays firmly in the tradition.

After choosing single-premiss rules, we are in a situation of extensive symmetry: derivations are sequences of inferences (top-down symmetry), par and copar are the same kind of structure with two different names, seq is self-dual, and we have a symmetric intuitive interpretation for communication, meant as the possibility of interaction. This state of affairs is represented in Figure 4, where arrows stand for the direction of growth of a derivation (please take this picture only as an informal, intuitive suggestion). Interaction itself will of course involve negation (in linear logic it is realised in the identity logical axiom, for example). For the time being we deal with the structural relations seq, par and copar, which do not involve negation.

| | | |
|---|---|---|
| $\Uparrow$ | $\Uparrow$ | $\Uparrow$ |
| $R$ and $T$ do not communicate | $R$ and $T$ communicate | $R$ and $T$ do not communicate |
| $\parallel$ | $\parallel$ | $\parallel$ |
| $\langle R; T\rangle$ | $[R, T]$ | $(R, T)$ |
| $\parallel$ | $\parallel$ | $\parallel$ |
| $R$ and $T$ do not communicate | $R$ and $T$ do not communicate | $R$ and $T$ communicate |
| $\Downarrow$ | $\Downarrow$ | $\Downarrow$ |

**Fig. 4** *Communication along the direction of growth of a derivation*

I will 'distil' logical rules out of conservation laws. Consider the following template of a rule, which I call *merge*:

$$\mathsf{g}{\downarrow}\,\frac{S\{Q\}}{S[R,T]} \quad ;$$

here we want to determine the laws which $Q$ must obey. The rule says that if the structures $R$ and $T$ communicate in the context $S\{\ \}$, then they can be replaced with the structure $Q$. How can we choose $Q$? Why not say that $Q$ is any structure made with all and only the atoms in $R$ and $T$, and such that $Q$ conserves all the structural relations that atoms have in $R$ and $T$, and otherwise is free? Here two things are conserved:

1 the atoms, their number and polarity, and

2 their space-temporal relations in the two communicating structures.

It turns out that this almost works, but if we want cut elimination we have to add a third law of conservation, which preserves a sort of integrity. The law says that

3 if $a \uparrow b$ in $R$ and $c \uparrow d$ in $T$, it cannot be in $Q$ that $a \uparrow d$, $b \uparrow c$, $\neg(a \uparrow c)$ and $\neg(b \uparrow d)$.

Later on I will get back to this condition, which is probably obscure now. For the sake of symmetry, I also add its 'cocondition', in which $\uparrow$ is replaced with $\downarrow$.

The preceding laws find a natural definition by relation webs. We can consider a set, denoted by $R \mathbin{\Diamond} T$ and called *merge set*, where we collect all the structures $Q$ that respect the conditions above. We know how to obtain these structures thanks to the characterisation by relation webs that we studied in the previous section. The rule $\mathsf{g}{\downarrow}$ is of course of limited practical use, because it requires picking up structures out of a large and difficult-to-compute set. Luckily, we can also characterise the merge set recursively, and then through a straightforward process of unfolding, we can obtain two simple rules that equivalently replace $\mathsf{g}{\downarrow}$: the *switch* ($\mathsf{s}$) and *seq* ($\mathsf{q}{\downarrow}$) rules.

There is another situation where communication occurs: between two structures $R$ and $T$ in a copar, while going downward in a derivation. I look then for explicit instances of this other rule, called *comerge*:

$$\mathsf{g}{\uparrow}\,\frac{S(R,T)}{S\{Q\}} \quad ,$$

where $Q$ is again any structure in $R \lozenge T$. After arguments symmetric to the ones used for the merge rule, I obtain two rules. One of them is again the switch rule, the other is new and is called *coseq* ($\mathsf{q}{\uparrow}$). Together the rules form system $\mathsf{SBVc} = \{\mathsf{q}{\downarrow}, \mathsf{q}{\uparrow}, \mathsf{s}\}$. In $\mathsf{SBVc}$, rule $\mathsf{q}{\downarrow}$ is the corule of $\mathsf{q}{\uparrow}$ and $\mathsf{s}$ is its own corule. In a corule the premiss and the conclusion are negated and exchanged with respect to the corresponding rule. Still perfect top-down symmetry, as expected.

At this point we just have to add interaction rules, to form system $\mathsf{SBV}$. They correspond to identity axioms and cut rules. There is no surprise in dealing with identity, but cut in the calculus of structures shows an interesting property. It is in fact possible, by using the switch and coseq rules, to replace its generic version by a version that only deals with atoms. This is a consequence of our insistence on maintaining a top-down symmetry. This fact entails two important properties:

1 it helps considerably in simplifying the cut elimination argument;

2 it separates the normal cut rule into two rules, one dealing with negation (so, interaction), and the other with order (or structure).

These properties are exhibited in all systems in the calculus of structures, and actually system $\mathsf{SBV}$ is the simplest example studied so far. The subsection on interaction insists on derivability results, which are responsible for the separation of the cut rule into 'subrules'.

Here is the plan of the section: the merge rules will be derived from relation webs in the first subsection. In the second and third subsections we will derive what I call the 'structure fragment' of $\mathsf{SBV}$. In the fourth subsection, we will study the 'interaction fragment' of $\mathsf{SBV}$ and some of its properties together with the structure one.

## 3.1 Merge Rules and Relation Webs

Intuitively, interacting means allowing the application of inference rules (with the final purpose of annihilating atoms). As we briefly mentioned already, the structures in a par context are allowed to interact along a derivation above the par, and cannot do so below the par. Dually, the structures in a copar context can interact going downward and cannot do so going upward. The structures in a seq context cannot interact either going up or going down in a derivation. The structures $(R,T)$ and $[R,T]$ are then at the extremes of the possibilities of interaction between $R$ and $T$. We can gather intermediate situations in the *merge set* of $R$ and $T$.

In this subsection I will give two definitions of merge set. The first one is more 'semantic' in nature than the second one. I will prove that they are in fact equivalent, and the second definition will pave the way to the extraction of simple inference rules that compute the merge set, which is the subject of the next subsection. This subsection can be skipped by readers ignoring relation webs.

Let us firstly consider the following notion of *immersion* of a structure in another.

**3.1.1 Definition** We say that $R$ is *immersed* in $Q$ if:

**1** the atom occurrences of $R$ are atom occurrences of $Q$: $\mathsf{occ}\,R \subseteq \mathsf{occ}\,Q$;

**2** the structural relations are respected: $\triangleleft_R \subseteq \triangleleft_Q$, $\downarrow_R \subseteq \downarrow_Q$, $\uparrow_R \subseteq \uparrow_Q$.

For example, all the structures immersed in $S = [\langle a; b\rangle, c]$ are: $\circ$, $a$, $b$, $c$, $\langle a; b\rangle$, $[a, c]$, $[b, c]$ and $[\langle a; b\rangle, c]$. The substructures of $S$ are only $\circ$, $a$, $b$, $c$, $\langle a; b\rangle$ and $[\langle a; b\rangle, c]$.

**3.1.2 Remark** Given the structure $Q$, we can find all the structures immersed in it simply by choosing their atoms in $Q$ in all the possible ways. Suppose the set $\xi \subseteq \mathsf{occ}\,Q$ is chosen, and be $\mathsf{w}\,Q = (\mathsf{occ}\,Q, \triangleleft, \downarrow, \uparrow)$. Consider the web candidate $\zeta = (\xi, \triangleleft\cap\xi^2, \downarrow\cap\xi^2, \uparrow\cap\xi^2)$: since $\mathsf{w}\,Q$ obeys $\mathsf{s}_1$–$\mathsf{s}_7$, so does $\zeta$. By Theorem 2.2.7, a structure $R$ exists such that $\mathsf{w}\,R = \zeta$; the structure $R$ is immersed in $Q$, by definition, and is unique modulo equivalence by Theorem 2.2.9.

How can we compose two structures? The following definition is in part very natural, and in part motivated by our need of getting a cut elimination theorem in the end.

**3.1.3 Definition** Given two structures $R$ and $T$ such that $\mathsf{occ}\,R \cap \mathsf{occ}\,T = \emptyset$, the *merge set* $R \Diamond T$ is the set of all the structures $Q$ such that $R$ and $T$ are immersed in $Q$ and:

$\mathsf{m}_1$ $\mathsf{occ}\,R \cup \mathsf{occ}\,T = \mathsf{occ}\,Q$;

$\mathsf{m}_2$ for all the distinct atom occurrences $a, b \in \mathsf{occ}\,R$, and $c, d \in \mathsf{occ}\,T$ the following hold in $Q$:

$$\mathsf{m}_2^{\downarrow} \qquad (a \downarrow b) \wedge (c \downarrow d) \wedge (a \downarrow d) \wedge (b \downarrow c) \Rightarrow (a \downarrow c) \vee (b \downarrow d) \quad ,$$

$$\mathsf{m}_2^{\uparrow} \qquad (a \uparrow b) \wedge (c \uparrow d) \wedge (a \uparrow d) \wedge (b \uparrow c) \Rightarrow (a \uparrow c) \vee (b \uparrow d) \quad .$$

When no confusion arises, we denote with $R \Diamond T$ both the merge set and any of its elements.

The merge set $R \Diamond T$ is obtained from $R$ and $T$ in a free way, provided that atoms, and the relations between them, are conserved. There are also further constraints imposed by the condition $\mathsf{m}_2$. I do not have at this time an intuitive, *a priori* understanding of this condition, but I can offer a very important technical one, *a posteriori*: removing the condition $\mathsf{m}_2$ leads to the impossibility of eliminating cuts.

It is maybe helpful visualising all the possibilities of interaction in the case, say, when $R = [a, b]$ and $T = [c, d]$. After ignoring symmetric cases due to permutations, four possibilities are allowed:

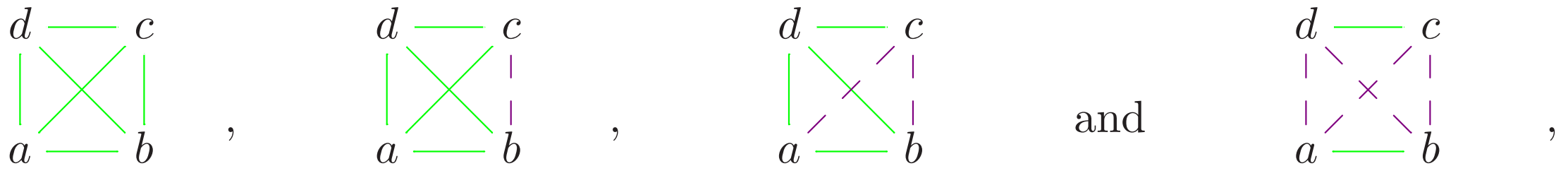

*e.g.*, $[a, b, c, d]$ *e.g.*, $[a, \langle b; c\rangle, d]$ *e.g.*, $[\langle [a, b]; c\rangle, d]$ *e.g.*, $\langle [a, b]; [c, d]\rangle$

where $b$ – – $c$ means $\neg(b \downarrow c)$; below the diagrams, examples of structures in $R \Diamond T$ are shown. The following case is ruled out by the square property $\mathsf{s}_7^{\downarrow}$ (see Theorem 2.2.4; there is no structure having this representation):

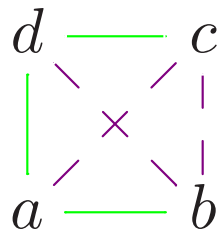

.

The only case ruled out by the condition $\mathsf{m}_2$ (specifically by $\mathsf{m}_2^{\downarrow}$) is this:

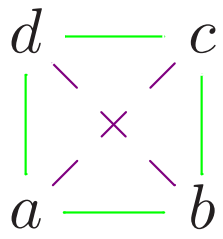

or, equivalently,

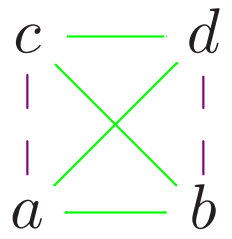

.

Therefore, for example, $[\langle a;c\rangle,\langle b;d\rangle]\notin R\lozenge T$.

Both $\mathsf{m}_1$ and $\mathsf{m}_2$ can be considered conservation laws. The axiom $\mathsf{m}_1$ enforces the strict application of the immersion relation: atoms are conserved together with the structural relations they had in the two structures that are merged.

To understand the axiom $\mathsf{m}_2$, consider for example $\mathsf{m}_2^{\uparrow}$ as it applies to $R=(a,b)$ and $T=(c,d)$: it forbids the inclusion of $([a,c],[b,d])$ into $R\lozenge T$. This means that, from the point of view of communication, $R$ and $T$ must be considered modules whose integrity can not be violated; either $R$ enters into $T$ or vice versa, but it cannot happen that a single module results from the communication of the two modules. This situation can be considered an *integrity conservation*.

**3.1.4 Remark** We could naturally extend the merge operator to sets of structures: Given the sets of structures $\phi$ and $\psi$, define $\phi\lozenge\psi$ as the set $\{\,Q\mid\exists R\in\phi.\exists T\in\psi.(Q\in R\lozenge T)\,\}$. Which are the properties of the law of composition $\lozenge$? It turns out immediately that the set $\{\circ\}$ is a unit and $\lozenge$ is commutative. Associativity does not hold: consider for instance $\{a\}\lozenge(\{b\}\lozenge\{(c,d)\})\ni([a,c],[b,d])\notin(\{a\}\lozenge\{b\})\lozenge\{(c,d)\}$.

Of course, Definition 3.1.3 is too implicit to be very helpful in a syntactic setting. I move one step more toward syntax with the following definition, which builds the merge set recursively.

**3.1.5 Definition** Given two structures $R$ and $T$, the *merge set* $R\lozenge T$ is recursively defined as the minimal set of structures such that:

**1** $\langle R;T\rangle,\langle T;R\rangle,[R,T],(R,T)\in R\lozenge T$;

**2** for every $R'$, $R''$, $T'$, $T''$ such that $R=\langle R';R''\rangle$ and $T=\langle T';T''\rangle$, where $R'\neq\circ\neq R''$ or $T'\neq\circ\neq T''$, it holds

$$\{\,\langle Q';Q''\rangle\mid Q'\in R'\lozenge T',\ Q''\in R''\lozenge T''\,\}\subseteq R\lozenge T\quad;$$

**3** for every $R'$, $R''$ and $T'$, $T''$ such that $R'\neq\circ\neq R''$ and $T'\neq\circ\neq T''$ the following hold:

**1** if $R=[R',R'']$ then $\{\,[Q',R'']\mid Q'\in R'\lozenge T\,\}\subseteq R\lozenge T$ ,

**2** if $R=(R',R'')$ then $\{\,(Q',R'')\mid Q'\in R'\lozenge T\,\}\subseteq R\lozenge T$ ,

**3** if $T=[T',T'']$ then $\{\,[T',Q'']\mid Q''\in R\lozenge T''\,\}\subseteq R\lozenge T$ ,

**4** if $T=(T',T'')$ then $\{\,(T',Q'')\mid Q''\in R\lozenge T''\,\}\subseteq R\lozenge T$ .

I invite the reader to check the examples given above against this new definition of merge set. Also note the possible use of the unit in Case 3.1.5.2: Given $R = \langle R'; R''\rangle$ and $T$, the merge set $R \between T$ contains, among others, the structures of the kind $\langle R' \between T; R''\rangle$ and $\langle R'; R'' \between T\rangle$.

The effect of the condition $\mathsf{m}_2$ on Definition 3.1.5 is to put a restriction on the way we build the merge set of a couple of structures whose external shape is that of a par or of a copar. Should the condition not be present, Definition 3.1.5 would be uniform for all the three possible contexts. Instead, given for example $R = (R', R'')$ and $T = (T', T'')$, we cannot put $(R' \between T', R'' \between T'')$ in $R \between T$; we only are allowed to consider $(R' \between T, R'')$, $(R', R'' \between T)$, $(R \between T', T'')$ and $(T', R \between T'')$.

As promised, I am going to prove the equivalence of Definitions 3.1.3 and 3.1.5.

**3.1.6 Theorem** *Definitions 3.1.3 and 3.1.5 are equivalent.*

**Proof** Let us indicate with $\between_1$ and $\between_2$, respectively, the merge operators defined in 3.1.3 and 3.1.5. We have to prove that $Q \in R \between_1 T \Leftrightarrow Q \in R \between_2 T$, for any structures $Q$, $R$ and $T$. Let us proceed by structural induction on $Q$.

**Base Cases**

**1** $Q = \circ$: Then $Q \in R \between_1 T \Leftrightarrow (R = \circ) \wedge (T = \circ) \Leftrightarrow Q \in R \between_2 T$.

**2** $Q = a$: Then $Q \in R \between_1 T \Leftrightarrow \big((R = a) \wedge (T = \circ)\big) \vee \big((R = \circ) \wedge (T = a)\big) \Leftrightarrow Q \in R \between_2 T$.

**Inductive Cases**

**3** $Q = \langle Q'; Q''\rangle$, where $Q' \neq \circ \neq Q''$:

$\Rightarrow$ There are structures $R'$, $R''$ and $T'$, $T''$ such that $R = \langle R'; R''\rangle$, $T = \langle T'; T''\rangle$, $\mathsf{occ}\, R' \cup \mathsf{occ}\, T' = \mathsf{occ}\, Q'$ and $\mathsf{occ}\, R'' \cup \mathsf{occ}\, T'' = \mathsf{occ}\, Q''$ (see Remark 3.1.2). Then $Q' \in R' \between_1 T'$ and $Q'' \in R'' \between_1 T''$. By induction hypothesis we have then $Q' \in R' \between_2 T'$ and $Q'' \in R'' \between_2 T''$, and then, by definition, $Q \in R \between_2 T$.

$\Leftarrow$ If $Q' = R$ and $Q'' = T$, or $Q' = T$ and $Q'' = R$, then Definition 3.1.3 holds, and then $Q \in R \between_1 T$. Otherwise there are $R'$, $R''$ and $T'$, $T''$ such that $R = \langle R'; R''\rangle$, $T = \langle T'; T''\rangle$ and $Q' \in R' \between_2 T'$, $Q'' \in R'' \between_2 T''$. By induction hypothesis, $Q' \in R' \between_1 T'$ and $Q'' \in R'' \between_1 T''$. Definition 3.1.3 then holds for $Q$, and we get $Q \in R \between_1 T$.

**4** $Q = [Q', Q'']$, where $Q' \neq \circ \neq Q''$:

$\Rightarrow$ There are $Q_1, \dots, Q_h$ such that $h \geqslant 2$, $Q = [Q_1, \dots, Q_h]$ and, for $1 \leqslant i \leqslant h$, it holds $Q_i \neq \circ$ and $Q_i$ is not a proper par. There can be no more than one $Q_k$ between $Q_1, \dots, Q_h$ such that $\mathsf{occ}\, Q_k \cap \mathsf{occ}\, R \neq \varnothing \neq \mathsf{occ}\, Q_k \cap \mathsf{occ}\, T$. In fact, suppose that there are two distinct structures, say $Q_k$ and $Q_l$, which enjoy this property: this leads to contradicting the condition $\mathsf{m}_2^{\downarrow}$. We can suppose that $Q_k = (Q'_k, Q''_k)$ and $Q_l = (Q'_l, Q''_l)$ (it would be the same assuming $Q_k = \langle Q'_k; Q''_k\rangle$ or $Q_l = \langle Q'_l; Q''_l\rangle$, independently) and we can take $a$ in $Q'_k$, $b$ in $Q'_l$, $c$ in $Q''_k$ and $d$ in $Q''_l$ such that $a$ and $b$ are in $R$ and $c$ and $d$ are in $T$. We would have then $a \downarrow b$, $c \downarrow d$, $a \downarrow d$, $b \downarrow c$, $a \uparrow c$ and $b \uparrow d$, and this is impossible. Therefore, there is no more than one $Q_k$ between $Q_1, \dots, Q_h$ whose atom occurrences are partly in $R$ and partly in $T$, and then, since $h \geqslant 2$, there is at least one $Q_m$ whose atom occurrences are all either in $R$ or in $T$. Let us suppose that they are in $R$ (same argument when they are in $T$), i.e., $\mathsf{occ}\, Q_m \subseteq \mathsf{occ}\, R$. Let $\hat{Q} = [Q_1, \dots, Q_{m-1}, Q_{m+1}, \dots, Q_h]$, so that $Q = [\hat{Q}, Q_m]$; there is then $R'$ such that $R = [R', Q_m]$ and $\mathsf{occ}\, R' \subseteq \mathsf{occ}\, \hat{Q}$, and by 3.1.3 it holds $\hat{Q} \in R' \between_1 T$. By induction hypothesis we have $\hat{Q} \in R' \between_2 T$ and then $Q \in R \between_2 T$ by Definition 3.1.5.

$\Leftarrow$ If $Q' = R$, $Q'' = T$ or $Q' = T$, $Q'' = R$ then Definition 3.1.3 holds trivially. Suppose instead that $R = [R', Q'']$ and $Q' \in R' \between_2 T$. By induction hypothesis $Q' \in R' \between_1 T$, and then $Q = [Q', Q''] \in R \between_1 T$. Consider in fact $a$ and $b$ in $R$ and $c$ and $d$ in $T$, and let us check the condition $\mathsf{m}_2^{\downarrow}$. If $a$ and $b$ are both in $R'$ the condition holds because $Q' \in R' \between_1 T$. If $a$ and $b$ are both in $Q''$ then $a \downarrow c$ and $b \downarrow d$ are true because $c$ and $d$ are in $Q'$. If $a$ is in $R'$ and $b$ is in $Q''$ then $b \downarrow d$. If $a$ is in $Q''$ and $b$ is in $R'$ then $a \downarrow c$. The condition $\mathsf{m}_2^{\uparrow}$ is easily checked to be valid. The same argument applies to the case when $T = [Q', T'']$ and $Q'' \in R \between_2 T''$.

**5** $Q = (Q', Q'')$, where $Q' \neq \circ \neq Q''$: Analogous to Case 4. □

The merge set behaves well with respect to negation:

**3.1.7 Theorem** *$S \in R \between T$ iff $\bar{S} \in \bar{R} \between \bar{T}$.*

**Proof** Negation exchanges par and copar, and they are used symmetrically in the definitions of merge set. Moreover, it does not matter to the definitions of merge whether atoms are negative or not. □

## 3.2 The Structure Fragment

Remember that we consider communication the possibility to interact. It is natural to say that two structures which communicate can be interleaved: an element in their merge set is chosen. Merging two structures $R$ and $T$ eliminates of course many possibilities of interaction; while the relations in $R$ and $T$ are conserved, each relation between an atom in $R$ and an atom in $T$ needs not be. Making a derivation grow means narrowing, in the direction of the growth, the freedom of interacting (i.e., the amount of communication) of the structures involved.

**3.2.1 Definition** The following inference rules $\mathsf{g}\downarrow$ and $\mathsf{g}\uparrow$ are called respectively *merge* and *comerge*:

$$\mathsf{g}\downarrow \frac{S\{R \between T\}}{S[R, T]} \qquad \text{and} \qquad \mathsf{g}\uparrow \frac{S(R, T)}{S\{R \between T\}} \quad .$$

The system $\{\mathsf{g}\downarrow, \mathsf{g}\uparrow\}$ is called *merge system* $\mathsf{V}$, or $\mathsf{MV}$.

Used together, the two inference rules give rise to this particular situation, which clearly shows the intermediate position of the structures in the merge set:

$$\begin{array}{l} \mathsf{g}\uparrow \dfrac{S(R, T)}{S\{R \between T\}} \\ \mathsf{g}\downarrow \dfrac{\phantom{S\{R \between T\}}}{S[R, T]} \end{array} \quad .$$

The rules $\mathsf{g}\downarrow$ and $\mathsf{g}\uparrow$ are not very friendly: they involve the merge set, which is hard to obtain, in general. Moreover, they hide simple relations between different kinds of context. They can be replaced with simpler rules, and we can obtain these rules through a process of unfolding, by Definition 3.1.5.

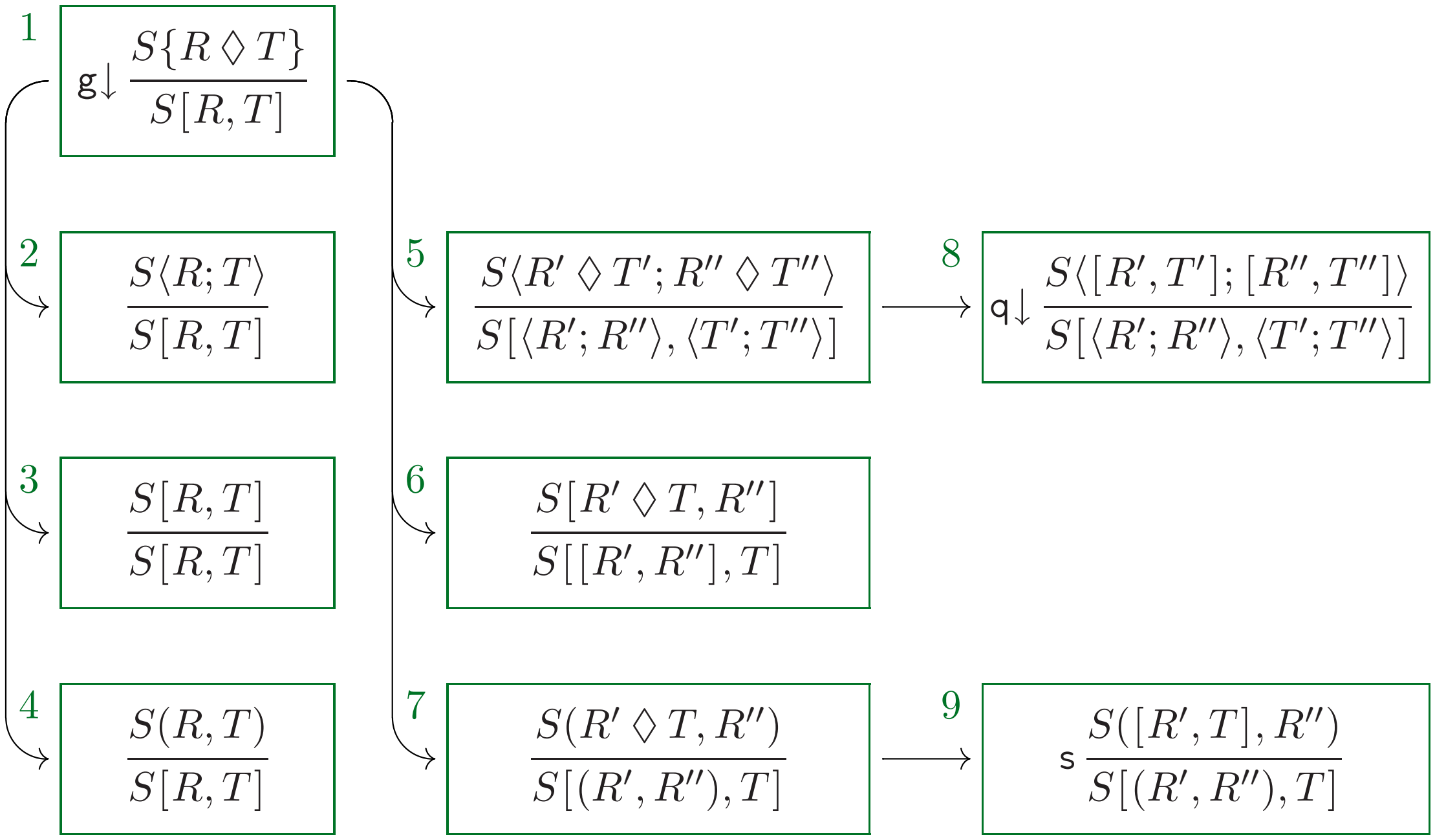

**Fig. 5** *Unfolding of the merge rule*

**3.2.2 Definition** An inference rule $\rho$ is *derivable* for the formal system $\mathscr{S}$ if $\rho \notin \mathscr{S}$ and for every instance $\rho\,\frac{T}{R}$ there exists a derivation $\begin{array}{c}T\\ \Vert\mathscr{S}\\ R\end{array}$.

In Figure 5 the unfolding of $\mathsf{g}\!\downarrow$ is shown. The following informal argument will be made rigorous later. The rule 1 ($\mathsf{g}\!\downarrow$) instantiates as rules 2, 3 and 4, according to the base in the inductive definition of merge set. The inductive cases are considered in instances 5, 6 and 7. The rules 8 and 9 are instances of 5 and 7, respectively. Now, the rules 2 and 4 are respectively instances of 8 and 9, so they are derivable for all the other ones; the rule 3 is vacuous; the rule 6 is equivalent to 1 (each is an instance of the other). In the end, $\mathsf{g}\!\downarrow$ can be replaced with 5, 7, 8 and 9; but, since 5 is also derivable for $\{1,8\}$, and 7 is derivable for $\{1,9\}$, it must be the case, to avoid an infinite regress, that 1 is derivable for $\{8,9\}$.

A dual argument holds for $\mathsf{g}\!\uparrow$, and is shown in Figure 6. Please note that the rule 9 in Figure 6 is the same as the rule 9 in Figure 5.

The following theorems make the unfolding argument rigorous, but let us before give a name to the newly discovered inference rules.

**3.2.3 Definition** The following inference rules $\mathsf{q}\!\downarrow$, $\mathsf{q}\!\uparrow$ and $\mathsf{s}$ are called respectively *seq*, *coseq* and *switch*:

$$\mathsf{q}\!\downarrow \frac{S\langle [R,T];[R',T']\rangle}{S[\langle R;R'\rangle,\langle T;T'\rangle]} \quad , \qquad \mathsf{q}\!\uparrow \frac{S(\langle R;T\rangle,\langle R';T'\rangle)}{S\langle (R,R');(T,T')\rangle} \qquad \text{and} \qquad \mathsf{s}\,\frac{S([R,T],R')}{S[(R,R'),T]} \quad .$$

The system $\{\mathsf{q}\!\downarrow,\mathsf{q}\!\uparrow,\mathsf{s}\}$ is called *core symmetric basic system* $\vee$, or SBVc.

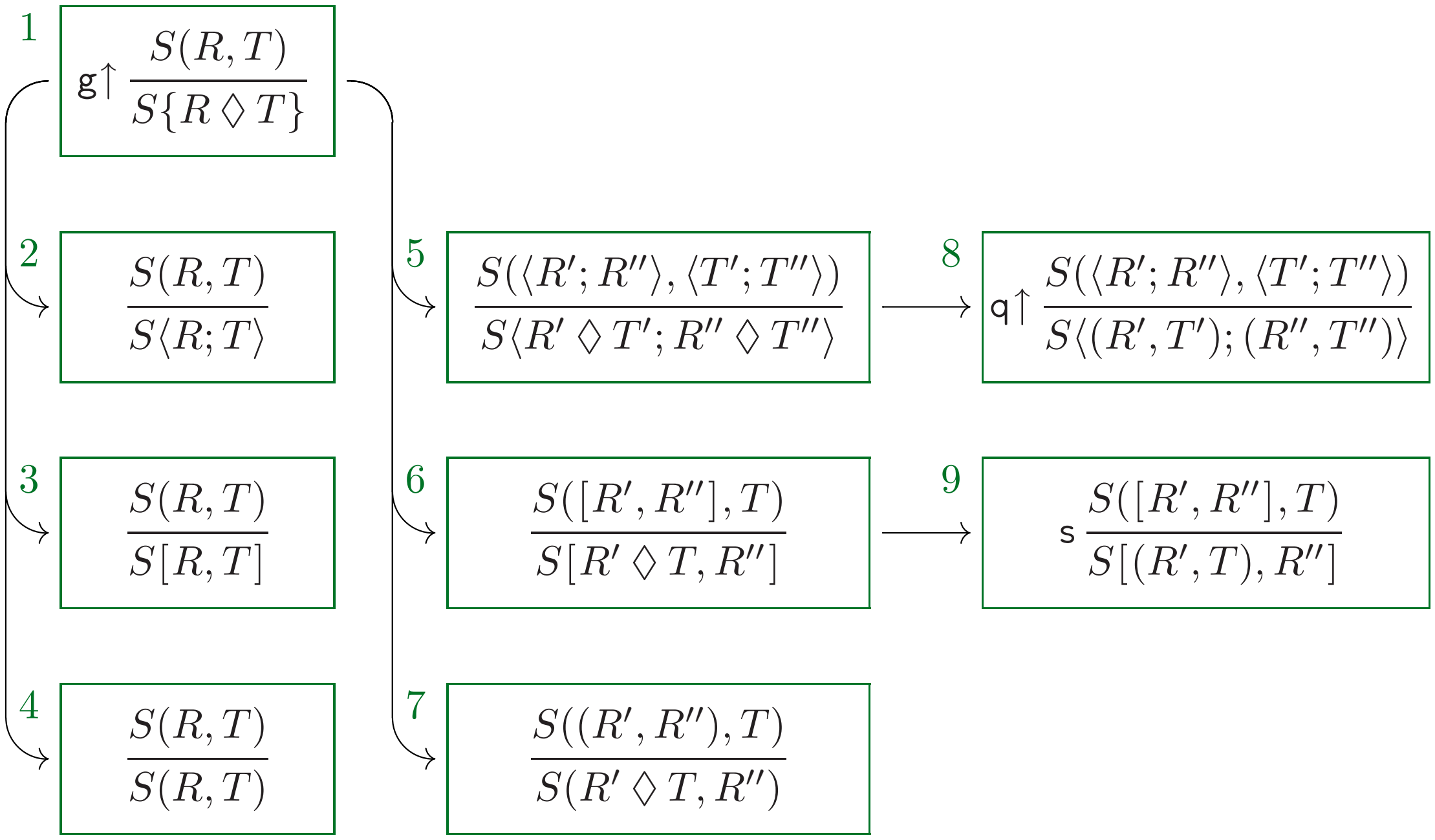

**Fig. 6** *Unfolding of the comerge rule*

The core fragment in the system we are developing corresponds precisely to the structure fragment. In other systems, the core is a proper fragment of the structure one, in particular this is true for the extension of system BV studied in [14]. I refer the reader to that paper for a definition of core fragment, which I do not show here, since in our present, small system it would not make any discrimination.

**3.2.4 Remark** All the three rules in system SBVc can be used in the trivial way in which premiss and conclusion are equivalent, for example:

$$\mathsf{q}\!\downarrow \frac{\langle [R,\circ];[R',\circ]\rangle}{[\langle R;R'\rangle,\langle \circ;\circ\rangle]} \quad , \qquad \mathsf{q}\!\uparrow \frac{(\langle \circ;T\rangle,\langle \circ;T'\rangle)}{\langle (\circ,\circ);(T,T')\rangle} \qquad \text{and} \qquad \mathsf{s}\,\frac{([R,T],\circ)}{[(R,\circ),T]} \quad .$$

Given three kinds of context (i.e., seq, par and copar), there are six possible ways in which a context is directly immersed into a different one. It turns out that each possible nesting appears as a premiss or as a conclusion in exactly one rule. The law for the creation of corules out of rules is clear at this point, and there is no surprise: premisses are exchanged with conclusions and submitted to negation. Please note that switch is its own corule.

The rule switch is common to the 'down' and 'up' fragments, and merge and comerge can be replaced with switch and the corresponding 'down' or 'up' seq rule.

**3.2.5 Theorem** *The rule* $\mathsf{g}\!\downarrow$ *is derivable for* $\{\mathsf{q}\!\downarrow,\mathsf{s}\}$.

**Proof** Let $\mathsf{g}\!\downarrow \dfrac{S\{Q\}}{S[R,T]}$ be an instance of $\mathsf{g}\!\downarrow$, where $Q \in R \between T$. Let us proceed by structural induction on $Q$ and use Definition 3.1.5.

**Base Cases**

**1** $Q = \circ$: It must be $R = T = \circ$, then $S\{Q\} = S[R,T]$.

**2** $Q = a$: It must be either $R = a$ and $T = \circ$, or $R = \circ$ and $T = a$; in either case $S\{Q\} = S[R,T]$.

**Inductive Cases**

**3** $Q = \langle R;T\rangle$: Consider $\mathsf{q}{\downarrow}\,\dfrac{S\langle [R,\circ];[\circ,T]\rangle}{S[\langle R;\circ\rangle,\langle\circ;T\rangle]}$. The same argument holds when $Q = \langle T;R\rangle$.

**4** $Q = [R,T]$: It holds $S\{Q\} = S[R,T]$.

**5** $Q = (R,T)$: Consider $\mathsf{s}\,\dfrac{S([\circ,R],T)}{S[(\circ,T),R]}$.

Suppose now that no case between the preceding ones applies. Suppose that there are $Q'$ and $Q''$ such that $Q' \neq \circ \neq Q''$:

**6** $Q = \langle Q';Q''\rangle$: There must be $R'$, $R''$, $T'$, $T''$ such that $Q' \in R' \lozenge T'$, $Q'' \in R'' \lozenge T''$ and $R = \langle R';R''\rangle$, $T = \langle T';T''\rangle$. Apply the induction hypothesis on

$$\mathsf{q}{\downarrow}\,\dfrac{\mathsf{g}{\downarrow}\,\dfrac{\mathsf{g}{\downarrow}\,\dfrac{S\langle Q';Q''\rangle}{S\langle Q';[R'',T'']\rangle}}{S\langle [R',T'];[R'',T'']\rangle}}{S[\langle R';R''\rangle,\langle T';T''\rangle]}\quad .$$

**7** $Q = [Q',Q'']$: There must be either $R'$ such that $R = [R',Q'']$ and $Q' \in R' \lozenge T$, or $T''$ such that $T = [Q',T'']$ and $Q'' \in R \lozenge T''$. Apply the induction hypothesis on

$$\mathsf{g}{\downarrow}\,\dfrac{S[Q',Q'']}{S[R',T,Q'']}\qquad\text{or}\qquad \mathsf{g}{\downarrow}\,\dfrac{S[Q',Q'']}{S[Q',R,T'']}\quad .$$

**8** $Q = (Q',Q'')$: There must be either $R'$ such that $R = (R',Q'')$ and $Q' \in R' \lozenge T$, or $T''$ such that $T = (Q',T'')$ and $Q'' \in R \lozenge T''$. Apply the induction hypothesis on

$$\mathsf{s}\,\dfrac{\mathsf{g}{\downarrow}\,\dfrac{S(Q',Q'')}{S([R',T],Q'')}}{S[(R',Q''),T]}\qquad\text{or}\qquad \mathsf{s}\,\dfrac{\mathsf{g}{\downarrow}\,\dfrac{S(Q',Q'')}{S(Q',[R,T''])}}{S[R,(Q',T'')]}\quad .$$

□

The ‘coproof’ of the last theorem is a proof of the following one.

**3.2.6 Theorem** *The rule* $\mathsf{g}{\uparrow}$ *is derivable for* $\{\mathsf{q}{\uparrow},\mathsf{s}\}$.

**3.2.7 Theorem** *The systems* MV *and* SBVc *are strongly equivalent.*

**Proof** It immediately follows from Theorems 3.2.5 and 3.2.6 and the fact that the rules of SBVc are instances of those of MV. □

Figure 7 shows the three rules in their graphical representation. To get an intuitive understanding of our rules, it is better to stick to a particular viewpoint: let us consider derivations from the bottom-up, proof-construction perspective.

The rule $\mathsf{q}{\downarrow}$ models the mutual behaviour of seq and par, i.e., sequentialisation and communication: two seq structures that can interact can do so in their parts, provided that the order is respected. The rule also puts a limitation in the way the context can interact with two structures in a seq. Consider $[\langle R;R'\rangle,\langle T;T'\rangle]$: structures $R$ and $R'$ do not communicate, but situations where $R$ communicates with $\langle T;T'\rangle$, or $R'$ with $\langle T;T'\rangle$, or $R$ with $T$ and $R'$ with $T'$ are possible:

$$\mathsf{q}{\downarrow}\,\frac{S\langle[R,\langle T;T'\rangle];[R',\circ]\rangle}{S[\langle R;R'\rangle,\langle\langle T;T'\rangle;\circ\rangle]}\quad,\qquad \mathsf{q}{\downarrow}\,\frac{S\langle[R,\circ];[R',\langle T;T'\rangle]\rangle}{S[\langle R;R'\rangle,\langle\circ;\langle T;T'\rangle\rangle]}\quad,\qquad \mathsf{q}{\downarrow}\,\frac{S\langle[R,T];[R',T']\rangle}{S[\langle R;R'\rangle,\langle T;T'\rangle]}\quad.$$

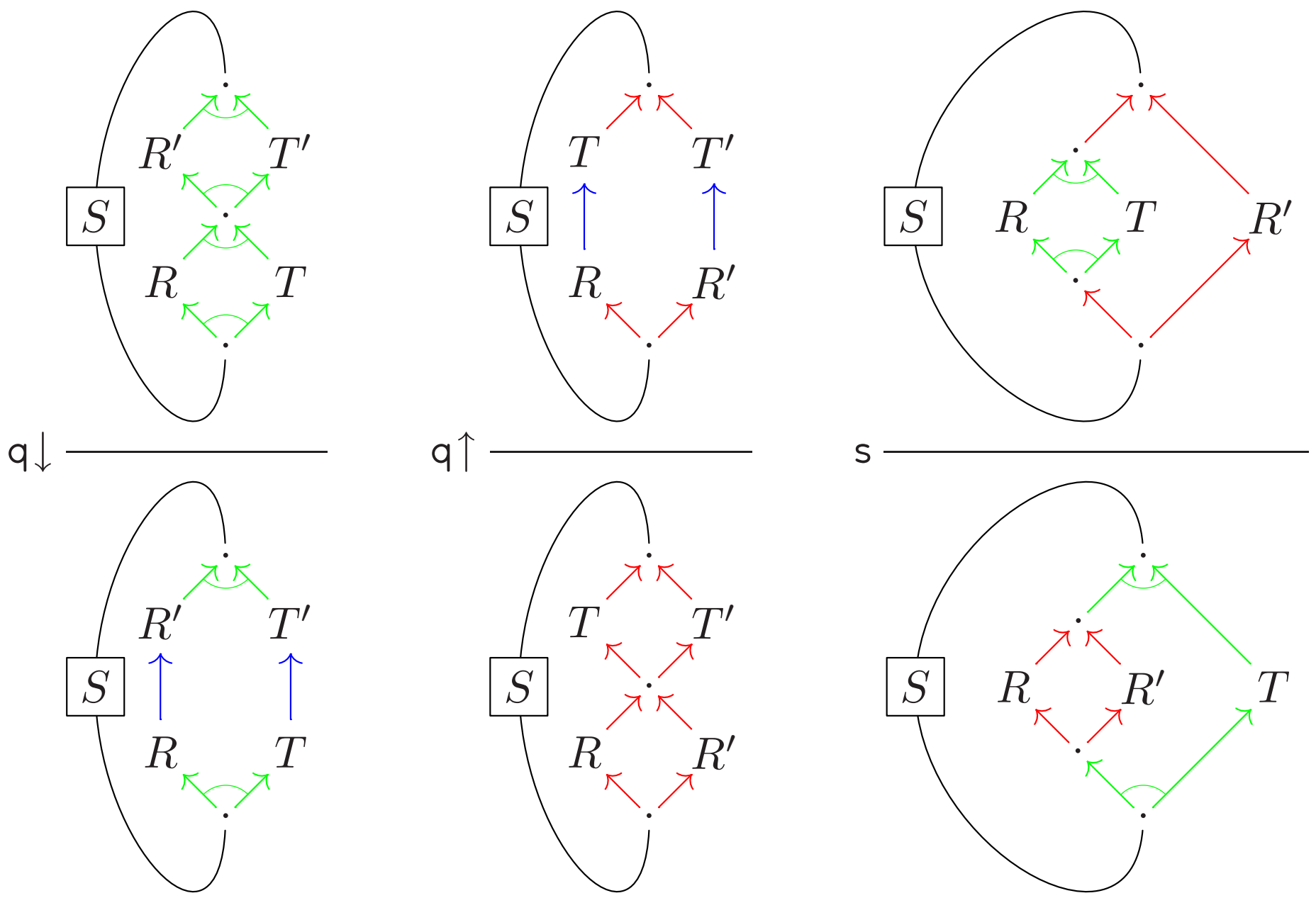

**Fig. 7** *Seq, coseq and switch rules* (*system* SBVc)

On the contrary, it is impossible for $R$ to interact with $T'$ and at the same time for $R'$ to interact with $T$. A notable instance of $\mathsf{q}{\downarrow}$ is

$$\mathsf{q}{\downarrow}\,\frac{S\langle R;T\rangle}{S[R,T]}\quad:$$

if two non-communicating structures $R$ and $T$ can be proved in $S\{\ \}$, *a fortiori* they can be proved in the same context when they communicate.

A good test for checking one's understanding of the $\mathsf{q}{\downarrow}$ rule is verifying that the following is one of its instances:

$$\mathsf{q}{\downarrow}\,\frac{[\langle[\langle R;T\rangle,P',U'];P''\rangle,U'']}{[\langle R;T\rangle,\langle P';P''\rangle,U',U'']}\quad.$$

We will meet again this and other similar difficult cases in the proof of the splitting theorem.

The rule $\mathsf{s}$ models the mutual behaviour of par and copar. A good way of understanding this rule is by considering the times rule of linear logic:

$$\otimes\,\frac{\vdash A,\Phi\quad\vdash B,\Psi}{\vdash A\otimes B,\Phi,\Psi}\quad,$$

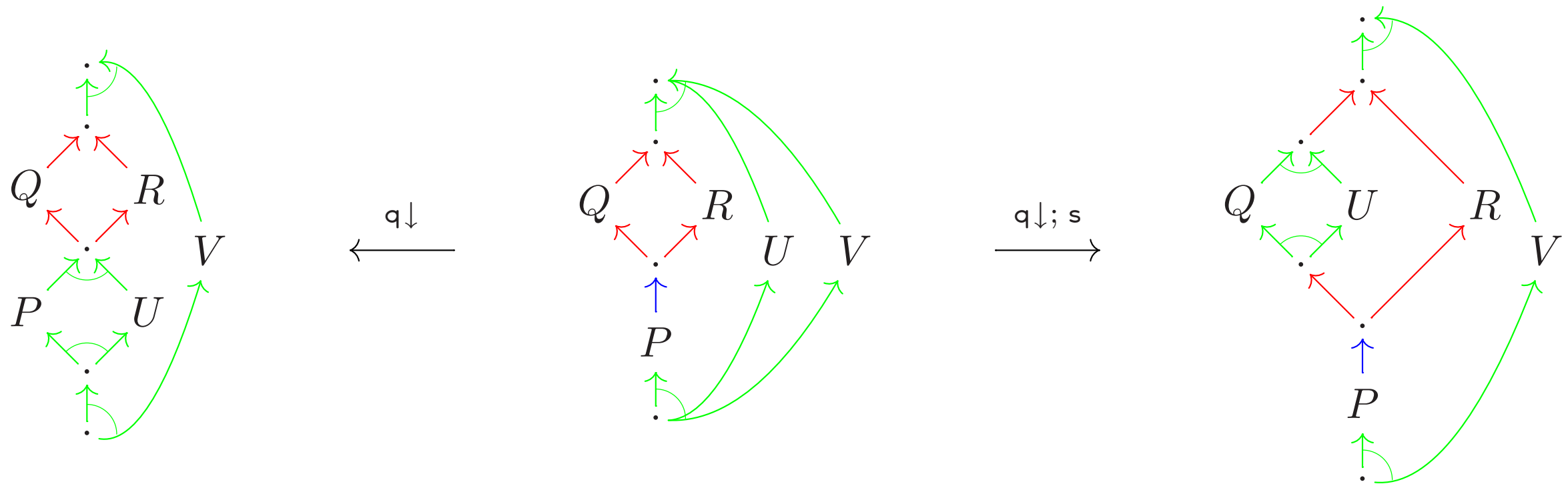

**Fig. 8** *Two possible rewritings of the central structure in a proof search*

where $A$ and $B$ are formulae and $\Phi$ and $\Psi$ are multisets of formulae. In our system the par context plays the role of the comma in linear logic's right side of sequents (so, it is essentially the connective $⅋$), and copar plays the role of the connective $\otimes$. The two branches in the derivation generated by an application of the rule $\otimes$ do not communicate in a proof construction. By applying the rule, one has to choose which part of the context communicates with which formula in the times relation: in our case $\Phi$ can interact with $A$ and $\Psi$ with $B$. The difference between linear logic and our case is that in linear logic the rule $\otimes$ compels an early choice of the partition of the context, while our system does not. The same situation in the times rule above is realised like this, by using switch:

$$\mathsf{s}\frac{\mathsf{s}\dfrac{([R_A, T_\Phi], [R'_B, U_\Psi])}{[([R_A, T_\Phi], R'_B), U_\Psi]}}{[(R_A, R'_B), T_\Phi, U_\Psi]} \quad .$$

Of course the two applications of $\mathsf{s}$ are independent, so a certain part of the context can be brought to communicate inside a copar on demand during the search for a proof, which is something that cannot be done inside the sequent system of linear logic. A notable instance of $\mathsf{s}$ is

$$\mathsf{s}\frac{S(R,T)}{S[R,T]} \quad :$$

it is of course reminiscent of the rule $\mathsf{mix}$ of linear logic, especially when $S\{\circ\} = \circ$. I will formally investigate the relation between our system and linear logic in Section 5.

The use of units makes these rules particularly flexible: the example in Figure 8 illustrates the use of the rules $\mathsf{q}{\downarrow}$ and $\mathsf{s}$ in proof construction, i.e., when considering the rewriting of structures going upward in a derivation. In the central structure of the figure, both $U$ and $V$ are applicable to the entire time span of the structure. Going upward in the derivation means *deciding*, *making choices* about which substructure interacts with which other. In the left structure, $U$ has been brought in communication with $P$ (after a single application of $\mathsf{q}{\downarrow}$); in the right one, it has been brought in communication with $Q$ (after an application of $\mathsf{q}{\downarrow}$ and one of $\mathsf{s}$). A seq maintains ordering, a copar acts as a switch for the structures coming from the outside.

At this point, a good test for the reader is trying to prove Proposition 4.2.2.

The rule $\mathsf{q}\uparrow$ models the mutual behaviour of copar and seq. In $S\langle(R,R');(T,T')\rangle$ no substructure between $R$, $R'$, $T$ and $T'$ communicates with any other one. Rewriting that structure to $S(\langle R;T\rangle,\langle R';T'\rangle)$ does not make a difference in this respect, but limits the ability of the substructures in $S\{\ \}$ to interact with $R$, $R'$, $T$ and $T'$. For example, given in $S\{\ \}$ the structure $\langle R'';T''\rangle$ that is in communication with $\langle(R,R');(T,T')\rangle$, it would be possible to reach a state where $R''$ communicates with $R$ and $T''$ communicates with $T'$. This is no longer possible after rewriting $\langle(R,R');(T,T')\rangle$ to $(\langle R;T\rangle,\langle R';T'\rangle)$: in the latter structure $R$ and $T'$ fall into a copar relation. A notable instance of $\mathsf{q}\uparrow$ is

$$\mathsf{q}\uparrow\frac{S(R,T)}{S\langle R;T\rangle}\quad .$$

After establishing that it is intuitively sound, we have to ask ourselves whether $\mathsf{q}\uparrow$ adds something to $\{\mathsf{g}\downarrow\}$ or is derivable.

It turns out immediately that $\mathsf{q}\uparrow$ is not derivable for $\{\mathsf{q}\downarrow,\mathsf{s}\}$. Just consider the following derivation, which cannot be mimicked in $\{\mathsf{q}\downarrow,\mathsf{s}\}$:

$$\mathsf{q}\uparrow\frac{(\langle a;c\rangle,b)}{\langle(a,b);c\rangle}\quad .$$

Analogously, the derivation

$$\mathsf{s}\frac{([a,c],b)}{[(a,b),c]}$$

can be used to show that $\mathsf{s}$ is not derivable for $\{\mathsf{q}\downarrow,\mathsf{q}\uparrow\}$.

We will achieve admissibility of $\mathsf{q}\uparrow$, for a stronger system than $\{\mathsf{q}\downarrow,\mathsf{s}\}$, but we have first to treat negation, by introducing rules for interaction. Until now we have been dealing with structures without looking at the polarity of atoms. We were not interested in knowing if a certain atom was $a$ or $b$ or $\bar{a}$, but just in the fact that a certain place contained an atom. Interaction will consider the whole information carried by atoms, by realising the communication between atoms that are dual through negation. I will deal with it in Subsection 3.4.

## 3.3 Comments on the Structure Fragment

System SBVc is the combinatorial core of the system we are building: it is about maintaining *order* (in a broader sense than the usual mathematical notion). In [28] Retoré studies properties of proof nets for pomset logic, an extension of linear logic with a non-commutative connective, derived from coherence semantics. He shows a rewriting system that yields cut elimination for certain proof nets. That rewriting system is essentially equivalent to system SBVc. Retoré got his rewriting rules from a study of the inclusion relation for series-parallel orders, in [4]. Since series-parallel orders are at the basis of my definition of structures, Retoré's result and mine are similar because they come from the same underlying combinatorial phenomena. The difference between his work and mine is in the use we make of the rewriting system: he applies its rules in the cut elimination procedure, I use them directly to obtain inference rules.

It would be interesting to characterise derivability in SBVc and MV in terms of structural relations. More precisely: There is a natural ordering relation, due to Retoré, $<$ between $\triangleleft$, $\downarrow$ and $\uparrow$, namely $\uparrow < \triangleleft < \downarrow$. The relation $<$ compares the 'energy' of structures, i.e., it compares the number of interactions that they make possible. The more one goes up in a derivation, the more choices one makes, the fewer possibilities of interaction are left. From this point of view the goal of a search for a proof is 'cooling down' a structure to zero.

**3.3.1 Definition** Between structures, the partial order relation $\leqslant$ is defined such that $T \leqslant R$ iff $\mathsf{occ}\, T = \mathsf{occ}\, R$, $\downarrow_T \subseteq \downarrow_R$ and $\uparrow_T \supseteq \uparrow_R$.

I think that this definition is adequate to characterise derivability in a weaker version of MV, one that disallows cut elimination:

**3.3.2 Definition** Let WMV (*weak merge system* V) be the system $\{\mathsf{wg}\downarrow, \mathsf{wg}\uparrow\}$, where $\mathsf{wg}\downarrow$ and $\mathsf{wg}\uparrow$ are like $\mathsf{g}\downarrow$ and $\mathsf{g}\uparrow$ except that the definition of merge set adopted is 3.1.3 where $\mathsf{m}_2$ is not required.

System WMV is then like MV, except that the integrity conservation of par and copar is not enforced. I could show that WMV is strongly equivalent to $\{\mathsf{q}\downarrow, \mathsf{q}\uparrow, \mathsf{ws}\}$, where ws is $\mathsf{ws}\,\dfrac{S([R,T],[R',T'])}{S[(R,R'),(T,T')]}$. I believe that the following statement is true:

**3.3.3 Conjecture** *$T \leqslant R$ iff there exists* $\begin{array}{c}T\\ \Vert\,\mathsf{WMV}\\ R\end{array}$.

At this time I do not see an easy argument to prove it, and this paper is already very long. Anyway, what is really interesting is:

**3.3.4 Problem** *Finding an improved definition of $\leqslant$, by adding conditions based only on webs, such that Conjecture* 3.3.3 *holds for* MV, *instead of* WMV.

There are chances that the necessary, new condition on webs would teach us something about cut elimination, since it would correspond to having the property $\mathsf{m}_2$, which has in turn been introduced for having cut elimination.

Definition 3.3.1 is already useful:

**3.3.5 Remark** In each non-trivial rule instance $\mathsf{q}\downarrow\,\dfrac{T}{R}$, $\mathsf{q}\uparrow\,\dfrac{T}{R}$ and $\mathsf{s}\,\dfrac{T}{R}$ we have $T < R$ (and $T = R$ holds if the rule instances are trivial, see Remark 3.2.4).

**3.3.6 Theorem** *For a given structure $S$, the set of derivations in* SBVc, *whose conclusion is $S$ and where no trivial rule instance appears, is finite*; *the same holds for the similar set of derivations whose premiss is $S$.*

**Proof** It immediately follows from Remark 3.3.5, the fact that the set of structures whose atom occurrences are the same as those of $S$ is finite, and the fact that only a finite number of instances of rules is applicable to each structure. (Remember that we consider structures as equivalence classes.) □

## 3.4 Interaction

**3.4.1 Definition** The following two inference rules $\mathsf{i}{\downarrow}$ and $\mathsf{i}{\uparrow}$ are called respectively *interaction* and *cointeraction* (or *cut*):

$$\mathsf{i}{\downarrow}\,\frac{S\{\circ\}}{S[R,\bar{R}]} \qquad \text{and} \qquad \mathsf{i}{\uparrow}\,\frac{S(R,\bar{R})}{S\{\circ\}} \quad .$$

The following two inference rules $\mathsf{ai}{\downarrow}$ and $\mathsf{ai}{\uparrow}$ are called respectively *atomic interaction* and *atomic cointeraction* (or *atomic cut*):

$$\mathsf{ai}{\downarrow}\,\frac{S\{\circ\}}{S[a,\bar{a}]} \qquad \text{and} \qquad \mathsf{ai}{\uparrow}\,\frac{S(a,\bar{a})}{S\{\circ\}} \quad .$$

Interaction is a natural law of conservation. In atomic interaction (which of course corresponds to an identity axiom in the sequent calculus), two complementary atoms in direct communication annihilate each other and disappear at the same time from the structure in which they are immersed. We can think of ‘charge’ as being conserved. We can add this rule to system SBVc with no hesitation, since we are now able to produce derivations for the structure $[a,b,(\bar{a},[(\bar{b},c),\bar{c}])]$ that correspond to its proof in linear logic, and without the context partitioning problem. Consider for example the following two possibilities:

$$\mathsf{s}\,\frac{\mathsf{ai}{\downarrow}\,\frac{\mathsf{s}\,\frac{\mathsf{ai}{\downarrow}\,\frac{\mathsf{ai}{\downarrow}\,\frac{\circ}{[a,\bar{a}]}}{[a,(\bar{a},[c,\bar{c}])]}}{[a,(\bar{a},c),\bar{c}]}}{[a,([b,\bar{b}],[(\bar{a},c),\bar{c}])]}}{[a,b,(\bar{b},[(\bar{a},c),\bar{c}])]} \qquad \text{and} \qquad \mathsf{s}\,\frac{\mathsf{ai}{\downarrow}\,\frac{\mathsf{s}\,\frac{\mathsf{ai}{\downarrow}\,\frac{\mathsf{ai}{\downarrow}\,\frac{\circ}{[b,\bar{b}]}}{[b,(\bar{b},[a,\bar{a}])]}}{[a,b,(\bar{b},\bar{a})]}}{[a,b,(\bar{b},\bar{a},[c,\bar{c}])]}}{[a,b,(\bar{b},[(\bar{a},c),\bar{c}])]} \quad .$$

In these two derivations the atoms to be annihilated are brought in direct communication on demand. Notice also that in the derivation at the right the first atoms that are annihilated are $c$ and $\bar{c}$, and they are nested inside the conclusion structure.

As in all the well-behaved systems, the rule $\mathsf{i}{\downarrow}$ can be replaced by $\mathsf{ai}{\downarrow}$, $\mathsf{q}{\downarrow}$ and $\mathsf{s}$, as we will see easily. This way interaction between complex structures is reduced to the atomic one by use of the combinatorial machinery we already have.

The cut rule of linear logic, represented in our syntax below at the left, can be realised by the derivation at the right:

$$\frac{([P,R],[\bar{R},Q])}{[P,Q]} \qquad \text{is obtained by} \qquad \mathsf{i}{\uparrow}\,\frac{\mathsf{s}\,\frac{\mathsf{s}\,\frac{([P,R],[\bar{R},Q])}{[Q,([P,R],\bar{R})]}}{[P,Q,(R,\bar{R})]}}{[P,Q]} \quad .$$

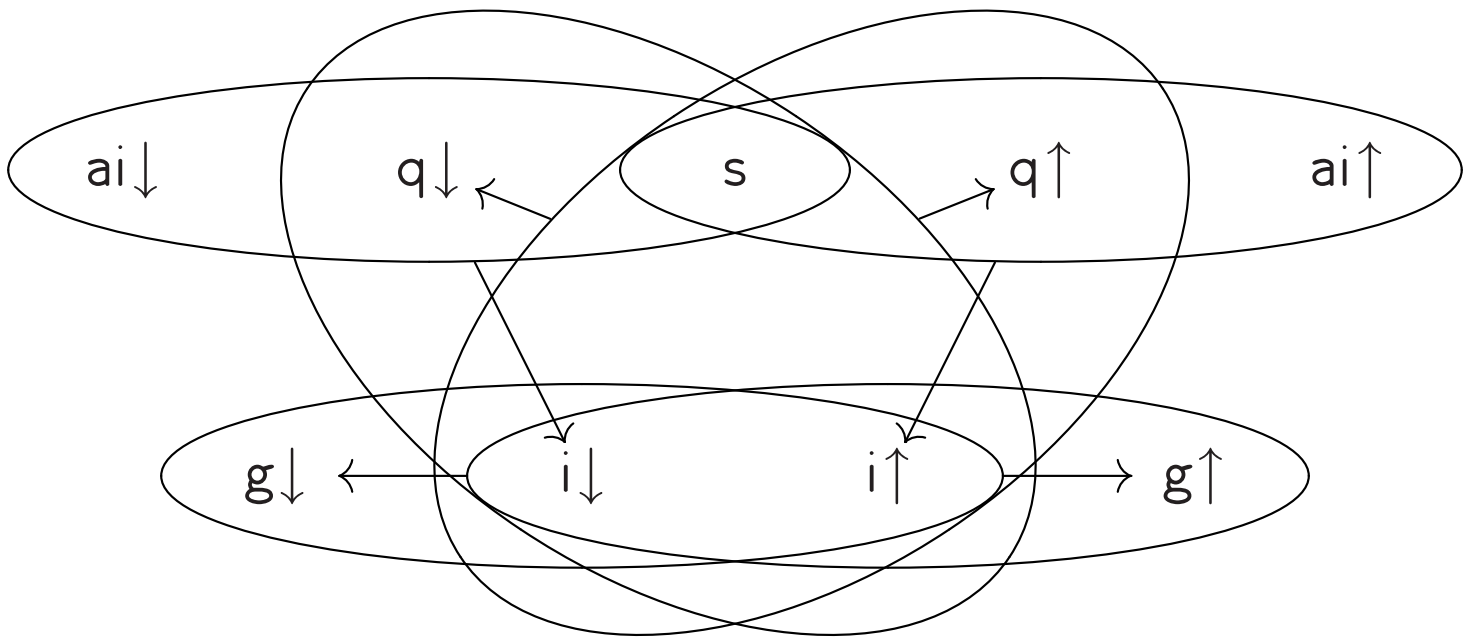

**Fig. 9** *Some derivability relations*

Since we live in a symmetric world, it is not surprising that cut admits a symmetric factorisation with respect to the interaction rule: cut can equivalently be replaced with the system $\{\mathsf{ai}\uparrow, \mathsf{q}\uparrow, \mathsf{s}\}$. In atomic cointeraction, at a certain point going forward in time, a particle and an anti-particle pop up out of empty space. Contrarily to what seems to be happening in nature, they cannot immediately annihilate each other in order for energy to be conserved. This process, which in our system would lead to not having cut elimination, is forbidden by the axiom $\mathsf{m}_2$ on integrity (see Definition 3.1.3). In fact, consider

$$\mathsf{ai}\uparrow \frac{\mathsf{ai}\uparrow \dfrac{\mathsf{ws}\dfrac{\mathsf{ai}\downarrow \dfrac{\mathsf{ai}\downarrow \dfrac{\circ}{[e_a, \bar{e}_c]}}{([e_a, \bar{e}_c], [\bar{e}_b, e_d])}}{[(e_a, \bar{e}_b), (\bar{e}_c, e_d)]}}{(e_a, \bar{e}_b)}}{\circ} \quad ,$$

where $\mathsf{ws}$ is a 'weak' switch instance that does not comply with the conservation law on integrity (the atoms have been indexed with $a$, $b$, $c$ and $d$ to help checking the condition). Here energy would be overall conserved, but there would also be a proof of $(e, \bar{e})$ that makes an essential use of cut: interaction between $e$ and $\bar{e}$ is indirectly allowed by the use of two auxiliary atoms, which must be created *ad hoc*. Therefore, it is mandatory, for having cut elimination, that copar structures are introduced as a whole into the cut's copar context, and this is precisely what the axiom $\mathsf{m}_2$ guarantees. What is conserved is the integrity of all the copars in a communication: either one enters another, or vice versa. It is forbidden that any two of them are broken and rebuilt into a larger copar, which prevents from communicating pieces of both the original copars, which are in turn in mutual communication (or sequential composition).

Of course, $\mathsf{ai}\downarrow$ and $\mathsf{ai}\uparrow$ are instances of $\mathsf{i}\downarrow$ and $\mathsf{i}\uparrow$. The first interesting question about $\mathsf{i}\downarrow$ and $\mathsf{i}\uparrow$ is whether they are derivable, respectively, for $\{\mathsf{ai}\downarrow, \mathsf{g}\downarrow\}$ and $\{\mathsf{ai}\uparrow, \mathsf{g}\uparrow\}$, i.e., whether they are derivable for $\{\mathsf{ai}\downarrow, \mathsf{q}\downarrow, \mathsf{s}\}$ and $\{\mathsf{ai}\uparrow, \mathsf{q}\uparrow, \mathsf{s}\}$. Several relations of derivability will be given in the following, therefore I provide a map in Figure 9, where $\textcircled{\mathscr{S}} \longrightarrow \rho$ means that the rule $\rho$ is derivable for the system $\mathscr{S}$.

**3.4.2 Theorem** *The rule* $\mathsf{i}\downarrow$ *is derivable for* $\{\mathsf{ai}\downarrow, \mathsf{q}\downarrow, \mathsf{s}\}$.

**Proof** Given $\mathsf{i}{\downarrow}\,\dfrac{S\{\circ\}}{S[R,\bar{R}]}$, let us perform a structural induction on $R$.

**Base Cases**

**1** $R=\circ$: In this case $S[R,\bar{R}]=S\{\circ\}$.

**2** $R$ is an atom: Then the given instance of $\mathsf{i}{\downarrow}$ is an instance of $\mathsf{ai}{\downarrow}$.

**Inductive Cases**

**3** $R=\langle P;Q\rangle$, where $P\neq\circ\neq Q$: Apply the induction hypothesis on $\mathsf{q}{\downarrow}\,\dfrac{\mathsf{i}{\downarrow}\,\dfrac{\mathsf{i}{\downarrow}\,\dfrac{S\{\circ\}}{S[Q,\bar{Q}]}}{S\langle[P,\bar{P}];[Q,\bar{Q}]\rangle}}{S[\langle P;Q\rangle,\langle\bar{P};\bar{Q}\rangle]}$.

**4** $R=[P,Q]$, where $P\neq\circ\neq Q$: Apply the induction hypothesis on $\mathsf{s}\,\dfrac{\mathsf{s}\,\dfrac{\mathsf{i}{\downarrow}\,\dfrac{\mathsf{i}{\downarrow}\,\dfrac{S\{\circ\}}{S[Q,\bar{Q}]}}{S([P,\bar{P}],[Q,\bar{Q}])}}{S[Q,([P,\bar{P}],\bar{Q})]}}{S[P,Q,(\bar{P},\bar{Q})]}$.

**5** $R=(P,Q)$, where $P\neq\circ\neq Q$: Analogous to Case 4. □

**3.4.3 Theorem** *The rule* $\mathsf{i}{\uparrow}$ *is derivable for* $\{\mathsf{ai}{\uparrow},\mathsf{q}{\uparrow},\mathsf{s}\}$.

**Proof** The argument is symmetric to the one in the proof of Theorem 3.4.2. Let us just see the two derivations involved in the inductive cases, where $P\neq\circ\neq Q$:

$$\mathsf{i}{\uparrow}\,\dfrac{\mathsf{i}{\uparrow}\,\dfrac{\mathsf{q}{\uparrow}\,\dfrac{S(\langle P;Q\rangle,\langle\bar{P};\bar{Q}\rangle)}{S\langle(P,\bar{P});(Q,\bar{Q})\rangle}}{S(Q,\bar{Q})}}{S\{\circ\}} \qquad\text{and}\qquad \mathsf{i}{\uparrow}\,\dfrac{\mathsf{i}{\uparrow}\,\dfrac{\mathsf{s}\,\dfrac{\mathsf{s}\,\dfrac{S(P,Q,[\bar{P},\bar{Q}])}{S(Q,[(P,\bar{P}),\bar{Q}])}}{S[(P,\bar{P}),(Q,\bar{Q})]}}{S(Q,\bar{Q})}}{S\{\circ\}} \quad .$$

□

The reduction of $\mathsf{i}{\downarrow}$ and $\mathsf{i}{\uparrow}$ to $\mathsf{ai}{\downarrow}$ and $\mathsf{ai}{\uparrow}$ through $\mathsf{g}{\downarrow}$ and $\mathsf{g}{\uparrow}$ is a first successful test of the internal symmetries of the system we are building. The ultimate test will be of course cut elimination.

The derivability of cut for a system containing an atomic cut is peculiar to our calculus of structures (one usually has just admissibility of cut, no matter whether atomic or not). For example, there is no such property in the sequent system of linear logic, while the same system enjoys the derivability of identity for a system containing the atomic identity. The reason of this is very simple: the polarity of the inference rules of the sequent calculus (introduction of the main connective going down in a derivation) goes in the right direction for the factorisation of identity, and in the wrong one for cut. The reduction to atomic cuts does take place in linear logic, but only when eliminating cuts in proofs. After performing appropriate permutations of rule instances in a proof, this is a typical reduction that could occur:

$$\mathsf{cut}\,\dfrac{\otimes\,\dfrac{\overset{\Pi}{\vdash A,\Phi}\quad \overset{\Pi'}{\vdash B,\Phi'}}{\vdash A\otimes B,\Phi,\Phi'}\qquad \parr\,\dfrac{\overset{\Pi''}{\vdash \bar{A},\bar{B},\Psi}}{\vdash \bar{A}\parr\bar{B},\Psi}}{\vdash \Phi,\Phi',\Psi} \quad\longrightarrow\quad \mathsf{cut}\,\dfrac{\mathsf{cut}\,\dfrac{\overset{\Pi}{\vdash A,\Phi}\quad \overset{\Pi''}{\vdash \bar{A},\bar{B},\Psi}}{\vdash \bar{B},\Phi,\Psi}\qquad \overset{\Pi'}{\vdash B,\Phi'}}{\vdash \Phi,\Phi',\Psi} \quad .$$

Compare this situation with the second derivation shown in the proof of Theorem 3.4.3, where $A$ and $B$ stand for $P$ and $Q$, and $\Phi$, $\Phi'$ and $\Psi$ constitute context $S\{\ \}$. A case like this, where the main connectives in principal formulae are introduced immediately above the cut, can only be guaranteed in proofs. In our system the scope of the reduction to atomic cuts is broader: the structures introduced by a cut can *always* be introduced atom by atom by atomic cuts.

I can now get back to the question of interdependence of $\mathsf{g}{\downarrow}$ and $\mathsf{g}{\uparrow}$, which I posed in the preceding section: by using the interaction rules, we can see that $\mathsf{g}{\downarrow}$ and $\mathsf{g}{\uparrow}$ are mutually derivable.

**3.4.4 Theorem** *The rule $\mathsf{g}{\uparrow}$ is derivable for $\{\mathsf{i}{\downarrow}, \mathsf{i}{\uparrow}, \mathsf{g}{\downarrow}\}$; symmetrically, the rule $\mathsf{g}{\downarrow}$ is derivable for $\{\mathsf{i}{\downarrow}, \mathsf{i}{\uparrow}, \mathsf{g}{\uparrow}\}$.*

**Proof** Let $\mathsf{g}{\uparrow}\,\dfrac{S(R,T)}{S\{Q\}}$ and $\mathsf{g}{\downarrow}\,\dfrac{S\{Q\}}{S[R,T]}$ be two rule instances such that $Q \in R \lozenge T$. By using Theorem 3.1.7 we can replace them, respectively, with

$$\mathsf{i}{\downarrow}\,\dfrac{\mathsf{g}{\downarrow}\,\dfrac{\mathsf{g}{\downarrow}\,\dfrac{S(R,T)}{S([Q,\bar{Q}],R,T)}}{S[Q,(\bar{Q},R,T)]}}{\mathsf{i}{\uparrow}\,\dfrac{S[Q,([\bar{R},\bar{T}],R,T)]}{S\{Q\}}} \qquad \text{and} \qquad \mathsf{i}{\downarrow}\,\dfrac{\mathsf{g}{\uparrow}\,\dfrac{\mathsf{g}{\uparrow}\,\dfrac{S\langle Q\rangle}{S(Q,[(\bar{R},\bar{T}),R,T])}}{S(Q,[\bar{Q},R,T])}}{\mathsf{i}{\uparrow}\,\dfrac{S[(Q,\bar{Q}),R,T]}{S[R,T]}} \quad .$$

$$\begin{array}{ll} \mathsf{i}{\downarrow} & S(R,T) \\ \hline & S([Q,\bar{Q}],R,T) \\ \mathsf{g}{\downarrow} & \\ \hline & S[Q,(\bar{Q},R,T)] \\ \mathsf{g}{\downarrow} & \\ \hline & S[Q,([\bar{R},\bar{T}],R,T)] \\ \mathsf{i}{\uparrow} & \\ \hline & S\{Q\} \end{array}$$

□

**3.4.5 Corollary** *The rule $\mathsf{q}{\uparrow}$ is derivable for $\{\mathsf{i}{\downarrow}, \mathsf{i}{\uparrow}, \mathsf{q}{\downarrow}, \mathsf{s}\}$; symmetrically, the rule $\mathsf{q}{\downarrow}$ is derivable for $\{\mathsf{i}{\downarrow}, \mathsf{i}{\uparrow}, \mathsf{q}{\uparrow}, \mathsf{s}\}$.*

Let $\rho$ be a rule and $\pi$ be its corule, i.e., $\pi$ is the scheme obtained by exchanging and negating premiss and conclusion in $\rho$. The rule $\pi$ is then derivable for the system $\{\mathsf{i}{\downarrow}, \mathsf{i}{\uparrow}, \mathsf{s}, \rho\}$, because each instance $\pi\,\dfrac{S\{Q\}}{S\{P\}}$ can be replaced by

$$\begin{array}{rc} \mathsf{i}{\downarrow} & \dfrac{S\{Q\}}{S(Q,[P,\bar{P}])} \\ \mathsf{s} & \dfrac{}{S[P,(Q,\bar{P})]} \\ \rho & \dfrac{}{S[P,(Q,\bar{Q})]} \\ \mathsf{i}{\uparrow} & \dfrac{}{S\{P\}} \end{array} \quad .$$

This means, basically, that all the imaginable, future corules in the 'up' fragment can be reduced to the corresponding rules via the cut rule. This is not a surprise, of course, and is perfectly analogous in spirit to what happens in the sequent calculus. The reader may then wonder about the reason why I emphasise so much dealing with the 'up' fragment. Consider the rule $\mathsf{q}{\uparrow}$: it is derivable for a system with $\mathsf{i}{\uparrow}$ and $\mathsf{q}{\downarrow}$, but at the same time $\mathsf{i}{\uparrow}$

$$\mathsf{ai}{\downarrow}\,\frac{S\{\circ\}}{S[a,\bar a]} \qquad \mathsf{ai}{\uparrow}\,\frac{S(a,\bar a)}{S\{\circ\}}$$

$$\mathsf{q}{\downarrow}\,\frac{S\langle[R,T];[R',T']\rangle}{S[\langle R;R'\rangle,\langle T;T'\rangle]} \qquad \mathsf{q}{\uparrow}\,\frac{S(\langle R;T\rangle,\langle R';T'\rangle)}{S\langle(R,R');(T,T')\rangle}$$

$$\mathsf{s}\,\frac{S([R,T],R')}{S[(R,R'),T]}$$

**Fig. 10** *Symmetric basic system* V (SBV)

itself is derivable for $\{\mathsf{ai}{\uparrow},\mathsf{q}{\uparrow},\mathsf{s}\}$. There is a *mutual* dependence between $\mathsf{q}{\uparrow}$ and $\mathsf{i}{\uparrow}$.

Let us put together our simplest rules and give a name to the resulting system.

**3.4.6 Definition** The system $\{\mathsf{ai}{\downarrow},\mathsf{ai}{\uparrow},\mathsf{q}{\downarrow},\mathsf{q}{\uparrow},\mathsf{s}\}$, shown in Figure 10, is called *symmetric basic system* V, or SBV; $\{\mathsf{ai}{\downarrow},\mathsf{q}{\downarrow},\mathsf{s}\}$ and $\{\mathsf{ai}{\uparrow},\mathsf{q}{\uparrow},\mathsf{s}\}$ are its *down* and *up fragments*, respectively.

The following easy theorem provides some systems equivalent to SBV.

**3.4.7 Theorem** *System* SBV *is strongly equivalent to the systems* $\{\mathsf{i}{\downarrow},\mathsf{i}{\uparrow},\mathsf{g}{\downarrow}\}$, $\{\mathsf{i}{\downarrow},\mathsf{i}{\uparrow},\mathsf{g}{\uparrow}\}$, $\{\mathsf{ai}{\downarrow},\mathsf{i}{\uparrow},\mathsf{g}{\downarrow}\}$ *and* $\{\mathsf{i}{\downarrow},\mathsf{ai}{\uparrow},\mathsf{g}{\uparrow}\}$.

**Proof** Let us prove strong equivalence of the system $\{\mathsf{i}{\downarrow},\mathsf{i}{\uparrow},\mathsf{g}{\downarrow}\}$ and SBV: $\mathsf{i}{\downarrow}$ can be replaced with $\{\mathsf{ai}{\downarrow},\mathsf{q}{\downarrow},\mathsf{s}\}$ (Theorem 3.4.2); $\mathsf{i}{\uparrow}$ with $\{\mathsf{ai}{\uparrow},\mathsf{q}{\uparrow},\mathsf{s}\}$ (Theorem 3.4.3); $\mathsf{g}{\downarrow}$ with $\{\mathsf{q}{\downarrow},\mathsf{s}\}$ (Theorem 3.2.5). Conversely, the rule $\mathsf{ai}{\downarrow}$ is an instance of $\mathsf{i}{\downarrow}$, the rule $\mathsf{ai}{\uparrow}$ of $\mathsf{i}{\uparrow}$, the rules $\mathsf{q}{\downarrow}$ and $\mathsf{s}$ are instances of $\mathsf{g}{\downarrow}$; by Corollary 3.4.5, the rule $\mathsf{q}{\uparrow}$ can be replaced with $\{\mathsf{i}{\downarrow},\mathsf{i}{\uparrow},\mathsf{q}{\downarrow},\mathsf{s}\}$, therefore with $\{\mathsf{i}{\downarrow},\mathsf{i}{\uparrow},\mathsf{g}{\downarrow}\}$. Proceed analogously for $\{\mathsf{i}{\downarrow},\mathsf{i}{\uparrow},\mathsf{g}{\uparrow}\}$. The system $\{\mathsf{ai}{\downarrow},\mathsf{i}{\uparrow},\mathsf{g}{\downarrow}\}$ is equivalent to $\{\mathsf{i}{\downarrow},\mathsf{i}{\uparrow},\mathsf{g}{\downarrow}\}$ because $\mathsf{ai}{\downarrow}$ is an instance of $\mathsf{i}{\downarrow}$, and by Theorem 3.4.2. Proceed analogously for $\{\mathsf{i}{\downarrow},\mathsf{ai}{\uparrow},\mathsf{g}{\uparrow}\}$. □

**3.4.8 Remark** Strong equivalence of SBV and $\{\mathsf{ai}{\downarrow},\mathsf{i}{\uparrow},\mathsf{g}{\downarrow}\}$, so of SBV and $\{\mathsf{ai}{\downarrow},\mathsf{i}{\uparrow},\mathsf{q}{\downarrow},\mathsf{s}\}$, is remarkable because the entire up fragment of SBV that is not common to the down one (i.e., the rules $\mathsf{ai}{\uparrow}$ and $\mathsf{q}{\uparrow}$) is concentrated in the rule $\mathsf{i}{\uparrow}$. See Remark 4.2.4.

**3.4.9 Remark** All the rules of SBV enjoy an important property: a certain notion of computational cost is constant for each of them. In $\mathsf{ai}{\downarrow}$ and $\mathsf{ai}{\uparrow}$ we have to check whether two atoms are dual through negation and erase them, or we have to introduce two dual atoms, depending on the interpretation. In $\mathsf{q}{\downarrow}$, $\mathsf{q}{\uparrow}$ and $\mathsf{s}$ we have to rearrange some pointers to the structures involved, without actually looking at the structures (think of the graphical representation we saw). None of the rules in $\{\mathsf{i}{\downarrow},\mathsf{i}{\uparrow},\mathsf{g}{\downarrow},\mathsf{g}{\uparrow}\}$ enjoys this property, because these rules involve actually inspecting the structures upon which they act.

Please note that there are no logical axioms in SBV, and derivations may grow indefinitely in both directions: $\mathsf{ai}{\downarrow}$ and $\mathsf{ai}{\uparrow}$ may always be used.

# 4 Breaking the Symmetry

In the end, is the cut rule admissible? If we just look at derivations, the cut rule itself is not replaceable with any other rule that we already have, except for the atomic cut rule, since it is the only one that produces matter, i.e., atoms. But of course we want to look at admissibility for *proofs*, i.e., special derivations in which all the matter is finally consumed.

A proof is an intrinsically asymmetric object, because at one side it has the formula to be proved, and at the other it has logical axioms—in our case emptiness. Since there are no logical axioms in SBV, we have no proofs. Let us add a logical axiom, and break the symmetry, by not adding its 'coaxiom'. (If I did so, I would get 'coprovability': a structure is 'coprovable' when its negation is provable. Since there is no point in carrying on this symmetry when I am interested in a non-symmetric idea, I leave the symmetry broken.)

**4.1 Definition** The following (logical axiom) rule $\circ\!\downarrow$ is called *unit*:

$$\circ\!\downarrow \frac{}{\circ} \quad .$$

A *proof* is a derivation whose topmost inference is an instance of the unit rule. Proofs are denoted by $\Pi$. A formal system $\mathscr{S}$ *proves* $R$ if there is in $\mathscr{S}$ a proof $\Pi$ whose conclusion is $R$, written $\underset{R}{{}^{\Pi}\|\mathscr{S}}$ (the name $\Pi$ can be omitted). Two systems are *(weakly) equivalent* if they prove the same structures.

Please note that $\circ\!\downarrow$ can only occur once in a derivation, and only at the top. Of course, I could have defined provability as the possibility of observing $\circ$ on top of a derivation, without introducing $\circ\!\downarrow$. I prefer following the traditional way, with logical axioms, because this will allow, in future extensions, to express more complex provability observations in formal systems, rather than in definitions.

**4.2 Definition** An inference rule $\rho$ is *admissible* for the formal system $\mathscr{S}$ if $\rho \notin \mathscr{S}$ and for every proof $\underset{R}{\|\mathscr{S}\cup\{\rho\}}$ there exists a proof $\underset{R}{\|\mathscr{S}}$.

Of course, derivability implies admissibility.

As a result of the break in the symmetry, the cut rule becomes superfluous for proofs, meaning that no production of matter is really necessary in order to annihilate the one we already have, if this is at all possible. It turns out that not only $\mathsf{i}\!\uparrow$ (and $\mathsf{ai}\!\uparrow$) is admissible, but the coseq rule $\mathsf{q}\!\uparrow$ is, too: the up fragment of SBV (excluding switch, which also belongs to the down fragment) is not necessary when observing provability. What remains is the formal system BV.

**4.3 Definition** The system $\{\circ\!\downarrow, \mathsf{ai}\!\downarrow, \mathsf{q}\!\downarrow, \mathsf{s}\}$, in Figure 11, is denoted BV and called *basic system* V.

This section is organised in two subsections: in the first one, the splitting theorem is proved, in the second, it is applied to proving that $\mathsf{q}\!\uparrow$ and $\mathsf{ai}\!\uparrow$ are admissible for BV. The splitting theorem is, perhaps as expected, technically complex. In exchange for the effort, it is also more general than ordinary cut elimination, and is applicable to other logical systems we are currently studying.

$$\circ\!\downarrow \frac{}{\circ} \qquad \mathsf{ai}\!\downarrow \frac{S\{\circ\}}{S[a,\bar{a}]} \qquad \mathsf{q}\!\downarrow \frac{S\langle [R,T];[R',T']\rangle}{S[\langle R;R'\rangle,\langle T;T'\rangle]} \qquad \mathsf{s}\,\frac{S([R,T],R')}{S[(R,R'),T]}$$

**Fig. 11** *Basic system* V (BV)

## 4.1 The Splitting Theorem

The classical arguments for proving cut elimination in the sequent calculus rely on the following property: when the principal formulae in a cut are active in both branches, they determine which rules are applied immediately above the cut. This is a consequence of the fact that formulae have a root connective, and logical rules only hinge on that, and nowhere else in the formula.

This property does not hold in the calculus of structures. Further, since rules can be applied anywhere deep inside structures, everything can happen above a cut. This complicates considerably the task of proving cut elimination. On the other hand, a great simplification is made possible in the calculus of structures by the reduction of cut to its atomic form. The remaining difficulty is actually understanding what happens, while going up in a proof, *around* the atoms produced by an atomic cut. The two atoms of an atomic cut can be produced inside any structure, and they do not belong to distinct branches, as in the sequent calculus: complex interactions with their context are possible. As a consequence, our techniques are largely different from the traditional ones.

Two approaches to cut elimination in the calculus of structures have been explored in other papers: in [13] we relied on permutations of rules, in [9] the authors relied on semantics. In this paper we use a third technique, called *splitting*, which has the advantage of being more uniform than the one based on permutations and which yields a much simpler case analysis. It also establishes a deep connection to the sequent calculus, at least for the fragments of systems that allow for a sequent calculus presentation (in this case, the commutative fragment). Since many systems are expressed in the sequent calculus, our method appears to be entirely general; still it is independent of the sequent calculus and of a complete semantics.

Splitting can be best understood by considering a sequent system with no weakening and contraction. Consider for example multiplicative linear logic: If we have a proof of the sequent

$$\vdash F\{A \otimes B\}, \Gamma \quad ,$$

where $F\{A \otimes B\}$ is a formula that contains the subformula $A \otimes B$, we know for sure that somewhere in the proof there is one and only one instance of the $\otimes$ rule that splits $A$ and

$B$ along with their context. We are in the following situation:

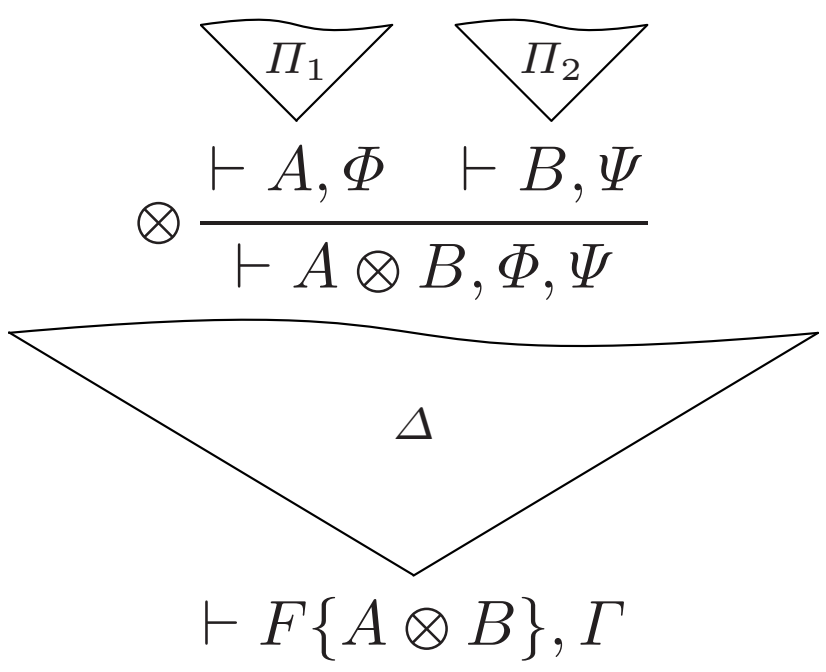

corresponds to

$$\begin{array}{c} \Pi_2 \Vert \\ [B,\Psi] \\ \Pi_1 \Vert \\ \mathsf{s}\,\dfrac{([A,\Phi],[B,\Psi])}{\mathsf{s}\,\dfrac{[([A,\Phi],B),\Psi]}{[(A,B),\Phi,\Psi]}} \\ \Delta \Vert \\ [F(A,B),\Gamma] \end{array} \quad .$$

We can consider, as shown at the left, the proof for the given sequent as composed of three pieces, $\Delta$, $\Pi_1$ and $\Pi_2$. In the calculus of structures, many different proofs correspond to the sequent calculus one: they differ for the different possible sequencing of rules, and because rules in the calculus of structures have smaller granularity and larger applicability. But, among all these proofs, there must also be one that fits the scheme at the right of the figure above. This precisely illustrates the idea behind the splitting technique.

The derivation $\Delta$ above implements a *context reduction* and a proper splitting. We can state, in general, these principles as follows:

1 Context reduction: If $S\{R\}$ is provable, then $S\{\ \}$ can be reduced, going up in a proof, to the structure $[\{\ \},U]$, such that $[R,U]$ is provable. In the example above, $[F\{\ \},\Gamma]$ is reduced to $[\{\ \},\Gamma']$, for some $\Gamma'$.

2 (Shallow) splitting: If $[(R,T),P]$ is provable, then $P$ can be reduced, going up in a proof, to $[P_1,P_2]$, such that $[R,P_1]$ and $[T,P_2]$ are provable. In the example above $\Gamma'$ is reduced to $[\Phi,\Psi]$.

Context reduction is in turn proved by splitting, which is then at the core of the matter. The biggest difficulty resides in proving splitting, and this mainly requires finding the right induction measure. We first need to establish two easy propositions, which express obvious properties induced by the rules of BV. I will often use them implicitly throughout the rest of the paper.

**4.1.1 Proposition** *The size of the premiss of a derivation in* BV *is not greater than the size of the conclusion.*

**4.1.2 Proposition** *In* BV, $\langle R;T\rangle$ *is provable if and only if* $R$ *and* $T$ *are provable and* $(R,T)$ *is provable if and only if* $R$ *and* $T$ *are provable.*

The calculus of structures is of course very similar to a term rewriting system, and especially so for those, like me, who have an intuitive bias toward considering derivations growing in a bottom-up fashion. In this case the direction in which rewriting occurs is clear, and it is natural to use some terminology of rewriting systems for inference rules.

**4.1.3 Definition** The inference rules of SBV are all of the kind $\rho\,\dfrac{S\{V\}}{S\{U\}}$: the structure $U$ is called the *redex* (and $V$ is the *contractum*) of the rule's instance.

**4.1.4 Theorem** (**Shallow Splitting**) *For all structures $R$, $T$ and $P$:*

**1** *if $[\langle R;T\rangle,P]$ is provable in* $\mathsf{BV}$ *then there exist $P_1$, $P_2$ and* $\begin{array}{c}\langle P_1;P_2\rangle\\ \Vert \mathsf{BV}\\ P\end{array}$ *such that $[R,P_1]$ and $[T,P_2]$ are provable in* $\mathsf{BV}$;

**2** *if $[(R,T),P]$ is provable in* $\mathsf{BV}$ *then there exist $P_1$, $P_2$ and* $\begin{array}{c}[P_1,P_2]\\ \Vert \mathsf{BV}\\ P\end{array}$ *such that $[R,P_1]$ and $[T,P_2]$ are provable in* $\mathsf{BV}$.

**Proof** All the derivations appearing in this argument are in $\mathsf{BV}$. Let us consider the lexicographic order $\prec$ on natural numbers defined by $(m',n')\prec(m,n)$ iff either $m'<m$, or $m'=m$ and $n'<n$. Consider the following two statements:

$$\begin{array}{rl}\mathsf{S}(m,n)=\forall m',n'.\forall R,T,P.\Big(\big((m',n')\preceq(m,n) & \\ & \wedge\, m'=|\mathrm{occ}\,[\langle R;T\rangle,P]| \\ & \wedge\,\text{there is a proof } \begin{array}{c}\overline{\Vert}\\ [\langle R;T\rangle,P]\end{array} \text{ of length } n'\big) \\ & \Rightarrow \exists P_1,P_2.\Big(\begin{array}{c}\langle P_1;P_2\rangle\\ \Vert\\ P\end{array} \wedge \begin{array}{c}\overline{\Vert}\\ [R,P_1]\end{array} \wedge \begin{array}{c}\overline{\Vert}\\ [T,P_2]\end{array}\Big)\Big) \quad ,\end{array}$$

$$\begin{array}{rl}\mathsf{C}(m,n)=\forall m',n'.\forall R,T,P.\Big(\big((m',n')\preceq(m,n) & \\ & \wedge\, m'=|\mathrm{occ}\,[(R,T),P]| \\ & \wedge\,\text{there is a proof } \begin{array}{c}\overline{\Vert}\\ [(R,T),P]\end{array} \text{ of length } n'\big) \\ & \Rightarrow \exists P_1,P_2.\Big(\begin{array}{c}[P_1,P_2]\\ \Vert\\ P\end{array} \wedge \begin{array}{c}\overline{\Vert}\\ [R,P_1]\end{array} \wedge \begin{array}{c}\overline{\Vert}\\ [T,P_2]\end{array}\Big)\Big) \quad .\end{array}$$

The statement of the theorem is equivalent to $\forall m,n.(\mathsf{S}(m,n)\wedge\mathsf{C}(m,n))$. We can consider $(m,n)$ a measure of $(\mathsf{S}(m,n)\wedge\mathsf{C}(m,n))$, and the proof is an induction, by $\prec$, on this measure. The base case being trivial, let us see the inductive ones. We will always assume $P\neq\circ$, since when $P=\circ$ the theorem is trivially proved by Proposition 4.1.2. For the same reason, we assume $R\neq\circ\neq T$. We will prove separately $\mathsf{S}(m,n)$ and $\mathsf{C}(m,n)$.

**1** $\forall m',n'.((m',n')\prec(m,n)\wedge\mathsf{S}(m',n')\wedge\mathsf{C}(m',n'))\Rightarrow\mathsf{S}(m,n)$: The size of $[\langle R;T\rangle,P]$ is $m$ and there is a proof of it of length $n$. Let us consider the bottom rule instance in this proof:

$$\rho\,\frac{\begin{array}{c}\overline{\Vert}\\ Q\end{array}}{[\langle R;T\rangle,P]} \quad ,$$

where we assume that $\rho$ is non-trivial, otherwise the induction hypothesis applies. Let us reason on the position of the redex of $\rho$ in $[\langle R;T\rangle,P]$. Here are the possibilities:

**1** $\rho=\mathsf{ai}{\downarrow}$: The following cases exhaust the possibilities:

**1** The redex is inside $R$:

$$\text{given} \qquad \mathsf{ai}{\downarrow}\,\frac{\begin{array}{c}\overline{\Vert}\\ [\langle R';T\rangle,P]\end{array}}{[\langle R;T\rangle,P]} \quad , \qquad \text{consider} \qquad \mathsf{ai}{\downarrow}\,\frac{\begin{array}{c}\overline{\Vert}\\ [R',P_1]\end{array}}{[R,P_1]} \quad .$$

**2** The redex is inside $T$: Analogous to the previous case.

**3** The redex is inside $P$:

$$\text{given}\qquad \begin{array}{c} \overline{\Vert} \\ \mathsf{ai}{\downarrow}\dfrac{[\langle R;T\rangle,\color{red}{P'}]}{[\langle R;T\rangle,\color{green}{P}]} \end{array} \quad,\qquad \text{consider}\qquad \begin{array}{c} \langle P_1;P_2\rangle \\ \Vert \\ \mathsf{ai}{\downarrow}\dfrac{\color{red}{P'}}{\color{green}{P}} \end{array} \quad.$$

**2** $\rho = \mathsf{q}{\downarrow}$: If the redex is inside $R$, $T$ or $P$, we have analogous situations to the ones seen above. The following cases exhaust the other possibilities:

**1** $R = \langle R';R''\rangle$, $P = [\langle P';P''\rangle, U]$ and

$$\begin{array}{c} \overline{\Vert} \\ \mathsf{q}{\downarrow}\dfrac{[\langle [R',P'];[\langle R'';T\rangle,P'']\rangle,U]}{[\langle R';R'';T\rangle,\langle P';P''\rangle,U]} \end{array} \quad:$$

We can apply the induction hypothesis, and we get:

$$\begin{array}{c} \langle U_1;U_2\rangle \\ \Vert \\ U \end{array} \quad,\qquad \begin{array}{c} \overline{\Vert} \\ [R',P',U_1] \end{array} \qquad\text{and}\qquad \begin{array}{c} \Pi\,\overline{\Vert} \\ [\langle R'';T\rangle,P'',U_2] \end{array} \quad.$$

Since $|\mathsf{occ}\,[\langle R'';T\rangle,P'',U_2]| < |\mathsf{occ}\,[\langle R';R'';T\rangle,\langle P';P''\rangle,U]|$ (otherwise the $\mathsf{q}{\downarrow}$ instance would be trivial), we can apply the induction hypothesis on $\Pi$ and get:

$$\begin{array}{c} \langle P_1';P_2\rangle \\ \Vert \\ [P'',U_2] \end{array} \quad,\qquad \begin{array}{c} \overline{\Vert} \\ [R'',P_1'] \end{array} \qquad\text{and}\qquad \begin{array}{c} \overline{\Vert} \\ [T,P_2] \end{array} \quad.$$

We can now take $P_1 = \langle [P',U_1];P_1'\rangle$ and build

$$\begin{array}{c} \langle [P',U_1];P_1';P_2\rangle \\ \Vert \\ \mathsf{q}{\downarrow}\dfrac{\langle [P',U_1];[P'',U_2]\rangle}{[\langle P';P''\rangle,\langle U_1;U_2\rangle]} \\ \Vert \\ [\langle P';P''\rangle,U] \end{array} \qquad\text{and}\qquad \begin{array}{c} \overline{\Vert} \\ \mathsf{q}{\downarrow}\dfrac{\langle [R',P',U_1];[R'',P_1']\rangle}{[\langle R';R''\rangle,\langle [P',U_1];P_1'\rangle]} \end{array} \quad.$$

A similar argument holds when $T = \langle T';T''\rangle$ and we are given a proof

$$\begin{array}{c} \overline{\Vert} \\ \mathsf{q}{\downarrow}\dfrac{[\langle [\langle R;T'\rangle,P'];[T'',P'']\rangle,U]}{[\langle R;T';T''\rangle,\langle P';P''\rangle,U]} \end{array} \quad.$$

**2** $P = [\langle P';P''\rangle,U',U'']$ and

$$\begin{array}{c} \overline{\Vert} \\ \mathsf{q}{\downarrow}\dfrac{[\langle [\langle R;T\rangle,P',U'];P''\rangle,U'']}{[\langle R;T\rangle,\langle P';P''\rangle,U',U'']} \end{array} \quad:$$

We can apply the induction hypothesis, and we get:

$$\begin{array}{c}\langle U_1;U_2\rangle\\ \|\\ U''\end{array}\quad,\quad \begin{array}{c}{\scriptstyle\Pi}\|\!\!\!\!-\\ {[\langle R;T\rangle,P',U',U_1]}\end{array}\quad\text{and}\quad \begin{array}{c}\|\!\!\!\!-\\ {[P'',U_2]}\end{array}\quad.$$

Since $|\mathsf{occ}\,[\langle R;T\rangle,P',U',U_1]| < |\mathsf{occ}\,[\langle R;T\rangle,\langle P';P''\rangle,U',U'']|$ (otherwise the $\mathsf{q}\!\downarrow$ instance would be trivial), we can apply the induction hypothesis on $\Pi$ and get:

$$\begin{array}{c}\langle P_1;P_2\rangle\\ \|\\ {[P',U',U_1]}\end{array}\quad,\quad \begin{array}{c}\|\!\!\!\!-\\ {[R,P_1]}\end{array}\quad\text{and}\quad \begin{array}{c}\|\!\!\!\!-\\ {[T,P_2]}\end{array}\quad.$$

We can now build

$$\begin{array}{c}\langle P_1;P_2\rangle\\ \|\\ {[P',U_1,U']}\\ \|\\ \mathsf{q}\!\downarrow\dfrac{[\langle[P',U_1];[P'',U_2]\rangle,U']}{[\langle P';P''\rangle,U',\langle U_1;U_2\rangle]}\\ \|\\ {[\langle P';P''\rangle,U',U'']}\end{array}\quad.$$

A similar argument holds when we are given a proof

$$\begin{array}{c}\|\!\!\!\!-\\ \mathsf{q}\!\downarrow\dfrac{[\langle P';[\langle R;T\rangle,P'',U']\rangle,U'']}{[\langle R;T\rangle,\langle P';P''\rangle,U',U'']}\end{array}\quad.$$

**3** $\rho=\mathsf{s}$: If the redex is inside $R$, $T$ or $P$, we have analogous situations to the ones seen in Case 1.1. The only other possibility is the following; let $P=[(P',P''),U',U'']$ and the given proof is

$$\begin{array}{c}\|\!\!\!\!-\\ \mathsf{s}\,\dfrac{[([\langle R;T\rangle,P',U'],P''),U'']}{[\langle R;T\rangle,(P',P''),U',U'']}\end{array}\quad.$$

We can apply the induction hypothesis, and we get:

$$\begin{array}{c}{[U_1,U_2]}\\ \|\\ U''\end{array}\quad,\quad \begin{array}{c}{\scriptstyle\Pi}\|\!\!\!\!-\\ {[\langle R;T\rangle,P',U',U_1]}\end{array}\quad\text{and}\quad \begin{array}{c}\|\!\!\!\!-\\ {[P'',U_2]}\end{array}\quad.$$

Since $|\mathsf{occ}\,[\langle R;T\rangle,P',U',U_1]| < |\mathsf{occ}\,[\langle R;T\rangle,(P',P''),U',U'']|$ (otherwise the $\mathsf{s}$ instance would be trivial), we can apply the induction hypothesis on $\Pi$ and get:

$$\begin{array}{c}\langle P_1;P_2\rangle\\ \|\\ {[P',U',U_1]}\end{array}\quad,\quad \begin{array}{c}\|\!\!\!\!-\\ {[R,P_1]}\end{array}\quad\text{and}\quad \begin{array}{c}\|\!\!\!\!-\\ {[T,P_2]}\end{array}\quad.$$

We can now build

$$\frac{\langle P_1; P_2\rangle \\ \| \\ [P', U', U_1] \\ \| }{} \quad \mathsf{s}\frac{[(P', [P'', U_2]), U', U_1]}{[(P', P''), U', U_1, U_2]} \quad \| \quad [(P', P''), U', U''] \quad .$$

2 $\forall m', n'.((m', n') \prec (m, n) \wedge \mathsf{S}(m', n') \wedge \mathsf{C}(m', n')) \Rightarrow \mathsf{C}(m, n)$: The size of $[(R, T), P]$ is $m$ and there is a proof of it of length $n$. Let us consider the bottom rule instance in this proof:

$$\rho\frac{Q}{[(R, T), P]} \quad ,$$

where we assume that $\rho$ is non-trivial, otherwise the induction hypothesis applies. Let us reason on the position of the redex of $\rho$ in $[(R, T), P]$. Here are the possibilities:

1 $\rho = \mathsf{ai}{\downarrow}$: Analogous to Case 1.1.

2 $\rho = \mathsf{q}{\downarrow}$: If the redex is inside $R$, $T$ or $P$, we have analogous situations to the ones seen in Case 1.1. The only other possibility is the following; let $P = [\langle P'; P''\rangle, U', U'']$ and the given proof is

$$\mathsf{q}{\downarrow}\frac{[\langle[(R, T), P', U']; P''\rangle, U'']}{[(R, T), \langle P'; P''\rangle, U', U'']} \quad .$$

We can apply the induction hypothesis, and we get:

$$\langle U_1; U_2\rangle \; \| \; U'' \quad , \quad \Pi \; [(R, T), P', U', U_1] \quad \text{and} \quad [P'', U_2] \quad .$$

Since $|\mathsf{occ}\,[(R, T), P', U', U_1]| < |\mathsf{occ}\,[(R, T), \langle P'; P''\rangle, U', U'']|$ (otherwise the $\mathsf{q}{\downarrow}$ instance would be trivial), we can apply the induction hypothesis on $\Pi$ and get:

$$[P_1, P_2] \; \| \; [P', U', U_1] \quad , \quad [R, P_1] \quad \text{and} \quad [T, P_2] \quad .$$

We can now build

$$[P_1, P_2] \quad \| \quad [P', U', U_1] \quad \| \quad \mathsf{q}{\downarrow}\frac{[\langle[P', U_1]; [P'', U_2]\rangle, U']}{[\langle P'; P''\rangle, U', \langle U_1; U_2\rangle]} \quad \| \quad [\langle P'; P''\rangle, U', U''] \quad .$$

A similar argument holds when we are given a proof

$$\begin{array}{c}\| \\ \mathsf{q}{\downarrow}\dfrac{[\langle P';[(R,T),P'',U']\rangle,U'']}{[(R,T),\langle P';P''\rangle,U',U'']}\end{array} \quad .$$

3 $\rho = \mathsf{s}$: If the redex is inside $R$, $T$ or $P$, we have analogous situations to the ones seen in Case 1.1. The following cases exhaust the other possibilities:

1 $R = (R', R'')$, $T = (T', T'')$, $P = [P', P'']$ and

$$\begin{array}{c}\| \\ \mathsf{s}\dfrac{[([(R',T'),P'],R'',T''),P'']}{[(R',R'',T',T''),P',P'']}\end{array} \quad :$$

We can apply the induction hypothesis, and we get:

$$\begin{array}{c}[P'_1,P'_2] \\ \| \\ P''\end{array} \quad , \quad \begin{array}{c}\Pi\| \\ [(R',T'),P',P'_1]\end{array} \quad \text{and} \quad \begin{array}{c}\Pi'\| \\ [(R'',T''),P'_2]\end{array} \quad .$$

$|\mathsf{occ}\,[(R',T'),P',P'_1]| \;<\; |\mathsf{occ}\,[(R',R'',T',T''),P',P'']| \;>\; |\mathsf{occ}\,[(R'',T''),P'_2]|$ holds (otherwise the $\mathsf{s}$ instance would be trivial), then we can apply twice the induction hypothesis on $\Pi$ and $\Pi'$ and get:

$$\begin{array}{c}[P''_1,P''_2] \\ \| \\ [P',P'_1]\end{array} \quad , \quad \begin{array}{c}\| \\ [R',P''_1]\end{array} \quad \text{and} \quad \begin{array}{c}\| \\ [T',P''_2]\end{array} \quad ;$$

$$\begin{array}{c}[P'''_1,P'''_2] \\ \| \\ P'_2\end{array} \quad , \quad \begin{array}{c}\| \\ [R'',P'''_1]\end{array} \quad \text{and} \quad \begin{array}{c}\| \\ [T'',P'''_2]\end{array} \quad .$$

We can now take $P_1 = [P''_1, P'''_1]$, $P_2 = [P''_2, P'''_2]$ and build

$$\begin{array}{c}[P''_1,P''_2,P'''_1,P'''_2] \\ \| \\ [P''_1,P''_2,P'_2] \\ \| \\ [P',P'_1,P'_2] \\ \| \\ [P',P'']\end{array} \quad , \quad \begin{array}{c}\| \\ [R'',P'''_1] \\ \| \\ \mathsf{s}\dfrac{[([R',P''_1],R''),P'''_1]}{[(R',R''),P''_1,P'''_1]}\end{array} \quad \text{and} \quad \begin{array}{c}\| \\ [T'',P'''_2] \\ \| \\ \mathsf{s}\dfrac{[([T',P''_2],T''),P'''_2]}{[(T',T''),P''_2,P'''_2]}\end{array} \quad .$$

2 $P = [(P', P''), U', U'']$ and

$$\begin{array}{c}\| \\ \mathsf{s}\dfrac{[([(R,T),P',U'],P''),U'']}{[(R,T),(P',P''),U',U'']}\end{array} \quad :$$

We can apply the induction hypothesis, and we get:

$$\begin{array}{c}[U_1,U_2]\\ \Vert\\ U''\end{array}\quad,\qquad \begin{array}{c}\Pi\,\overline{\Vert}\\ [(R,T),P',U',U_1]\end{array}\qquad\text{and}\qquad \begin{array}{c}\overline{\Vert}\\ [P'',U_2]\end{array}\quad.$$

Since $|\mathsf{occ}\,[(R,T),P',U',U_1]| < |\mathsf{occ}\,[(R,T),(P',P''),U',U'']|$ (otherwise the $\mathsf{s}$ instance would be trivial), we can apply the induction hypothesis on $\Pi$ and get:

$$\begin{array}{c}[P_1,P_2]\\ \Vert\\ [P',U',U_1]\end{array}\quad,\qquad \begin{array}{c}\overline{\Vert}\\ [R,P_1]\end{array}\qquad\text{and}\qquad \begin{array}{c}\overline{\Vert}\\ [T,P_2]\end{array}\quad.$$

We can now build

$$\begin{array}{c}[P_1,P_2]\\ \Vert\\ [P',U',U_1]\\ \Vert\\ \mathsf{s}\,\dfrac{[(P',[P'',U_2]),U',U_1]}{[(P',P''),U',U_1,U_2]}\\ \Vert\\ [(P',P''),U',U'']\end{array}\quad.$$

□

**4.1.5 Theorem** (**Context Reduction**) *For all structures $R$ and for all contexts $S\{\ \}$ such that $S\{R\}$ is provable in* $\mathsf{BV}$, *there exists a structure $U$ such that for all structures $X$ there exist derivations:*

$$\begin{array}{c}[X,U]\\ \Vert\,\mathsf{BV}\\ S\{X\}\end{array}\qquad\textit{and}\qquad \begin{array}{c}\overline{\Vert}\,\mathsf{BV}\\ [R,U]\end{array}\quad.$$

**Proof** By induction on the size of $S\{\circ\}$. The base case is trivial: $U = \circ$. There are three inductive cases:

**1** $S\{\ \} = \langle S'\{\ \};P\rangle$, for some $P \neq \circ$: By Proposition 4.1.2 there are proofs in $\mathsf{BV}$ of $S'\{R\}$ and of $P$. By applying the induction hypothesis, we can find $U$ and build, for all $X$:

$$\begin{array}{c}[X,U]\\ \Vert\,\mathsf{BV}\\ S'\{X\}\\ \Vert\,\mathsf{BV}\\ \langle S'\{X\};P\rangle\end{array}\quad,$$

such that $[R,U]$ is provable in $\mathsf{BV}$. The same argument applies when $S\{\ \} = \langle P;S'\{\ \}\rangle$, for some $P \neq \circ$.

**2** $S\{\ \} = [S'\{\ \},P]$, for some $P \neq \circ$ such that $S'\{\ \}$ is not a proper par: If $S'\{\circ\} = \circ$ then the theorem is proved; otherwise there are two possibilities:

**1** $S'\{\ \} = \langle S''\{\ \}; P'\rangle$, for some $P' \neq \circ$: By Theorem 4.1.4 there exist:

$$\begin{array}{c}\langle P_1; P_2\rangle \\ \Vert\,\mathsf{BV} \\ P\end{array} \quad , \quad \begin{array}{c}\Pi \Vert\!\!\!\!\top\,\mathsf{BV} \\ [S''\{R\}, P_1]\end{array} \quad \text{and} \quad \begin{array}{c}\Vert\!\!\!\!\top\,\mathsf{BV} \\ [P', P_2]\end{array} \quad .$$

By applying the induction hypothesis to $\Pi$, we get:

$$\begin{array}{c}[X, U] \\ \Vert\,\mathsf{BV} \\ [S''\{X\}, P_1] \\ \Vert\,\mathsf{BV} \\ \mathsf{q}{\downarrow}\dfrac{\langle [S''\{X\}, P_1]; [P', P_2]\rangle}{[\langle S''\{X\}; P'\rangle, \langle P_1; P_2\rangle]} \\ \Vert\,\mathsf{BV} \\ [\langle S''\{X\}; P'\rangle, P]\end{array} \quad \text{and} \quad \begin{array}{c}\Vert\!\!\!\!\top\,\mathsf{BV} \\ [R, U]\end{array} \quad .$$

We can proceed analogously when $S'\{\ \} = \langle P'; S''\{\ \}\rangle$, for some $P' \neq \circ$.

**2** $S'\{\ \} = (S''\{\ \}, P')$, for some $P' \neq \circ$: Analogous to the previous case.

**3** $S\{\ \} = (S'\{\ \}, P)$, for some $P \neq \circ$: Analogous to Case 1. □

**4.1.6 Corollary** (**Splitting**) *For all structures $R$, $T$ and for all contexts $S\{\ \}$:*

**1** *if $S\langle R; T\rangle$ is provable in $\mathsf{BV}$ then there exist structures $S_1$ and $S_2$ such that, for all structures $X$, there exists a derivation*

$$\begin{array}{c}[X, \langle S_1; S_2\rangle] \\ \Vert\,\mathsf{BV} \\ S\{X\}\end{array} \quad ;$$

**2** *if $S(R, T)$ is provable in $\mathsf{BV}$ then there exist structures $S_1$ and $S_2$ such that, for all structures $X$, there exists a derivation*

$$\begin{array}{c}[X, S_1, S_2] \\ \Vert\,\mathsf{BV} \\ S\{X\}\end{array} \quad ;$$

*and, in both cases, there are proofs* $\begin{array}{c}\Vert\!\!\!\!\top\,\mathsf{BV} \\ [R, S_1]\end{array}$ *and* $\begin{array}{c}\Vert\!\!\!\!\top\,\mathsf{BV} \\ [T, S_2]\end{array}$.

**Proof** First apply Theorem 4.1.5 and then apply Theorem 4.1.4. □

## 4.2 Admissibility of the Up Fragment

Let us start with admissibility of $\mathsf{q}{\uparrow}$. Since I broke the symmetry, I can now informally consider interaction a phenomenon that only occurs inside par contexts while going upward in a derivation. Take the conclusion of a $\mathsf{q}\!\uparrow$ instance:

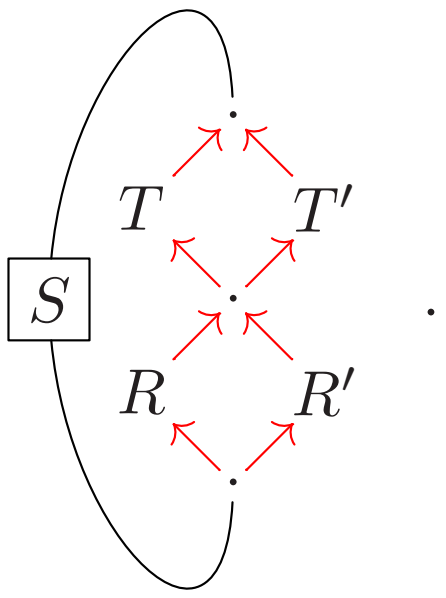

At the end of the process of rewriting this structure toward the unit axiom, the structures $R$, $R'$, $T$ and $T'$ must vanish, i.e., all the atoms inside them have to participate in $\mathsf{ai}\!\downarrow$ interactions. Consider $R$: where are atoms that can annihilate its atoms? There are two possibilities: inside $R$ itself or inside $S\{\ \}$, and nowhere else. Now, we have to ask ourselves whether the application of $\mathsf{q}\!\uparrow$ to the conclusion above contributes in an essential way to the vanishing of $R$. Through $\mathsf{q}\!\uparrow$ we get

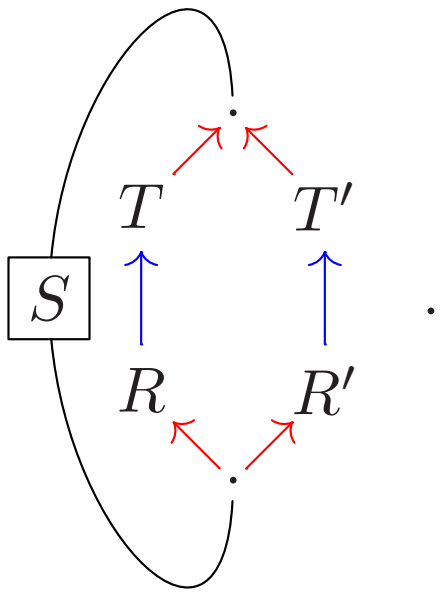

The application of $\mathsf{q}\!\uparrow$ just introduces new constraints: we cannot now use a structure $\langle R''; T''\rangle$ in $S\{\ \}$ to obtain $[R, R'']$ and $[T', T'']$ as we could before. Since the same argument holds for $R'$, $T$ and $T'$, we can conclude that, as far as provability is concerned, $\mathsf{q}\!\uparrow$ should be disposable.

This argument, which is essentially, but not entirely, correct (does the reader see the problem?), seems reasonable and amenable to a direct implementation in an admissibility proof. It is in fact what I did in the beginning, through the study of permutabilities inside $\mathsf{BV}$. Unfortunately, in certain special cases, the argument requires an unbearable complexity in the case analysis. The following proof uses splitting and is instead much more direct, and, in my opinion, more illuminating.

**4.2.1 Theorem** *The rule $\mathsf{q}\!\uparrow$ is admissible for $\mathsf{BV}$.*

**Proof** Consider the proof

$$\mathsf{q}\!\uparrow\frac{\overset{\displaystyle\prod\mathsf{BV}}{S(\langle R;T\rangle,\langle R';T'\rangle)}}{S\langle (R,R');(T,T')\rangle} \quad .$$

By Corollary 4.1.6 there exist $S_1$ and $S_2$ such that there are derivations:

$$\begin{array}{c}[\langle (R,R');(T,T')\rangle, S_1, S_2]\\ \Vert\,\mathsf{BV}\\ S\langle (R,R');(T,T')\rangle\end{array} \quad , \quad \begin{array}{c}\prod\mathsf{BV}\\ [\langle R;T\rangle, S_1]\end{array} \quad \text{and} \quad \begin{array}{c}\prod\mathsf{BV}\\ [\langle R';T'\rangle, S_2]\end{array} \quad .$$

By Theorem 4.1.4 we have, for some $S_R$, $S_T$, $S_{R'}$, $S_{T'}$:

$$\begin{array}{c}\langle S_R;S_T\rangle\\ \Vert\,\mathsf{BV}\\ S_1\end{array}\ ,\qquad \begin{array}{c}\overline{\Vert}\,\mathsf{BV}\\ [R,S_R]\end{array}\ ,\qquad \begin{array}{c}\overline{\Vert}\,\mathsf{BV}\\ [T,S_T]\end{array}\ ,\qquad \begin{array}{c}\langle S_{R'};S_{T'}\rangle\\ \Vert\,\mathsf{BV}\\ S_2\end{array}\ ,\qquad \begin{array}{c}\overline{\Vert}\,\mathsf{BV}\\ [R',S_{R'}]\end{array}\ \text{and}\ \begin{array}{c}\overline{\Vert}\,\mathsf{BV}\\ [T',S_{T'}]\end{array}\ .$$

We can then build

$$\begin{array}{c}
\overline{\Vert}\,\mathsf{BV}\\
\mathsf{q}{\downarrow}\,\dfrac{\langle[R',S_{R'}];[T',S_{T'}]\rangle}{[\langle R';T'\rangle,\langle S_{R'};S_{T'}\rangle]}\\
\Vert\,\mathsf{BV}\\
\mathsf{s}\,\dfrac{[\langle R';([T,S_T],T')\rangle,\langle S_{R'};S_{T'}\rangle]}{[\langle R';[(T,T'),S_T]\rangle,\langle S_{R'};S_{T'}\rangle]}\\
\Vert\,\mathsf{BV}\\
\mathsf{s}\,\dfrac{[\langle([R,S_R],R');[(T,T'),S_T]\rangle,\langle S_{R'};S_{T'}\rangle]}{\mathsf{q}{\downarrow}\,\dfrac{[\langle[(R,R'),S_R];[(T,T'),S_T]\rangle,\langle S_{R'};S_{T'}\rangle]}{[\langle(R,R');(T,T')\rangle,\langle S_R;S_T\rangle,\langle S_{R'};S_{T'}\rangle]}}\\
\Vert\,\mathsf{BV}\\
S\langle(R,R');(T,T')\rangle
\end{array}\quad ,$$

which is a proof in $\mathsf{BV}$.

We can repeat inductively this argument, starting from the top, for any proof in $\mathsf{BV}\cup\{\mathsf{q}{\uparrow}\}$, eliminating all the $\mathsf{q}{\uparrow}$ instances one by one. □

Let us now consider a structure $S\{\circ\}$ that, at a certain point while going up building a proof in $\mathsf{BV}\cup\{\mathsf{ai}{\uparrow}\}$, creates two complementary atoms $a$ and $\bar{a}$, by a cut:

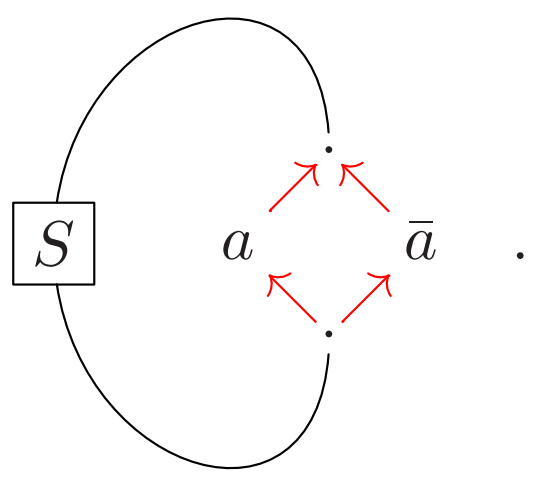

Suppose that this is the last creation of atoms, i.e., the proof of $S(a,\bar{a})$ is in $\mathsf{BV}$. The atoms $a$ and $\bar{a}$ do not communicate; therefore, since they will eventually be annihilated, there must be two other atoms $\bar{a}$ and $a$ that will do the job of killing them. Proceeding toward the top, nothing can leave the copar context that hosts the newly created atoms $a$ and $\bar{a}$. This means that there are, up in the proof, two structures $R\{\bar{a}\}$ and $T\{a\}$ that, at some point going upward, enter the copar, carrying the killer atoms with them. So, essentially, we have to consider the case where two structures $R\{\circ\}$ and $T\{\circ\}$ are to be proved in this situation:

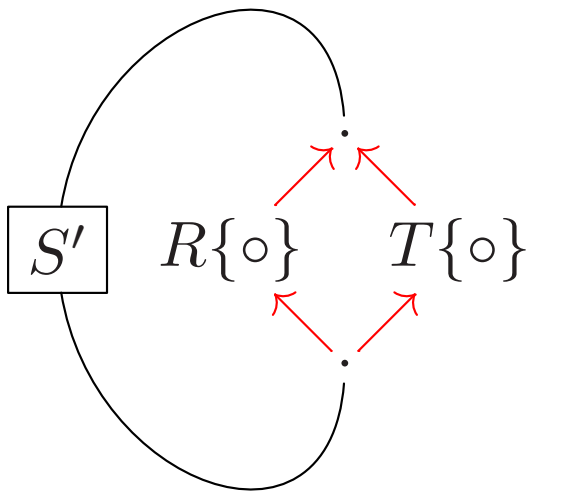

Since $R\{\circ\}$ and $T\{\circ\}$ do not communicate, all the resources needed to annihilate them must come either from inside themselves or from $S'\{\ \}$.

Consider now the alternative situation in which the original atoms $a$ and $\bar{a}$ have not been created at all. Since two structures $R\{\bar{a}\}$ and $T\{a\}$ are inside $S\{\ \}$ and can both enter the same copar, they must live in a par structural relation. Therefore, from $S\{\circ\}$, we can reach the structure:

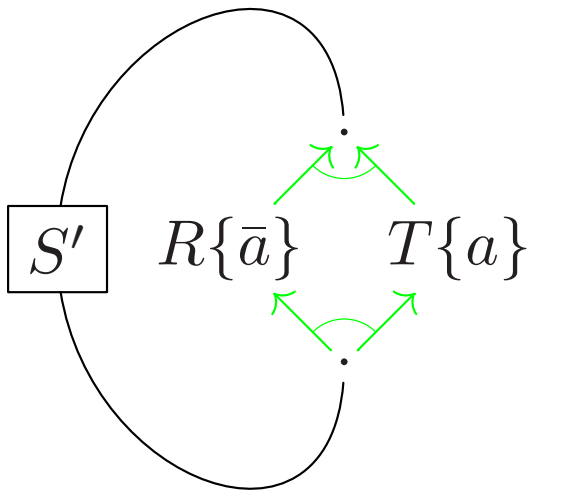

In fact, $R\{\bar{a}\}$ and $T\{a\}$ could enter the original copar context $(a, \bar{a})$ only by distinct uses of the rule s. If one replaces $(a, \bar{a})$ with $\circ$ and does not apply s, then $R\{\bar{a}\}$ and $T\{a\}$ remain in the same par context. But then one of $R\{\bar{a}\}$ and $T\{a\}$ can be merged into the other until $\bar{a}$ and $a$ come into direct communication, for example:

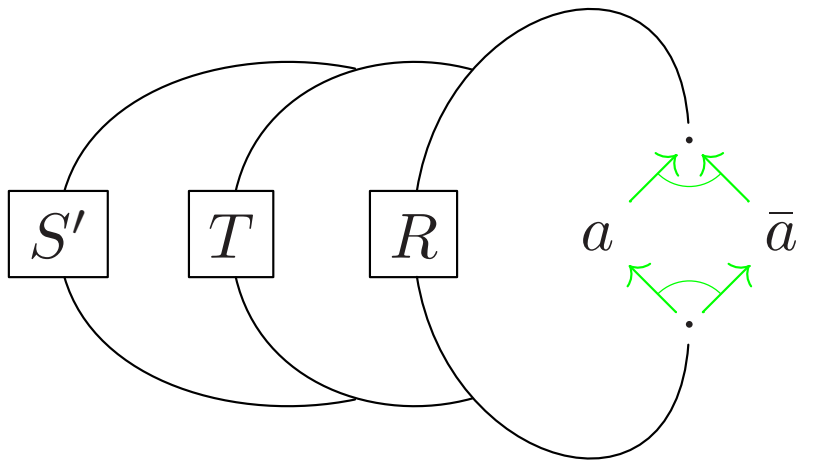

Now the two complementary atoms $a$ and $\bar{a}$ can be eliminated by an application of ai↓. Since $S'(R\{\circ\}, T\{\circ\})$ was provable, then $R\{\circ\}$ and $T\{\circ\}$ can be annihilated, each of them by use of its own resources, or of those coming from $S'\{\ \}$. This shows that the original cut was superfluous, and we can continue applying this argument on the other cut instances.

We can implement this idea in the following theorem, but before that I have to state an easy proposition.

**4.2.2 Proposition** *For every context $S\{\ \}$ and structures $R$, $T$, there exists a derivation*

$$\begin{array}{c} S[R,T] \\ \| \{\mathsf{q}\!\downarrow,\mathsf{s}\} \\ [S\{R\},T] \end{array} .$$

**4.2.3 Theorem** *The rule* $\mathsf{ai}\uparrow$ *is admissible for* $\mathsf{BV}$.

**Proof** Consider the proof

$$\mathsf{ai}\uparrow \frac{\begin{array}{c}\| \mathsf{BV} \\ S(a,\bar{a})\end{array}}{S\{\circ\}} \quad .$$

By Corollary 4.1.6 there exist $S_1$ and $S_2$ such that there are derivations:

$$\begin{array}{c}[S_1,S_2] \\ \| \mathsf{BV} \\ S\{\circ\}\end{array} \quad , \qquad \Pi_1 \| \mathsf{BV} \atop [a,S_1] \qquad \text{and} \qquad \Pi_2 \| \mathsf{BV} \atop [\bar{a},S_2] \quad .$$

Consider the proof $\Pi_1$: there must be a context $S'_1\{\ \}$ such that $S_1 = S'_1\{\bar{a}\}$ and

$$\Pi_1 = \begin{array}{c}\| \mathsf{BV} \\ \mathsf{ai}\downarrow \dfrac{S''\{\circ\}}{S''[a,\bar{a}]} \\ \| \mathsf{BV} \\ [a,S'_1\{\bar{a}\}]\end{array} \quad ,$$

for some $S''\{\ \}$, in which we individuate the instance of $\mathsf{ai}\downarrow$ where the $a$ occurrence interacts with the $\bar{a}$ occurrence coming from $S'_1\{\bar{a}\}$. We can replace in $\Pi_1$ every occurrence of $a$ and $\bar{a}$ with $\circ$, and we obtain a proof in $\mathsf{BV}$ of $S'_1\{\circ\}$. Analogously, we can transform $\Pi_2$ into a proof in $\mathsf{BV}$ of an $S'_2\{\circ\}$ such that $S_2 = S'_2\{a\}$. We can then build the following proof

$$\begin{array}{c}\| \mathsf{BV} \\ S'_1\{\circ\} \\ \| \mathsf{BV} \\ \mathsf{ai}\downarrow \dfrac{S'_1\{S'_2\{\circ\}\}}{S'_1\{S'_2[a,\bar{a}]\}} \\ \| \mathsf{BV} \\ [S'_1\{\bar{a}\},S'_2\{a\}] \\ \| \mathsf{BV} \\ S\{\circ\}\end{array} \quad ,$$

where we used twice Proposition 4.2.2.

We can repeat inductively this argument, starting from the top, for any proof in $\mathsf{BV} \cup \{\mathsf{ai}\uparrow\}$, eliminating all the $\mathsf{ai}\uparrow$ instances one by one. □

**4.2.4 Remark** The admissible fragment of $\mathsf{SBV} \cup \{\circ\downarrow\}$ is $\{\mathsf{ai}\uparrow, \mathsf{q}\uparrow\}$. By Theorems 3.4.7 and 3.2.5, system $\mathsf{SBV} \cup \{\circ\downarrow\}$ is strongly equivalent to the system $\{\circ\downarrow, \mathsf{ai}\downarrow, \mathsf{i}\uparrow, \mathsf{q}\downarrow, \mathsf{s}\} = \mathsf{BV} \cup \{\mathsf{i}\uparrow\}$.

**4.2.5 Theorem** $\mathsf{SBV} \cup \{\circ\downarrow\}$ *and* $\mathsf{BV}$ *are equivalent.*

**Proof** Given a proof in $\mathsf{SBV} \cup \{\circ\downarrow\}$, eliminate the $\mathsf{q}\uparrow$ and $\mathsf{ai}\uparrow$ instances one by one starting from the top. □

Since any atom $a$ cannot be proved in $\mathsf{BV}$, then $\mathsf{SBV} \cup \{\circ\downarrow\}$ and $\mathsf{BV}$ are consistent. This consistency property can be reinforced as in the following theorem, whose proof is particularly interesting:

**4.2.6 Theorem** *If $S$ is provable in* $\mathsf{BV}$ *then $\bar{S}$ is not provable, provided $S \neq \circ$.*

**Proof** A proof of $S$ must be as

$$\begin{array}{c} \mathsf{ai}\downarrow \dfrac{\circ\downarrow \dfrac{}{\circ}}{[a,\bar{a}]} \\ \Big\| \mathsf{BV} \\ S \end{array} .$$

Get $\begin{array}{c}\bar{S} \\ \| \mathsf{SBV} \\ (a,\bar{a})\end{array}$ by flipping the given proof. If $\bar{S}$ is provable, then $(a,\bar{a})$ is provable in $\mathsf{SBV} \cup \{\circ\downarrow\}$ and then in $\mathsf{BV}$: impossible. □

**4.2.7 Theorem** *It is decidable whether a given structure is provable in* $\mathsf{SBV} \cup \{\circ\downarrow\}$ *and in* $\mathsf{BV}$.

**Proof** The set of derivations in $\mathsf{BV}$ that have a given conclusion, for which to look, is finite (if no trivial instances of rules are applied, which is not a limitation); see Subsection 3.3. □

Most cut-free sequent systems enjoy the *subformula property*: the premisses of rule instances only contain subformulae of formulae in their conclusions. Our case is different, because we do not use connectives to split formulae in conclusions into parts that go into premisses, but it is analogous. Apart from the cut rule, all the other rules lower the 'energy' of a formula (see Subsection 3.3), either by the property mentioned in Remark 3.3.5 (with $\mathsf{q}\downarrow$, $\mathsf{q}\uparrow$ and $\mathsf{s}$) or because atom occurrences diminish in number (with $\mathsf{ai}\downarrow$). In this sense, we get simpler structures while going up in the building of a derivation.

A consequence of the subformula property is that systems are modular with respect to connectives: Given a system, one can remove a connective and its dual through negation and their relative rules, without affecting the provability of the formulae that do not use those connectives. A similar situation occurs in $\mathsf{BV}$. Consider not to employ the seq structural relation, i.e., just use flat structures. For them, the following system is appropriate:

**4.2.8 Definition** System $\mathsf{FBV}$ (called *flat system* $\mathsf{BV}$) is the system $\{\circ\downarrow, \mathsf{ai}\downarrow, \mathsf{s}\}$.

Deducing over flat structures with system $\mathsf{FBV}$ means remaining in a flat horizontal line where all the structural relations are commutative.

**4.2.9 Theorem** *If a flat structure is provable in* $\mathsf{BV}$ *then it is provable in* $\mathsf{FBV}$.

**Proof** Let $\Pi$ be a proof in $\mathsf{BV}$ of a given flat structure $S$. Transform $\Pi$ into an object $\Pi'$, by changing every seq context appearing in each structure in $\Pi$ into a par context. $\Pi'$ is still a proof, where all the instances of $\mathsf{q}\downarrow$ become trivial and can be eliminated, this way obtaining a proof of $S$ in $\mathsf{FBV}$. □

This is a key result in the correspondence between $\mathsf{BV}$ and linear logic, as we are going to see in the next section.

# 5 Relation with Linear Logic

Linear logic is, of course, the main source of inspiration of this paper. In this section I show that $\mathsf{BV}$ extends the multiplicative fragment of linear logic while conserving provability

$$\mathsf{id}\,\frac{}{\vdash A,\bar{A}} \qquad \parr\,\frac{\vdash A,B,\Phi}{\vdash A\parr B,\Phi} \qquad \otimes\,\frac{\vdash A,\Phi \quad \vdash B,\Psi}{\vdash A\otimes B,\Phi,\Psi} \qquad \mathsf{mix}\,\frac{\vdash \Phi \quad \vdash \Psi}{\vdash \Phi,\Psi}$$

**Fig. 12** *System* MLL+mix

in its commutative fragment. The version of linear logic about which I will talk is the multiplicative fragment plus mix, without units. As we will see later on, the rule mix is necessary and fits naturally in our scheme, and its use is consistent with the results in [27], where it is employed.

**5.1 Definition** The system of *multiplicative linear logic without units and with the rule* mix is denoted by MLL+mix and is such that:

**1** *Formulae*, denoted by $A$, $B$ and $C$ are built over atoms according to

$$A ::= a \mid A \parr A \mid A \otimes A \mid \bar{A} \quad ,$$

where the binary connectives $\parr$ and $\otimes$ are called respectively *par* and *times* and $\bar{A}$ is the *negation* of $A$. When necessary, brackets are used to disambiguate expressions. Negation is defined by De Morgan laws:

$$\overline{A \parr B} = \bar{A} \otimes \bar{B} \quad ,$$
$$\overline{A \otimes B} = \bar{A} \parr \bar{B} \quad ,$$

and formulae are considered equivalent modulo the relation $=$.

**2** *Sequents*, denoted by $\Sigma$, are expressions of the kind

$$\vdash A_1, \dots, A_h \quad ,$$

where $h \geqslant 0$ and the comma between the formulae $A_1, \dots, A_h$ stands for multiset union. Multisets of formulae are denoted by $\Phi$ and $\Psi$. For example, $\vdash \Phi, \Psi, A$ is a sequent whose formulae are those in $\Phi \uplus \Psi \uplus \{A\}_+$.

**3** *Derivations*, denoted by $\Delta$, are trees represented like

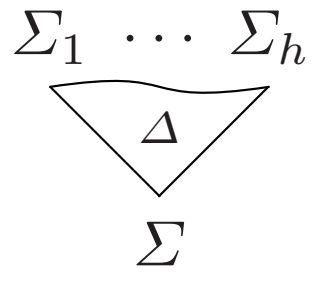

,

where $h \geqslant 0$, the sequents $\Sigma_1, \dots, \Sigma_h$ are called *premisses*, $\Sigma$ is the *conclusion*, and a finite number (possibly zero) of instances of the inference rules in Figure 12 are applied. A derivation with no premisses is a *proof*, denoted by $\Pi$.

Linear logic formulae correspond to flat structures (i.e., structures where seq contexts do not appear) different from the unit.

**5.2 Definition** The function $\underline{\cdot}_\vee$ transforms the formulae of MLL+mix into flat structures according to the following inductive definition:

$$\underline{a}_\vee = a \quad ,$$
$$\underline{A \mathbin{⅋} B}_\vee = [\underline{A}_\vee, \underline{B}_\vee] \quad ,$$
$$\underline{A \otimes B}_\vee = (\underline{A}_\vee, \underline{B}_\vee) \quad .$$

The domain of $\underline{\cdot}_\vee$ is extended to sequents this way:

$$\underline{\vdash A_1, \dots, A_h}_\vee = [\underline{A_1}_\vee, \dots, \underline{A_h}_\vee] \quad ,$$

where $h > 0$; let $\underline{\vdash}_\vee = \circ$.

The first, fairly obvious result of this section is about the ability of BV to mimic the derivations in MLL+mix. System FBV is, of course, enough. We need firstly to establish that if we restrict ourselves to using interaction rules with flat structures, then seq and coseq rules are no more necessary.

**5.3 Lemma** *If the rules* $\mathsf{i}{\downarrow}\,\dfrac{S\{\circ\}}{S[R,\bar{R}]}$ *and* $\mathsf{i}{\uparrow}\,\dfrac{S(R,\bar{R})}{S\{\circ\}}$ *are restricted to the case in which $R$ is flat, then they are derivable for* $\{\mathsf{ai}{\downarrow}, \mathsf{s}\}$ *and* $\{\mathsf{ai}{\uparrow}, \mathsf{s}\}$*, respectively.*

**Proof** The proof is just a special case of the ones of Theorems 3.4.2 and 3.4.3. □

**5.4 Theorem** *For every derivation* $\begin{array}{c}\Sigma_1 \ \cdots\ \Sigma_h \\ \bigtriangledown \\ \Sigma\end{array}$ *in* MLL+mix *there exists* $\begin{array}{c}(\underline{\Sigma_1}_\vee, \dots, \underline{\Sigma_h}_\vee) \\ \big\| \,\mathsf{FBV} \\ \underline{\Sigma}_\vee\end{array}$ .

**Proof** Let $\Delta$ be a given derivation in MLL+mix: let us proceed by structural induction on $\Delta$. The following cases are possible:

1 If $\Delta$ is $\Sigma$, take $\underline{\Sigma}_\vee$.

2 If $\Delta$ is $\mathsf{id}\,\dfrac{}{\vdash A, \bar{A}}$, for some formula $A$, take the proof in FBV obtained from $\mathsf{i}{\downarrow}\,\dfrac{\circ{\downarrow}\,\dfrac{}{\circ}}{[\underline{A}_\vee, \underline{\bar{A}}_\vee]}$ by unfolding the instance of $\mathsf{i}{\downarrow}$, by Lemma 5.3.

3 If $\Delta$ has shape $⅋\,\dfrac{\begin{array}{c}\Sigma_1 \ \cdots\ \Sigma_h \\ \bigtriangledown \\ \vdash A, B, \Phi\end{array}}{\vdash A \mathbin{⅋} B, \Phi}$, take $\begin{array}{c}(\underline{\Sigma_1}_\vee, \dots, \underline{\Sigma_h}_\vee) \\ \big\| \,\mathsf{FBV} \\ \underline{\vdash A, B, \Phi}_\vee\end{array}$, which exists by induction hypothesis.

4 Suppose that $\Delta$ has shape

$$\otimes\,\dfrac{\begin{array}{c}\Sigma'_1 \ \cdots\ \Sigma'_k \\ \bigtriangledown \\ \vdash A, \Phi'\end{array} \qquad \begin{array}{c}\Sigma''_1 \ \cdots\ \Sigma''_l \\ \bigtriangledown \\ \vdash B, \Psi'\end{array}}{\vdash A \otimes B, \Phi', \Psi'} \qquad \text{or} \qquad \mathsf{mix}\,\dfrac{\begin{array}{c}\Sigma'_1 \ \cdots\ \Sigma'_k \\ \bigtriangledown \\ \vdash \Phi\end{array} \qquad \begin{array}{c}\Sigma''_1 \ \cdots\ \Sigma''_l \\ \bigtriangledown \\ \vdash \Psi\end{array}}{\vdash \Phi, \Psi} \quad ,$$

where $k, l \geqslant 0$ and, for the derivation at the left, we can suppose that $\vdash \Phi = \vdash A, \Phi'$ and $\vdash \Psi = \vdash B, \Psi'$. Let us also suppose that $\Phi \neq \varnothing_+ \neq \Psi$, otherwise the instance of mix would be trivial. In both cases there are, by induction hypothesis, two derivations

$$\begin{array}{c}(\underline{\Sigma'_1}_{\mathsf{V}},\dots,\underline{\Sigma'_k}_{\mathsf{V}})\\ \Delta'\Big\Vert\,\mathsf{FBV}\\ \underline{\vdash\Phi}_{\mathsf{V}}\end{array}\qquad\text{and}\qquad\begin{array}{c}(\underline{\Sigma''_1}_{\mathsf{V}},\dots,\underline{\Sigma''_l}_{\mathsf{V}})\\ \Delta''\Big\Vert\,\mathsf{FBV}\\ \underline{\vdash\Psi}_{\mathsf{V}}\end{array}\quad .$$

Let $\Delta_1=(\Delta',\underline{\Sigma''_1}_{\mathsf{V}},\dots,\underline{\Sigma''_l}_{\mathsf{V}})$, where with $(\Delta',S)$ we denote the derivation obtained from $\Delta'$ by immersing every structure appearing in it in the hole of $(\ ,S)$; analogously, let $\Delta_2=(\underline{\vdash\Phi}_{\mathsf{V}},\Delta'')$. We can take, respectively, the derivations

$$\begin{array}{c}(\underline{\Sigma'_1}_{\mathsf{V}},\dots,\underline{\Sigma'_k}_{\mathsf{V}},\underline{\Sigma''_1}_{\mathsf{V}},\dots,\underline{\Sigma''_l}_{\mathsf{V}})\\ \Delta_1\Big\Vert\,\mathsf{FBV}\\ (\underline{\vdash A,\Phi'}_{\mathsf{V}},\underline{\Sigma''_1}_{\mathsf{V}},\dots,\underline{\Sigma''_l}_{\mathsf{V}})\\ \Delta_2\Big\Vert\,\mathsf{FBV}\\ \mathsf{s}\,\dfrac{\mathsf{s}\,\dfrac{(\underline{\vdash A,\Phi'}_{\mathsf{V}},\underline{\vdash B,\Psi'}_{\mathsf{V}})}{[(\underline{\vdash A,\Phi'}_{\mathsf{V}},\underline{B}_{\mathsf{V}}),\underline{\vdash\Psi'}_{\mathsf{V}}]}}{[(\underline{A}_{\mathsf{V}},\underline{B}_{\mathsf{V}}),\underline{\vdash\Phi'}_{\mathsf{V}},\underline{\vdash\Psi'}_{\mathsf{V}}]}\end{array}\qquad\text{or}\qquad\begin{array}{c}(\underline{\Sigma'_1}_{\mathsf{V}},\dots,\underline{\Sigma'_k}_{\mathsf{V}},\underline{\Sigma''_1}_{\mathsf{V}},\dots,\underline{\Sigma''_l}_{\mathsf{V}})\\ \Delta_1\Big\Vert\,\mathsf{FBV}\\ (\underline{\vdash\Phi}_{\mathsf{V}},\underline{\Sigma''_1}_{\mathsf{V}},\dots,\underline{\Sigma''_l}_{\mathsf{V}})\\ \Delta_2\Big\Vert\,\mathsf{FBV}\\ \mathsf{s}\,\dfrac{(\underline{\vdash\Phi}_{\mathsf{V}},\underline{\vdash\Psi}_{\mathsf{V}})}{[\underline{\vdash\Phi}_{\mathsf{V}},\underline{\vdash\Psi}_{\mathsf{V}}]}\end{array}\quad .$$

□

**5.5 Corollary** *If $\vdash A$ is provable in* MLL+mix *then $\underline{A}_{\mathsf{V}}$ is provable in* FBV*.*

System FBV produces more, inherently different derivations from those of MLL+mix, as we saw in the introduction, because it is not bound to the application of rules to a main connective. A natural question at this point is whether it *proves more* than what MLL+mix does, and the answer is *no*. We need the inverse transformation of $\underline{\cdot}_{\mathsf{V}}$.

**5.6 Definition** The function $\underline{\cdot}_{\mathsf{L}}$ transforms non-unit normal flat structures into the formulae of MLL+mix according to the following inductive definition:

$$\begin{aligned}\underline{a}_{\mathsf{L}}&=a\quad ,\\ \underline{[R,T]}_{\mathsf{L}}&=\underline{R}_{\mathsf{L}}\parr\underline{T}_{\mathsf{L}}\quad ,\\ \underline{(R,T)}_{\mathsf{L}}&=\underline{R}_{\mathsf{L}}\otimes\underline{T}_{\mathsf{L}}\quad ,\end{aligned}$$

where the formulae in the range of $\underline{\cdot}_{\mathsf{L}}$ are to be considered modulo the equivalence relation generated by commutativity and associativity of $\parr$ and $\otimes$ (so that equivalent structures in the domain of $\underline{\cdot}_{\mathsf{L}}$ are brought to equivalent formulae). This definition is extended to non-normal, non-unit flat structures in the obvious way.

**5.7 Theorem** *If a non-unit flat structure $P$ is provable in* FBV *then $\vdash\underline{P}_{\mathsf{L}}$ is provable in* MLL+mix*.*

**Proof** In the proof we will use the following known properties of provability in MLL+mix:

**1** If $\vdash\bar{B},C$ is provable then $\vdash\overline{A\{B\}},A\{C\}$ is provable (where $A\{B\}$ is a context formula $A\{\ \}$ whose hole has been filled with $B$).

**2** The rule $\mathsf{cut}\ \dfrac{\vdash A,\Phi\quad\vdash\bar{A},\Psi}{\vdash\Phi,\Psi}$ is admissible for MLL+mix.

The first property is easily provable by induction on the structure of $A\{\ \}$, the second by a standard first order cut elimination argument.

To prove the theorem let us proceed by induction on the length of the proof $\Pi$ of $P$. In the base case $P = [a, \bar{a}]$:

$$\text{when}\qquad \Pi = \mathsf{ai}{\downarrow}\,\dfrac{\circ{\downarrow}\,\dfrac{}{\circ}}{[a,\bar a]}\qquad\text{take}\qquad \parr\,\dfrac{\mathsf{id}\,\dfrac{}{\vdash a,\bar a}}{\vdash a\parr\bar a}\quad .$$

There are two inductive cases, corresponding to the bottom rule instance in $\Pi$; in both of them $\Pi''$ is the proof in $\mathsf{MLL{+}mix}$ corresponding, by induction hypothesis, to a proof $\Pi'$ in $\mathsf{FBV}$:

**3** $\quad P = S[a, \bar{a}]$, where $S\{\circ\} \neq \circ$, and the bottom inference rule instance is $\mathsf{ai}{\downarrow}$: Let us firstly prove the following fact about provability in $\mathsf{MLL{+}mix}$ (very similar to the property 1 above): $\vdash \overline{\underline{S\{\circ\}}_{\mathsf L}}, \underline{S[a,\bar a]}_{\mathsf L}$ is provable. Let us proceed by induction on the structure of $S\{\circ\}$. The inductive cases are

$$\parr\,\dfrac{\otimes\,\dfrac{\mathsf{id}\,\dfrac{}{\vdash \overline{\underline{T}_{\mathsf L}}, \underline{T}_{\mathsf L}} \qquad \dfrac{\nabla}{\vdash \overline{\underline{U\{\circ\}}_{\mathsf L}}, \underline{U[a,\bar a]}_{\mathsf L}}}{\vdash \overline{\underline{T}_{\mathsf L}} \otimes \overline{\underline{U\{\circ\}}_{\mathsf L}}, \underline{T}_{\mathsf L}, \underline{U[a,\bar a]}_{\mathsf L}}}{\vdash \overline{\underline{T}_{\mathsf L}} \otimes \overline{\underline{U\{\circ\}}_{\mathsf L}}, \underline{T}_{\mathsf L} \parr \underline{U[a,\bar a]}_{\mathsf L}}\qquad\text{and}\qquad \parr\,\dfrac{\otimes\,\dfrac{\mathsf{id}\,\dfrac{}{\vdash \overline{\underline{T}_{\mathsf L}}, \underline{T}_{\mathsf L}} \qquad \dfrac{\nabla}{\vdash \overline{\underline{U\{\circ\}}_{\mathsf L}}, \underline{U[a,\bar a]}_{\mathsf L}}}{\vdash \overline{\underline{T}_{\mathsf L}}, \overline{\underline{U\{\circ\}}_{\mathsf L}}, \underline{T}_{\mathsf L} \otimes \underline{U[a,\bar a]}_{\mathsf L}}}{\vdash \overline{\underline{T}_{\mathsf L}} \parr \overline{\underline{U\{\circ\}}_{\mathsf L}}, \underline{T}_{\mathsf L} \otimes \underline{U[a,\bar a]}_{\mathsf L}}\quad ,$$

where $U\{\circ\} \neq \circ$. The base cases are

$$\parr\,\dfrac{\mathsf{mix}\,\dfrac{\mathsf{id}\,\dfrac{}{\vdash \overline{\underline{T}_{\mathsf L}}, \underline{T}_{\mathsf L}} \qquad \parr\,\dfrac{\mathsf{id}\,\dfrac{}{\vdash a,\bar a}}{\vdash a\parr\bar a}}{\vdash \overline{\underline{T}_{\mathsf L}}, \underline{T}_{\mathsf L}, a\parr\bar a}}{\vdash \overline{\underline{T}_{\mathsf L}}, \underline{T}_{\mathsf L} \parr \underline{[a,\bar a]}_{\mathsf L}}\qquad\text{and}\qquad \otimes\,\dfrac{\mathsf{id}\,\dfrac{}{\vdash \overline{\underline{T}_{\mathsf L}}, \underline{T}_{\mathsf L}} \qquad \parr\,\dfrac{\mathsf{id}\,\dfrac{}{\vdash a,\bar a}}{\vdash a\parr\bar a}}{\vdash \overline{\underline{T}_{\mathsf L}}, \underline{T}_{\mathsf L} \otimes \underline{[a,\bar a]}_{\mathsf L}}\quad .$$

Coming back to our main proof:

$$\text{when}\qquad \Pi = \mathsf{ai}{\downarrow}\,\dfrac{\begin{array}{c}\Pi'\,\Vert\,\mathsf{FBV}\\ S\{\circ\}\end{array}}{S[a,\bar a]}\qquad\text{take}\qquad \mathsf{cut}\,\dfrac{\dfrac{\Pi''}{\vdash \underline{S\{\circ\}}_{\mathsf L}} \qquad \dfrac{\nabla}{\vdash \overline{\underline{S\{\circ\}}_{\mathsf L}}, \underline{S[a,\bar a]}_{\mathsf L}}}{\vdash \underline{S[a,\bar a]}_{\mathsf L}}\quad .$$

**4** $\quad P = S[(R, R'), T]$ and the bottom inference rule instance is $\mathsf{s}$:

$$\text{when}\qquad \Pi = \mathsf{s}\,\dfrac{\begin{array}{c}\Pi'\,\Vert\,\mathsf{FBV}\\ S([R,T],R')\end{array}}{S[(R,R'),T]}\qquad\text{take}\qquad \mathsf{cut}\,\dfrac{\dfrac{\Pi''}{\vdash \underline{S([R,T],R')}_{\mathsf L}} \qquad \dfrac{\nabla}{\vdash \overline{\underline{S([R,T],R')}_{\mathsf L}}, \underline{S[(R,R'),T]}_{\mathsf L}}}{\vdash \underline{S[(R,R'),T]}_{\mathsf L}}\quad ,$$

where we applied the property 1 above with the proof

$$\parr\,\dfrac{\parr\,\dfrac{\otimes\,\dfrac{\otimes\,\dfrac{\mathsf{id}\,\dfrac{}{\vdash \bar R, R} \qquad \mathsf{id}\,\dfrac{}{\vdash \bar R', R'}}{\vdash \bar R, \bar R', R\otimes R'} \qquad \mathsf{id}\,\dfrac{}{\vdash \bar T, T}}{\vdash \bar R\otimes\bar T, \bar R', R\otimes R', T}}{\vdash \bar R\otimes\bar T, \bar R', (R\otimes R')\parr T}}{\vdash (\bar R\otimes\bar T)\parr\bar R', (R\otimes R')\parr T}\quad .$$

Since cuts can be eliminated, by the property 2, the two proofs obtained in the inductive cases can be transformed into cut-free proofs of MLL+mix. □

The correspondence between BV and MLL+mix, with respect to provability, is then complete on their common language:

**5.8 Theorem** *If a non-unit flat structure $S$ is provable in* BV *then* $\vdash \underline{S}_{\mathsf{L}}$ *is provable in* MLL+mix*.*

**Proof** It follows from Theorems 4.2.9 and 5.7. □

**5.9 Remark** The correspondence of FBV to linear logic can be extended to multiplicative units as follows. Let $\bot$ and $1$ be, respectively, the units for the connectives $⅋$ and $\otimes$. Consider adding to MLL+mix the following three inference rules, where the first two are the usual rules for $\bot$ and $1$ and the third, called mix0, as Abramsky and Jagadeesan do in [1], is a nullary version of mix:

$$\bot \frac{\vdash \Phi}{\vdash \bot, \Phi} \quad , \qquad 1 \frac{}{\vdash 1} \quad , \qquad \mathsf{mix0} \frac{}{\vdash} \quad .$$

These rules make $\bot$ and $1$ collapse: both $\vdash \bot, \bot$ and $\vdash 1, 1$ are provable. This means that we are left, essentially, with a single unit for both $⅋$ and $\otimes$, and we can map this unit into $\circ$ with no effort.

## 6 Conclusions

The question answered by this paper is: Is it possible to design a formal system in the tradition of proof theory, only based on simple properties that are universally recognised as fundamental ones in the concurrent management of information? This system should only come out of the analysis of some local properties of computation, like sending or receiving an atomic message, or preventing information to be exchanged by turning a finite, bounded switch. Not to float in a purely philosophical realm, I decided to apply these principles to the concrete problem of extending the multiplicative core of linear logic with a non-commutative operator. This investigation took the form of an experiment, of which I will give an account that also summarises the results achieved.

I departed from the rigid scheme of the sequent calculus and conceived what I call the 'calculus of structures'. This is the tool I used for designing various formal systems, and it has the following differences from the sequent calculus:

1 Structures take the place of formulae and sequents, and the difference is only in the way we look at them. While formulae are trees built by binary connectives, structures are collections of atom occurrences where each couple of occurrences is bound by one and only one structural relation. Certain other laws hold for structures, whose purpose is to establish some good modularity properties. I have characterised the class of structures in terms of properties that structural relations must possess, and this move frees us from the idiosyncrasies of syntax: we can say that structures are *essentially* determined by their inner structural relations.

2 Since we have no more the idea of *main connective*, it is natural to think of inference rules that can act (i.e., rewrite) anywhere in structures. We can also limit ourselves to employing single-premiss rules only. This move only makes sense if we can show that we get an analogous feature to the *subformula property*. This in fact happens with our derivations, which can be read, bottom-up-wise, as progressively limiting the range of choices for interaction to happen. Vice versa, they can be read, top-down-wise, as making weaker and weaker assertions about times and places where interactions occur.

Then I used the calculus of structures for building a commutative/non-commutative system, in a very implicit way. The idea was: let us define the structural relations in which we are interested, and let us think of simple conservation laws that could make sense, and then let us try to find rules, whose complexity is minimal and which can implement our conservation laws.

The starting point was using system WMV (mentioned in Definition 3.3.2) on structures built over the three structural relations discussed in the introduction. System WMV is only based on the idea that two structures can freely merge provided that in doing so the original structural relations are conserved. I cannot think of a derivability notion simpler than this, which possesses our analogues of the subformula property. After having unfolded the two symmetric rules that I had and having added rules for interaction, I discovered that the cut rule was not admissible.

Interaction rules were untouchable, since what they do is simply to recognise that two structures are dual through negation, and this is primitive for me. Then I had to modify how I dealt with order, i.e., with structural relations, having in mind the only, limited objective of getting cut elimination. For some reason, solving this problem has been very difficult. The result is Definition 3.1.3, which adds to the previous merging law an axiom, which I call *integrity conservation*. Besides giving us cut elimination, this definition has the surprising property of producing exactly multiplicative linear logic (with mix) plus the desired non-commutative extension. This is remarkable because we get the core part of linear logic out of a very simple-minded, natural, purely syntactic and finitistic approach, essentially based on ideas of symmetry and conservation. Moreover, the resulting system BV is itself simple: order is managed by two rules, interaction by one, and it has an intuitive semantics which I do not hesitate to call *beautiful*.

The success of the calculus of structures in dealing with non-commutativity stems from a simple fact. Consider the seq rule $\mathsf{q}{\downarrow}\,\dfrac{S\langle [R,T];[R',T']\rangle}{S[\langle R;R'\rangle,\langle T;T'\rangle]}$. In the sequent calculus we are only able to mimic $\dfrac{S([R,T],[R',T'])}{S[\langle R;R'\rangle,\langle T;T'\rangle]}$, i.e., we get a derivation where the two substructures in the premiss are in a copar instead of a seq. This is too strong, because it prevents certain communications, between $S\{\circ\}$ and $[R,T]$ and between $S\{\circ\}$ and $[R',T']$, that are allowed in $\mathsf{q}{\downarrow}$. The problem with the sequent calculus is that double-premiss rules always perform a conjunction between premisses, and we need something finer than that.

This does not mean that it is impossible to devise sequent systems that deal with non-commutativity together with commutativity. The recent works [2, 29], by Abrusci and Ruet, are in fact a successful attempt at this task, and the first, as far as I know. The authors manage to build a multiplicative linear sequent system of four connectives, two commutative and two non-commutative, where each couple contains both a conjunction and a disjunction. Their system can be expanded to contain all of linear logic, therefore all of classical logic. The price they pay, in my opinion, is simplicity. Despite recent simplifications, their sequents are provided with a technically sophisticated but quite complex order structure. This conflicts with my belief that systems should be *simple* to be useful, but of course this is a very subjective matter.

Polakow and Pfenning, in [24], devise a natural deduction and an intuitionistic sequent system for a form of non-commutative linear logic, obtained as a refinement of intuitionistic linear logic. Their approach is essentially aimed at logic programming, with no attempt at an intimate coexistence of commutative and non-commutative relations, and their logic captures non-commutativity in hypotheses. The intrinsic, intuitionistic asymmetry of their systems prevents having an involutive negation. Yet, having a perfect symmetry, through negation, between two dual entities that communicate, is an absolute prerequisite, from the point of view of the logical foundations of concurrency. Then, I would say that the systems of Polakow and Pfenning are not a satisfying solution to our specific, foundational problem.

These two papers are the only that I know related to this work, apart from Retoré's ones, discussed in several places elsewhere in this paper. I do not know how to compare my system to those of others, if not on the basis of their promise of being groundwork for future applications and research. As a matter of fact, besides the formal systems that I discussed, there is the issue of having introduced a new proof-theoretical *methodology*.

Since the cut rule is admissible, our system is in principle a good candidate for such applications like automated deduction, logic programming, etc. The most direct comparison should be done with linear logic, of course, and then with the study of its uniform proofs. In this respect, the reference paper is Miller's [20], where it is defined the formal system $\mathsf{Forum}$, which is equivalent to linear logic.

The biggest problem in the proof-construction field is eliminating as much as possible unnecessary, spurious non-determinism in the bottom-up search for a proof. $\mathsf{Forum}$ succeeds in eliminating all the non-determinism due to permutability of inference rules in linear logic, and employs a clever technique, called *focusing* and due to Andreoli [3], to keep the search for a proof nailed to certain choices done during a computation. Unfortunately, being intimately tied to the sequent calculus, $\mathsf{Forum}$ suffers from the very same context partitioning problem of the times rule. Therefore, it will be interesting for me to see whether my system can be subjected to the same treatment that Andreoli and Miller performed on linear logic, since this could lead to a system with all the good properties of $\mathsf{Forum}$, with first-class non-commutativity, and without the context partitioning problem.

To this purpose, the situation in $\mathsf{BV}$ and its extensions is quite different from the one in the sequent calculus. As one of the examples in the introduction shows, reductions may occur everywhere in the structure, while building a derivation. There is not just a single active area corresponding to the main connective, or a few active areas corresponding to each of the formulae in a sequent. In fact, this is a big difference from the sequent calculus, and a definite improvement from the point of view of concurrency. On the other hand, non-determinism can be undesirable, and the calculus of structures is much more non-deterministic than the sequent calculus.

Non-determinism in the sequent calculus is tamed by an analysis of rule permutations and their recombination. It's true that the calculus of structures yields more permutations (non-determinism), but at the same time it also offers more possibilities of recombination. This is testified by what we call *decomposition theorems*: derivations can be rearranged such that they are composed of a fixed sequence of segments, each of which is carried on by one of several disjoint subsystems of the original system. For example, multiplicative exponential linear logic exhibits seven disjoint subsystems which enjoy this property in several different rearrangements [14, 33]. This flexibility has no counterpart in other formalisms, like the sequent calculus, natural deduction or proof nets. A first success toward the taming of nondeterminism has been obtained by Bruscoli in [10], where she is able faithfully to capture a fragment of CCS [21] containing the prefix operator. So, I would say that it is far from obvious whether the calculus of structures is weak (or strong) at applications in proof construction.

The question is: Is there a *logical* reason to confine rewriting to certain, selected places of logical structures, just because they happen to be external? In the calculus of structures, all atoms in suitable structural relations may participate to rewriting at any time. This, in principle, could be important also for applications.

The conservation principles outlined in the paper are of course semantic ideas. Their application led me to the discovery of inference rules. It should be possible to work on these semantic ideas and to get a characterisation of derivability in larger and larger fragments. A problem is stated (3.3.4) whose solution would lead to a semantic characterisation of derivability in the combinatorial core of our system. The interest of the characterisation envisaged is in the fact that it would only involve checking a natural ordering relation induced by structural relations. I do not know if this investigation could give us also an *inspiring* semantics, even if the starting point seems good.

Games semantics is definitely something to explore. The paper [1] gives semantics to MLL+mix (multiplicative linear logic plus mix), which is a fragment of linear logic strictly included in my system. The semantics given there is fully complete for proof nets: every winning strategy denotes a unique cut-free proof net. Since derivations in the calculus of structures would interleave parts of their corresponding, hypothetical proof nets, as happens in the sequent calculus, it seems obvious firstly to establish proof nets for our system. In fact, one proof net usually stands for many derivations differing in the sequentialisation of the information in the net; therefore, proof nets are closer objects to semantics than derivations. The natural candidate for our case is Retoré's notion of proof net in [28], which, again, includes MLL+mix and features a non-commutative connective. That connective appears to be very similar to mine, and I hope the third projected paper on system BV will finally resolve the question.

Is it justified to introduce a new calculus, together with a new proof-theoretical methodology, simply to solve the problem of non-commutativity? The calculus of structures does more than just solving this problem. So far, the people in my group and myself have dealt with classical logic [7, 9, 5, 8, 6] ([5] contains a stunningly concise and elegant cut elimination proof), linear logic [32, 13, 31, 33, 34], a conservative extension of both multiplicative exponential linear logic and system BV [14], modal logics [30] and Yetter's non-commutative logic [11]. We are now able to present all the above mentioned logics in an entirely local way, meaning that each application of a rule has a bounded computational cost, including contraction and the managing of additive contexts. More in general, we obtain properties of atomicity, locality and modularity, like the aforementioned decomposition theorems, that are not possible in any other proof-theoretical formalism. We are currently implementing deductive systems in the calculus of structures, starting from those in this paper for system BV [16, 17].

I would like to express the following belief, if the reader forgives me some mysticism. I think that, in the construction of a formal system for computer science, *local* notions of semantics like those that I employed, for example simple laws of conservation, are more important than *global* ones, like a denotational semantics for derivations, say. The global perspective is in the end always enforced by cut elimination. In other words, I believe that we should start from the bottom, from the languages and the properties we want to model, and only later try to make a sense of it all. We can do so because cut elimination would anyway prevent us from making the worst mistakes. By doing this way I have been rewarded by the merge rules and their 'space-temporal' interpretation: it has been an outcome of unexpected beauty.

## Acknowledgments

This work has been supported in part by a Marie Curie TMR grant that I used for a post-doc at INRIA, Nancy, in 1997/8. I started this project in 1994, encouraged by Dale Miller, who taught me a lot about proof theory. Horst Reichel and Enno Sandner listened to me and put interesting questions about preliminary versions of this paper. Giorgio Levi has been very encouraging throughout this endeavour. This work would not be so minimalist in spirit had I not listened to and enjoyed so much the music of Michael Nyman and Philip Glass. Jim Lipton carefully read the introduction and discussed with me several related issues; he has been very helpful. Pietro Di Gianantonio has been a great source of encouragement, especially because he liked this paper; he directly inspired the splitting theorem. Alwen Tiu, Janis Voigtländer, Kai Brünnler, Lutz Straßburger, Paola Bruscoli and Steffen Hölldobler checked many details; Alwen, Kai and Lutz also found several mistakes in early versions of the cut elimination argument, and this eventually led me to discover splitting. I warmly thank Christian Retoré for having introduced me to the fascinating problem of non-commutativity in linear logic; he deeply influenced my ideas on this subject. Finally, I thank Roy Dyckhoff and François Lamarche for the unbelievable amount of work they dedicated to my paper; its present shape owes much to their careful reading and checking.

# Index of Symbols

Typeset with Xy-pic